\documentclass[10pt,twocolumn,letterpaper]{article}

\usepackage[pagenumbers]{iccv} % To force page numbers, e.g. for an arXiv version

\usepackage{multirow}
\usepackage{graphicx}
\usepackage{amsmath}
\usepackage{color} 
\usepackage{colortbl}
\usepackage{booktabs}
\usepackage{arydshln}
\usepackage{tabularx}
\usepackage{adjustbox}
\usepackage{makecell}
\usepackage{amssymb}
\usepackage{amsmath}
\usepackage{amsthm}

\newtheorem{lemma}{Lemma}

\newcommand{\red}[1]{{\color{red}#1}}

\newlength{\vspacelength}
\definecolor{iccvblue}{rgb}{0.21,0.49,0.74}
\usepackage[pagebackref,breaklinks,colorlinks,allcolors=iccvblue]{hyperref}

\def\paperID{6212} % *** Enter the Paper ID here
\def\confName{ICCV}
\def\confYear{2025}

\title{Uncover Treasures in DCT: Advancing JPEG Quality Enhancement by Exploiting Latent Correlations}

\def\spaces{~~~~~~}
\author{Jing~Yang\thanks{Equal contribution. ~~\textsuperscript{\dag}Corresponding author.}\spaces{}Qunliang~Xing\footnotemark[1]\spaces{}Mai~Xu\(^{\dag}\)\spaces{}Minglang~Qiao\\
Beihang University}

\begin{document}
% \pagecolor{black}
% \color{white}
\maketitle
\begin{abstract} 
    Joint Photographic Experts Group (JPEG) achieves data compression by quantizing Discrete Cosine Transform (DCT) coefficients, which inevitably introduces compression artifacts. Most existing JPEG quality enhancement methods operate in the pixel domain, suffering from the high computational costs of decoding. Consequently, direct enhancement of JPEG images in the DCT domain has gained increasing attention. However, current DCT-domain methods often exhibit limited performance. To address this challenge, we identify two critical types of correlations within the DCT coefficients of JPEG images. Building on this insight, we propose an Advanced DCT-domain JPEG Quality Enhancement (AJQE) method that fully exploits these correlations. The AJQE method enables the adaptation of numerous well-established pixel-domain models to the DCT domain, achieving superior performance with reduced computational complexity. Compared to the pixel-domain counterparts, the DCT-domain models derived by our method demonstrate a 0.35 dB improvement in PSNR and a 60.5\% increase in enhancement throughput on average.
\end{abstract}
    
\begin{figure}
  \centering
  \includegraphics[trim={28mm 4mm 39mm 24.5mm}, clip, width=\linewidth]{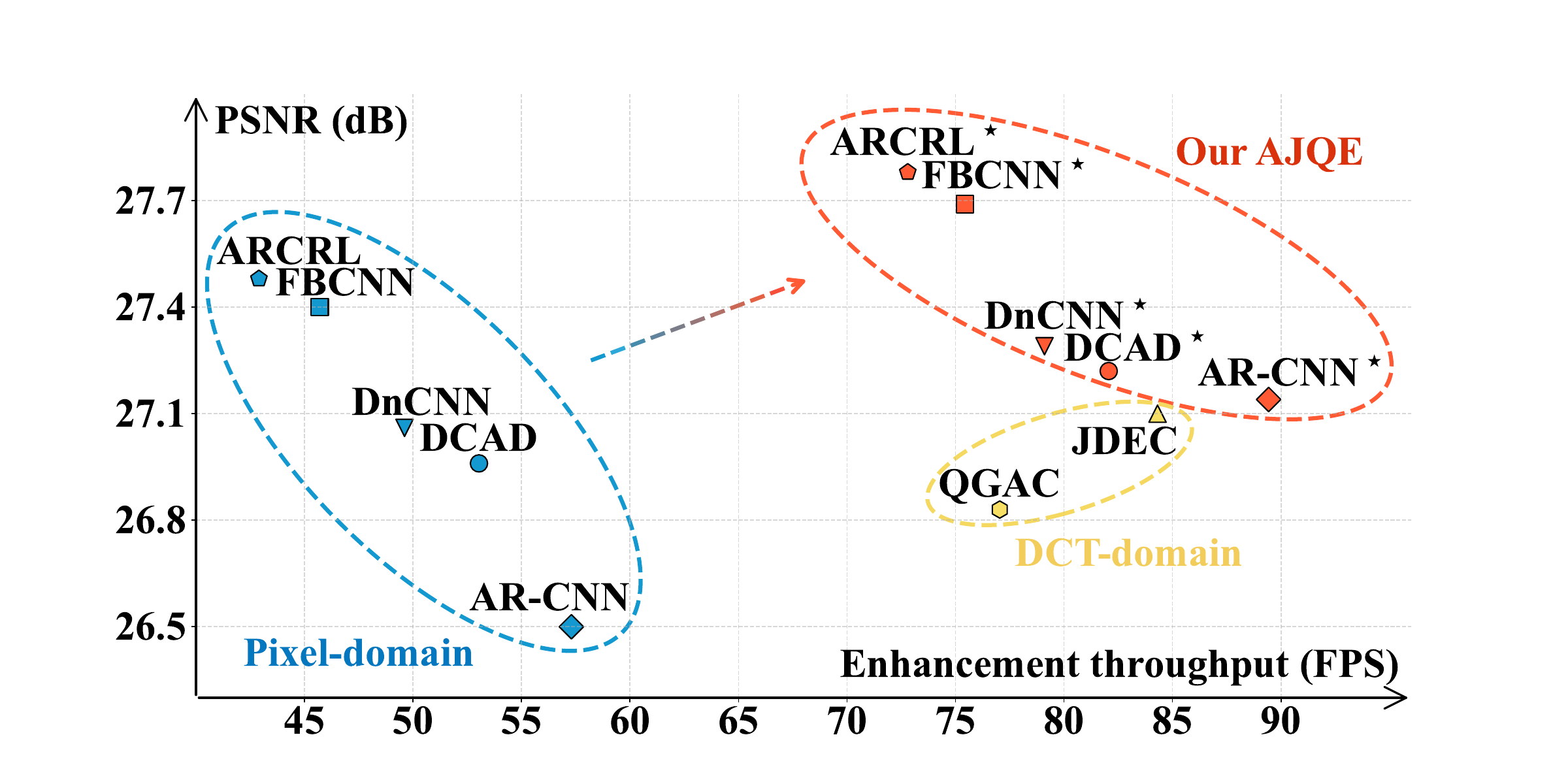}
  \includegraphics[trim={12mm 1mm 13mm 14mm}, clip, width=\linewidth]{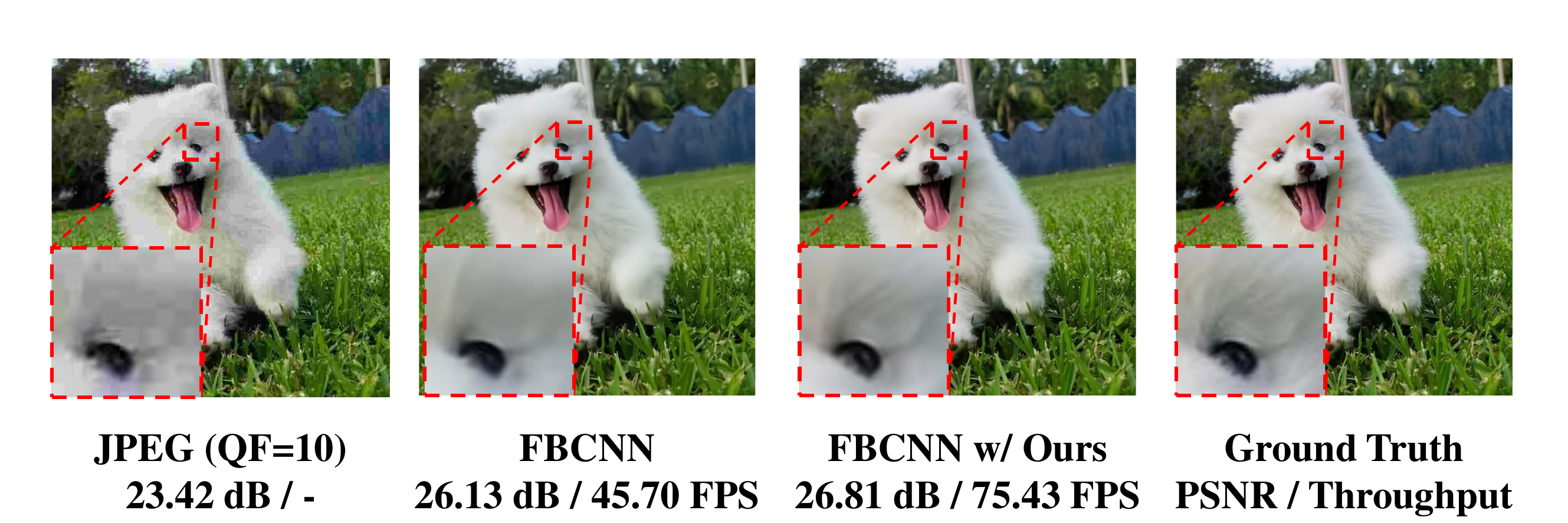}
  \caption{\textbf{Enhancement quality vs. efficiency among pixel-domain, DCT-domain, and our AJQE method\textsuperscript{*}.}
  Comparison includes the latest JDEC~\cite{Han_2024_CVPR}, FBCNN~\cite{Jiang_2021_ICCV}, and more.
  Top: Objective performance averaged over the BSDS500 dataset~\cite{arbelaezContourDetectionHierarchical2011a} with Quality Factor (QF) being 10.
  Bottom: Subjective performance.}
  \label{fig:Figure1}
\end{figure}

\section{Introduction}
\label{sec:introduction}

Joint Photographic Experts Group (JPEG)~\cite{wallaceJPEGStillPicture1991} remains one of the most widely used image compression standards today.
As of March 2025, more than 75\% of websites rely on JPEG as their primary image format, according to W\(^3\)Techs statistics~\cite{w3techs2025image}.
JPEG's efficiency in reducing file sizes while maintaining visually acceptable image quality is crucial for storage and transmission.
The core strength of JPEG lies in its use of the Discrete Cosine Transform (DCT)~\cite{ahmedDiscreteCosineTransform1974} to convert image pixel values into the DCT domain, followed by quantization to achieve effective compression.
However, this quantization introduces compression artifacts, such as blocking, blurring, and ringing, which can degrade the Quality of Experience (QoE) for users~\cite{seshadrinathanStudySubjectiveObjective2010,tanVideoQualityEvaluation2016} and impair downstream performance in various applications~\cite{Fu_2021_ICCV,Parmar_2022_CVPR}.

To mitigate compression artifacts, numerous methods have been developed~\cite{dongCompressionArtifactsReduction2015a,
Wang_2016_CVPR,
xingEarlyExitNot2020a,jiangLearningParallaxTransformer2022,wangJPEGArtifactsRemoval2022,xingDAQEEnhancingQuality2023}.
Mainstream methods typically take JPEG-decoded image pixels as input and enhance them in the pixel domain.
However, since JPEG quantization occurs in the DCT domain, resulting in quantized DCT coefficients stored as lossy bitstream on disk, recent methods~\cite{ehrlichQuantizationGuidedJPEG2020a,Han_2024_CVPR,ouyangJPEGQuantizedCoefficient2024} have emerged that directly enhance JPEG images in the DCT domain.
These methods claim superior performance over pixel-domain methods.

Unfortunately, we observe significant performance limitations in these DCT-domain methods, as shown in \cref{fig:Figure1}.
One reason is that image pixel values exhibit strong spatial continuity due to the texture continuity within each object~\cite{rudermanStatisticsNaturalImages1993}.
However, the DCT operation transforms these spatially-correlated pixels into spatially-decorrelated DCT coefficients using orthogonal basis functions~\cite{ahmedDiscreteCosineTransform1974}, resulting in weak correlations among the DCT coefficients.
Consequently, existing DCT-domain methods, which rely on spatially-invariant convolution filters and spatial attention operators, struggle to process these weakly-correlated DCT coefficients, leading to suboptimal enhancement performance.

To address these limitations, we propose an Advanced DCT-domain JPEG Quality Enhancement (AJQE) method.
We begin by demonstrating the presence of two types of correlations within the DCT coefficients: block-based and point-based correlations, as detailed in \cref{sec:observation}.
Building on this observation, we design two enhancement modules---the Enhancement Module powered by Block-based Correlation (EMBC) and the Enhancement Module powered by Point-based Correlation (EMPC)---to capture these correlations in the DCT domain.
With our proposed EMBC and EMPC modules, we can adapt numerous well-established pixel-domain models into DCT-domain models, enabling them to perform JPEG enhancement directly in the DCT domain and improving their enhancement performance.
Moreover, by eliminating the need to decode JPEG images into the pixel domain during enhancement, our method bypasses the costly decoding process and significantly increases enhancement throughput.
As illustrated in \cref{fig:Figure1}, the DCT-domain models derived by our method demonstrate superior image quality and enhancement efficiency compared to their original pixel-domain counterparts.

In summary, our contributions are as follows:
\begin{itemize}
  \item {We identify two types of correlations within the DCT coefficients of JPEG images.
  These correlations are essential for enhancing the weakly-correlated DCT coefficients using prevalent convolution and attention operators.}
  \item {We propose the AJQE method with two enhancement modules that effectively extract and utilize these correlations.
  With these modules, our AJQE method facilitates the adaptation of numerous well-established pixel-domain models into their DCT-domain counterparts, significantly improving their performance.}
  \item {We conduct extensive experiments and demonstrate that the DCT-domain models derived by our method achieve a 0.35 dB improvement in PSNR and a 60.5\% increase in enhancement throughput on average compared to their pixel-domain counterparts.}
\end{itemize}

\section{Related Work}
\label{sec:related}

\textbf{Pixel-domain JPEG quality enhancement.}
The majority of methods enhance JPEG image quality in the pixel domain by leveraging neural networks and their powerful nonlinear learning capabilities~\cite{dongCompressionArtifactsReduction2015a,zhangGaussianDenoiserResidual2017,Liu_2018_CVPR_Workshops,DBLP:conf/iclr/ZhangLLZF19,Fu_2019_ICCV,xingEarlyExitNot2020a,Jiang_2021_ICCV,wangUformerGeneralUShaped2022,jiangLearningParallaxTransformer2022,zhaoComprehensiveDelicateEfficient2023,wangJPEGArtifactsRemoval2022}.
Dong~\etal~\cite{dongCompressionArtifactsReduction2015a} pioneered learning-based methods for JPEG quality enhancement, utilizing a lightweight four-layer neural network.
Zhang~\etal~\cite{zhangGaussianDenoiserResidual2017} further advanced this field by incorporating residual learning~\cite{He_2016_CVPR} and batch normalization~\cite{ioffeBatchNormalizationAccelerating2015}, improving both stability and enhancement performance.
Xing~\etal~\cite{xingEarlyExitNot2020a} addressed the challenge of blind JPEG quality enhancement with unknown compression factors by introducing a dynamic resource-efficient network that progressively enhances images through an early-exit mechanism.
Most recently, Jiang~\etal~\cite{Jiang_2021_ICCV} proposed a blind enhancement method that predicts quantization factors (QFs), effectively balancing artifact removal with detail preservation.

\textbf{DCT-domain JPEG quality enhancement.}
While the above methods focus exclusively on the pixel (RGB) domain, other methods consider the DCT domain, recognizing that JPEG images are stored as lossy bitstreams containing DCT coefficients.
Dual-domain methods~\cite{guoBuildingDualDomainRepresentations2016,Wang_2016_CVPR,zhangDmcnnDualDomainMultiScale2018a,Chen_2018_CVPR_Workshops} combine information from both the pixel and DCT domains to facilitate the enhancement of JPEG images.
Ehrlich~\etal~\cite{ehrlichQuantizationGuidedJPEG2020a} proposed enhancing DCT coefficients of JPEG images with the assistance of a quantization-matrix parameterization network.
More recently, Han~\etal~\cite{Han_2024_CVPR} proposed transforming DCT coefficients into enhanced RGB values using a cosine spectrum estimator as a ``decoder'', bypassing the computationally expensive JPEG decoding process.
However, we observe significant performance limitations in these methods, as they struggle to effectively handle weakly-correlated DCT coefficients.
In this paper, we propose a novel DCT-domain method that overcomes this limitation by capturing two critical correlations, allowing the adaptation of numerous pixel-domain models into their DCT-domain counterparts and yielding superior performance.
We shall delve into these correlations in the following sections.

\begin{table}[t]
    \centering
    \tiny
    \renewcommand{\arraystretch}{1.1} % Adjust row spacing
    \begin{adjustbox}{width=.75\linewidth}
    \begin{tabular}{@{\extracolsep{0pt}} c | c c | c c@{}}
        \Xhline{0.7pt}
        \multirow{2}{*}{Metric} & \multicolumn{2}{c|}{DIV2K} & \multicolumn{2}{c}{BSDS500} \\
        \cline{2-5}
        & Pixel & DCT & Pixel & DCT \\
        \hline
        \textit{MI}\(\uparrow\) & 0.945 & 0.007 & 0.915 & 0.006 \\
        \textit{GC}\(\downarrow\) & 0.055 & 1.003 & 0.085 & 0.999 \\  
        \Xhline{0.7pt}
    \end{tabular}
    \end{adjustbox}
    \caption{\textbf{Correlations within pixel values and DCT coefficients.}
    Note that \textit{MI} ranges from \(-1\) to \(1\), where values close to \(1\) and \(0\) represent strong positive and negligible correlations, respectively.
    Conversely, \textit{GC} ranges from \(0\) to \(2\), where values close to \(0\) and \(1\) represent strong positive and negligible correlations, respectively.}
    \label{tab:corre}
\end{table}

\section{Correlations within DCT Coefficients}
\label{sec:observation}

\begin{figure}[t]
    \centering
    \begin{subfigure}{\linewidth}
        \centering
        \includegraphics[trim={3mm -8mm 5mm 5mm}, clip, width=\linewidth]{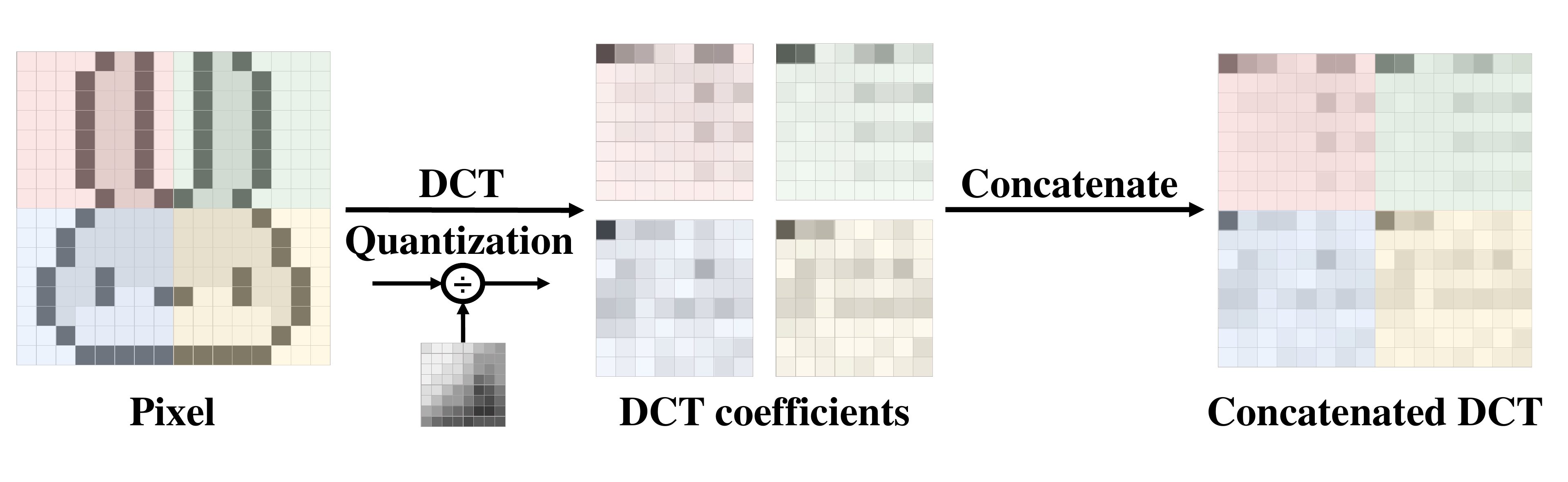}
        \caption{Transformation from pixel values to concatenated DCT coefficients.}
        \label{fig:pixel-contact-rearrange:a}
    \end{subfigure}
    \begin{subfigure}{.61\linewidth}
        \centering
        \includegraphics[trim={18mm -5mm 15mm -30mm}, clip, width=\linewidth]{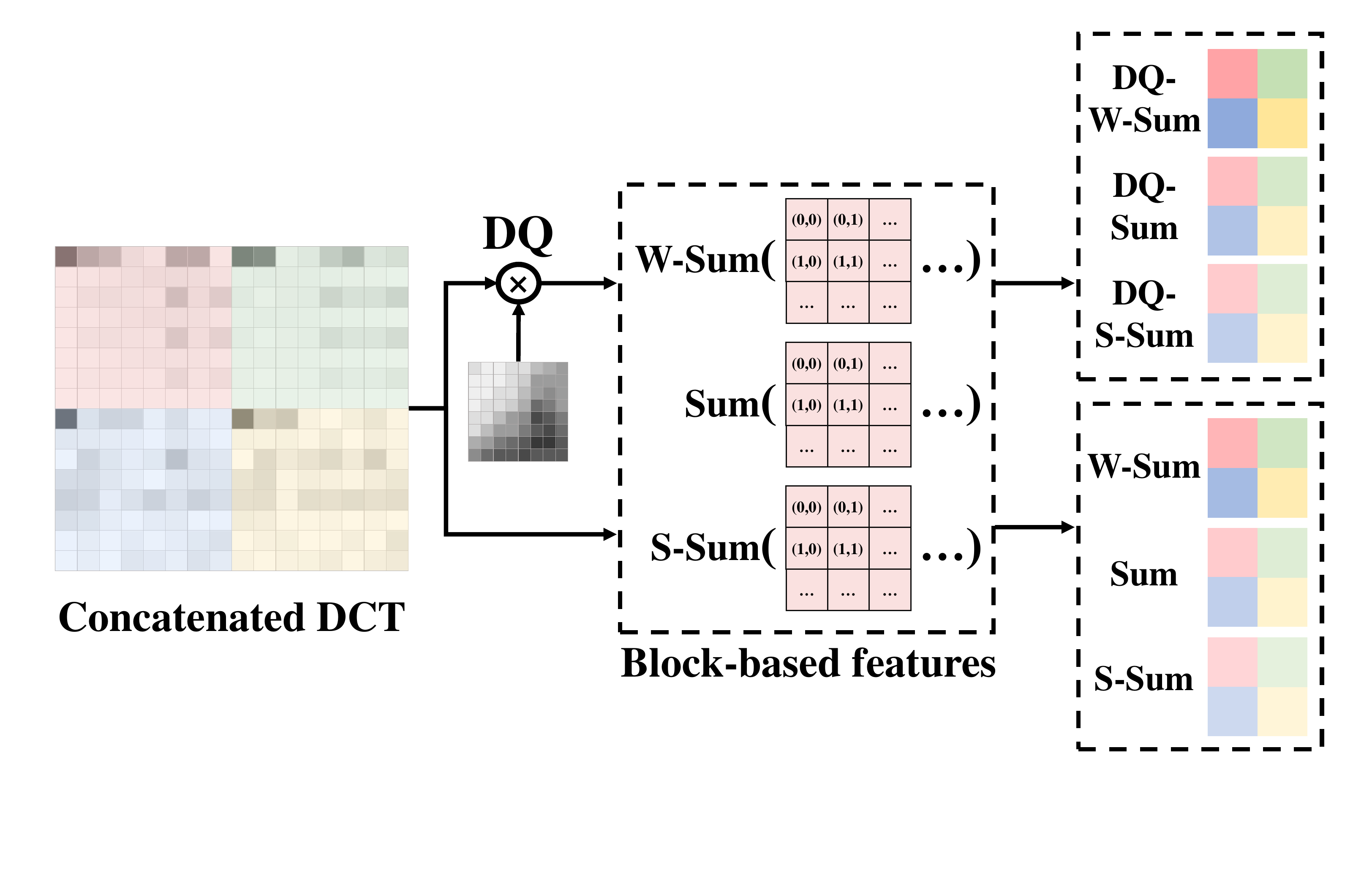}
        \caption{Block-based correlations using block-based features.
        ``DQ'': dequantization.}
        \label{fig:pixel-contact-rearrange:b}
    \end{subfigure}
    \begin{subfigure}{.38\linewidth}
        \centering
        \includegraphics[trim={10mm -50mm 1mm -50mm}, clip, width=\linewidth]{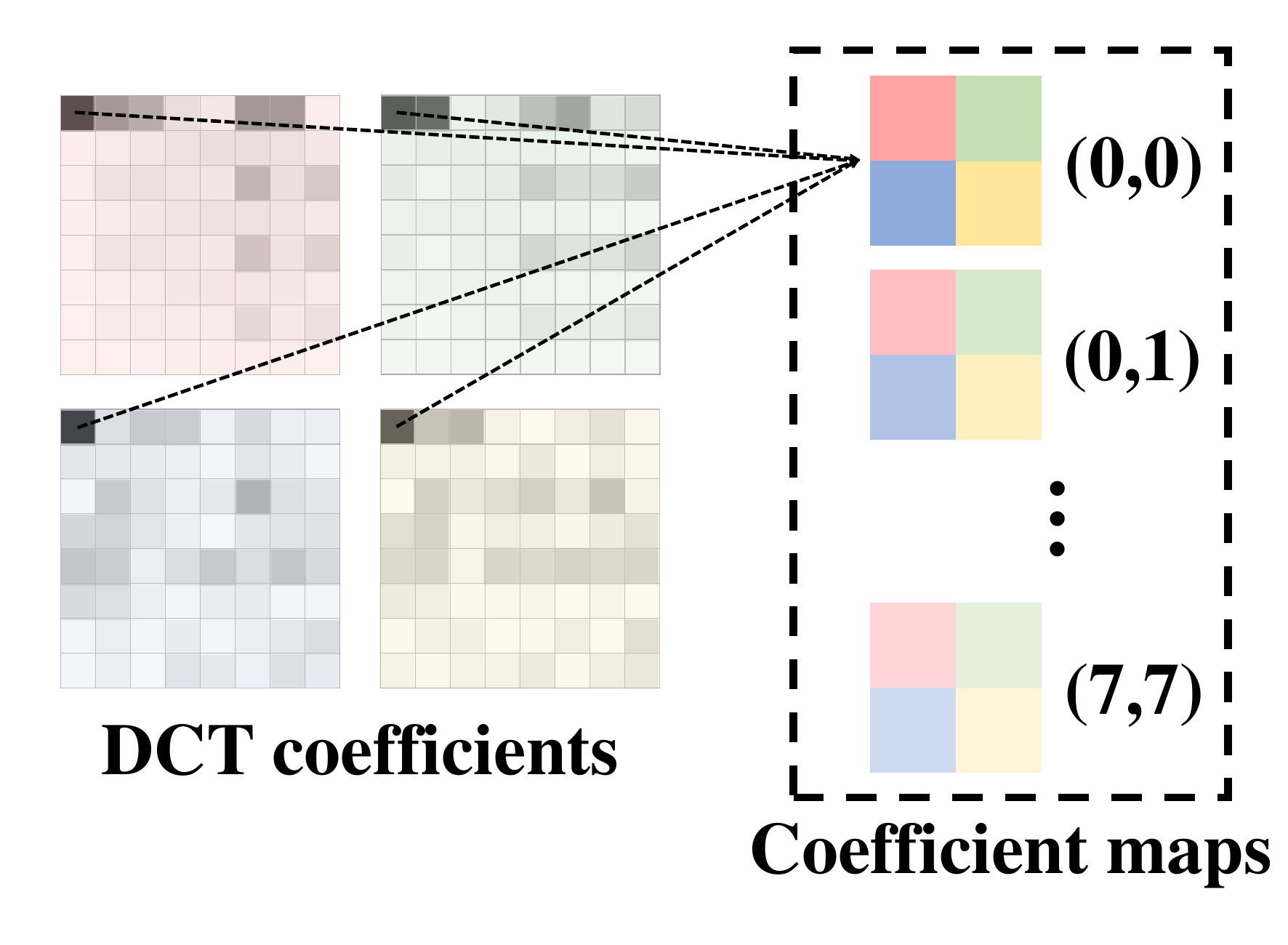}
        \caption{Point-based correlations using coefficient maps.}
        \label{fig:pixel-contact-rearrange:c}
    \end{subfigure}
    \caption{\textbf{Toy examples illustrating different correlations within DCT coefficients.}
    The color intensity per image indicates the strength of the correlation within this image.}
    \label{fig:pixel-contact-rearrange}
\end{figure}

\textbf{Baseline: Spatial correlations within DCT coefficients.}
DCT coefficients exhibit weak spatial correlations due to the block-based orthogonal transform and quantization, as illustrated in \cref{fig:pixel-contact-rearrange:a}.
Specifically, this type of correlation refers to spatial auto-correlation, which reflects the tendency for geographically proximate locations to have similar or dissimilar values~\cite{toblerComputerMovieSimulating1970}.
To quantify the spatial auto-correlation of DCT coefficients, we employ two widely-used metrics, \ie, Moran's I (\textit{MI})~\cite{moranNotesContinuousStochastic1950} and Geary's C (\textit{GC})~\cite{gearyContiguityRatioStatistical1954}.
We conduct our analysis on the DIV2K~\cite{Agustsson_2017_CVPR_Workshops} and BSDS500 datasets~\cite{arbelaezContourDetectionHierarchical2011a}.
Raw images from these datasets are compressed into JPEG images using OpenCV~\cite{bradski2000opencv}, with a QF ranging from 10 to 100.
\cref{tab:corre} presents the results for the luminance component with QF set to 50\footnote{Results for chroma components and other QF values yield consistent observations and are detailed in our supplementary.}.
As shown, pixel values exhibit strong spatial correlations (with \textit{MI} values close to \(1\) and \textit{GC} values close to \(0\)), while DCT coefficients display negligible spatial correlations (with \textit{MI} values close to \(0\) and \textit{GC} values close to \(1\)).
This presents a significant challenge for current DCT-domain enhancement methods, which typically rely on spatial correlations.
To overcome this, we identify two other types of correlations within the DCT coefficients.

\begin{figure}
    \includegraphics[trim={9mm 102mm 22mm 109mm}, clip, width=\linewidth]{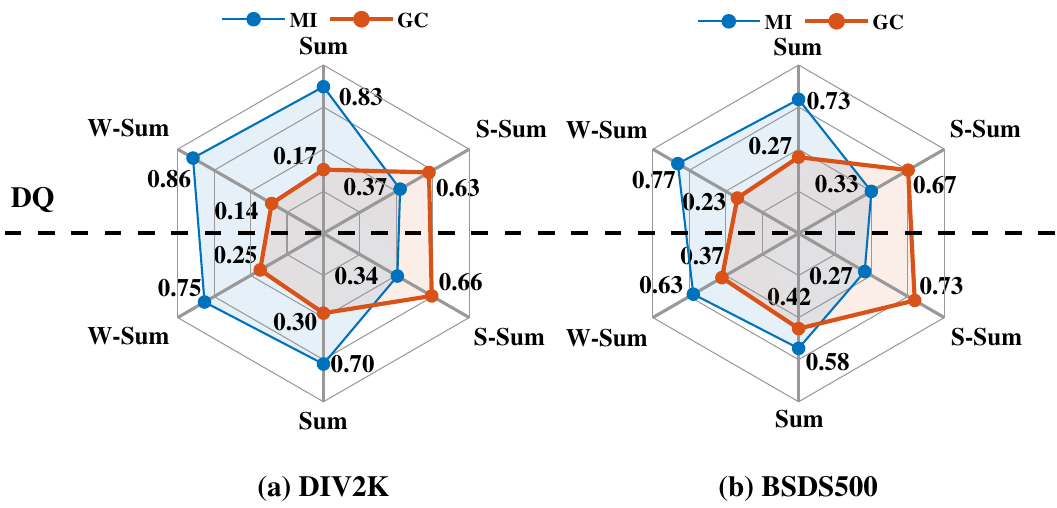}
    \caption{\textbf{Block-based correlations using different block-based features.}
    Upper: DCT blocks are dequantized before calculating feature values.
    Lower: DCT blocks remain quantized.}
    \label{fig:inter-curve}
\end{figure}

\textbf{Block-based correlations within DCT coefficients.}
During JPEG compression, images are partitioned into \(8 \times 8\) blocks before undergoing DCT transform and quantization.
As illustrated in \cref{fig:pixel-contact-rearrange:b}, each DCT block can be represented by a block-based feature.
Specifically, ``Sum'', ``S-Sum'', and ``W-Sum'' refer to the sum (intensity), squared sum (energy), and weighted sum (convolution) of DCT coefficients, respectively.
Considering DCT blocks are represented mainly by low-frequency coefficients, the convolution kernel for ``W-Sum'' is a Gaussian matrix centered at the top-left corner.
Before calculating feature values, each DCT block can either remain quantized or be dequantized by multiplying it with the quantization matrix.
We then evaluate the block-based auto-correlations based on these feature values.
As shown in \cref{fig:inter-curve}, among all types of block-based features, the ``DQ-W-Sum'' feature exhibits the strongest correlations, with an \textit{MI} of 0.86 and a \textit{GC} of 0.14 on the DIV2K dataset.
Similar results are observed on the BSDS500 dataset.
In summary, we identify strong block-based correlations when DCT coefficients are dequantized and summarized for each block, significantly stronger than the spatial correlations.

\begin{figure}[t]
    \centering
    \includegraphics[trim={0mm -10mm 0mm -5mm}, clip, width=.9\linewidth]{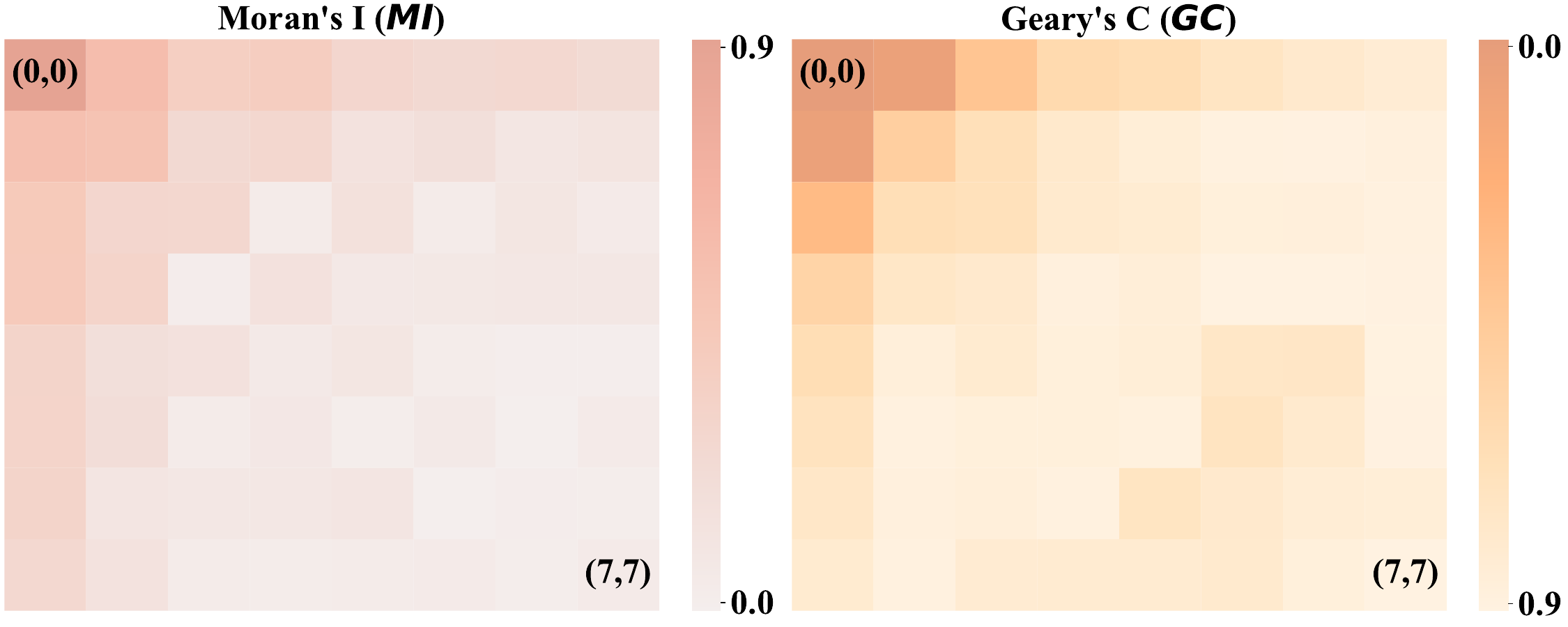}
    \caption{\textbf{Point-based correlations using coefficient maps on the DIV2K dataset.}
    Similar results on the BSDS500 dataset are provided in our supplementary.}
    \label{fig:freq}
\end{figure}

\begin{figure*}[t]
    \centering
    \includegraphics[trim={6mm 13mm 6mm 10mm}, clip, width=\linewidth]{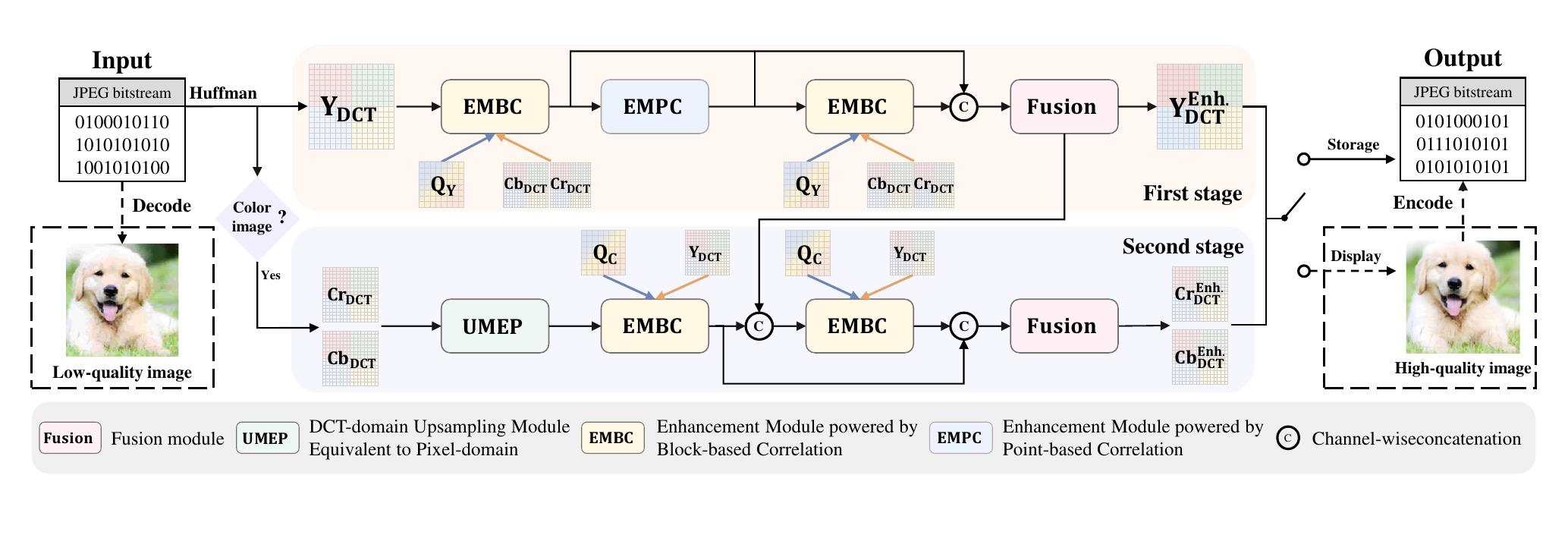}
    \caption{\textbf{Overview of the proposed AJQE method.}
    Our method directly accesses the JPEG bitstream as input, enhances the quality of the DCT coefficients within the bitstream, and outputs the enhanced coefficients.}
    \label{fig:framework}
\end{figure*}

\textbf{Point-based correlations within DCT coefficients.}
Each \(8 \times 8\) DCT block contains 64 points, which represent the spatial frequency components of the block.
As illustrated in \cref{fig:pixel-contact-rearrange:c}, we extract coefficient maps~\cite{Xu_2020_CVPR,ouyangJPEGQuantizedCoefficient2024} from a concatenated DCT image by mapping the same component from each DCT block into a single map.
This results in 64 maps for a concatenated DCT image, corresponding to 64 coefficients per DCT block.
We then evaluate the spatial auto-correlations for these maps, as shown in \cref{fig:freq}.
We refer to this type of correlation as point-based correlations among DCT blocks.
Notably, all coefficient maps demonstrate positive correlations.
Additionally, the correlations among low-frequency (top-left) components are stronger than those among high-frequency (bottom-right) components.
In summary, we identify significant point-based correlations through the formation of coefficient maps, significantly stronger than the spatial correlations.

\section{Advanced DCT-domain JPEG Quality Enhancement (AJQE)}
\label{sec:method}

In this paper, we propose an Advanced DCT-domain JPEG Quality Enhancement (AJQE) method, as showed in \cref{fig:framework}.

\textbf{Input.}
Our method directly accesses the JPEG bitstream from disk, performs entropy decoding (\ie, Huffman decoding) to obtain quantized YCbCr DCT blocks, and then dequantizes them.
Images composed of these blocks are denoted by \(\mathbf{Y}_{\text{DCT}}\), \(\mathbf{Cb}_{\text{DCT}}\), and \(\mathbf{Cr}_{\text{DCT}}\).
In comparison, pixel-domain methods require an additional inverse-DCT transformation to convert these blocks into YCbCr blocks, followed by a color transformation to RGB.
This process is computationally expensive, particularly during training, where millions of decodings are required (accounting for approximately 44\% of JPEG reading time).

\textbf{DCT-domain enhancement.}
The enhancement process is divided into two stages.
In the first stage, \(\mathbf{Y}_{\text{DCT}}\) is enhanced to \(\mathbf{Y}_{\text{DCT}}^{\text{Enh.}}\) utilizing \(\mathbf{Cb}_{\text{DCT}}\), \(\mathbf{Cr}_{\text{DCT}}\), and the quantization matrix for DCT-Y blocks (\(\mathbf{Q}_{\text{Y}}\)).
In the second stage, \(\mathbf{Cb}_{\text{DCT}}\) and \(\mathbf{Cr}_{\text{DCT}}\) are enhanced to \(\mathbf{Cb}_{\text{DCT}}^{\text{Enh.}}\) and \(\mathbf{Cr}_{\text{DCT}}^{\text{Enh.}}\), respectively, utilizing \(\mathbf{Y}_{\text{DCT}}^{\text{Enh.}}\) and the quantization matrix for DCT-CbCr blocks (\(\mathbf{Q}_{\text{C}}\)).
This results in enhanced DCT coefficients: \(\mathbf{Y}_{\text{DCT}}^{\text{Enh.}}\), \(\mathbf{Cb}_{\text{DCT}}^{\text{Enh.}}\), and \(\mathbf{Cr}_{\text{DCT}}^{\text{Enh.}}\).

\textbf{Output.}
With the enhanced DCT coefficients, our method can directly store the results in a JPEG file by applying entropy encoding.
This capability is highly efficient for storage and transmission, with JPEG being used by more than 75\% of websites~\cite{w3techs2025image}.
For display, the coefficients can be further decoded into the RGB space.

In the following sections, we detail the two modules used for DCT-domain enhancement: the Enhancement Module powered by Block-based Correlation (EMBC) in \cref{sec:method:embc} and the Enhancement Module powered by Point-based Correlation (EMPC) in \cref{sec:method:empc}.
These modules are built upon our findings and are effective in capturing correlations within DCT coefficients.
Additionally, we explain the fusion mechanism of the output features from these modules in \cref{sec:method:fusion} and the upsampling mechanism for DCT-CbCr blocks: the DCT-domain Upsampling Module Equivalent to Pixel-domain (UMEP) in \cref{sec:method:upsampling}.

\begin{figure}[t]
    \centering
    \includegraphics[trim={49mm 27mm 52mm 26mm}, clip, width=\linewidth]{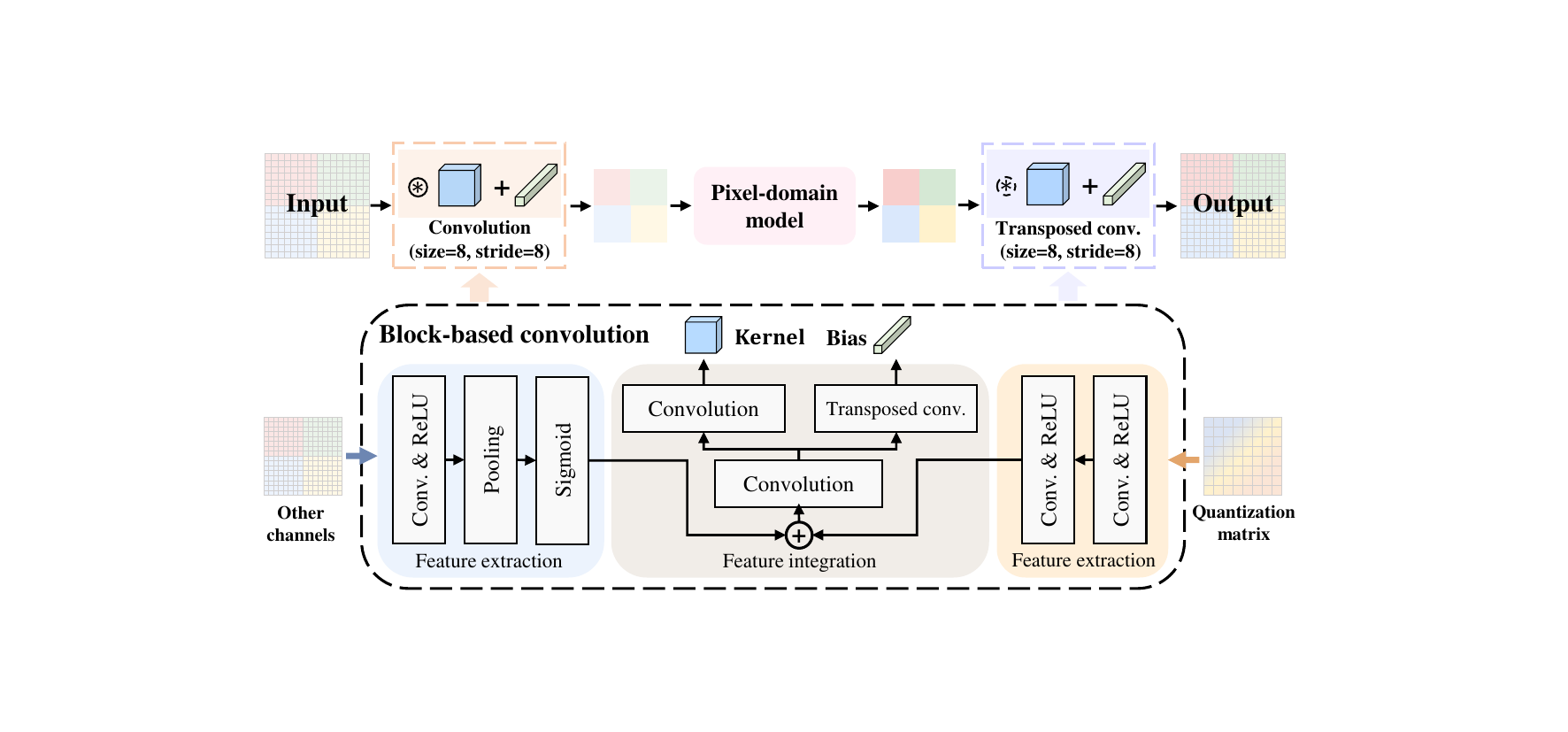}
    \caption{\textbf{Architecture of EMBC.}}
    \label{fig:EMBC}
\end{figure}

\subsection{Enhancement Module powered by Block-based Correlation (EMBC)}
\label{sec:method:embc}

As identified in \cref{sec:observation}, strong block-based correlations exist when DCT coefficients are dequantized and weighted-summarized for each block.
This process can be viewed as a block-based convolution~\cite{ehrlichQuantizationGuidedJPEG2020a,deguerreFastObjectDetection2019a} to the dequantized DCT blocks.
Building on this concept, we make the kernel and bias learnable parameters, which are generated through a few convolutional layers.

As illustrated in \cref{fig:framework,fig:EMBC}, when enhancing the quality of \(\mathbf{Y}_{\text{DCT}}\), the block-based convolution employs \(\mathbf{Q}_{\text{Y}}\), \(\mathbf{Cb}_{\text{DCT}}\), and \(\mathbf{Cr}_{\text{DCT}}\) to generate its kernel and bias.
Note that the block-based convolutions of EMBCs in the first stage learn independent kernels and biases, which are generated through the same process.
Similarly, in the second stage, the block-based convolution utilizes \(\mathbf{Q}_{\text{C}}\) and \(\mathbf{Y}_{\text{DCT}}\) to produce its kernel and bias.
This block-based convolution effectively extracts the block-based correlations in a learnable manner.
Consequently, we apply an existing pixel-domain model following the block-based convolution.
Finally, a second block-based convolution is applied to adapt the pixel-domain enhancement results back to the DCT domain, completing the inference of EMBC.

\subsection{Enhancement Module powered by Point-based Correlation (EMPC)}
\label{sec:method:empc}

In addition to block-based correlations, we have also identified significant point-based correlations among the DCT coefficients, as detailed in \cref{sec:observation}.
Recall that we extract 64 coefficient maps by mapping the same component from each DCT block into a single map.
These coefficient maps exhibit positive spatial correlations, making them suitable for enhancement using pixel-domain models.

\begin{figure}[t]
    \centering
    \includegraphics[trim={75mm 38mm 68mm 24mm}, clip, width=\linewidth]{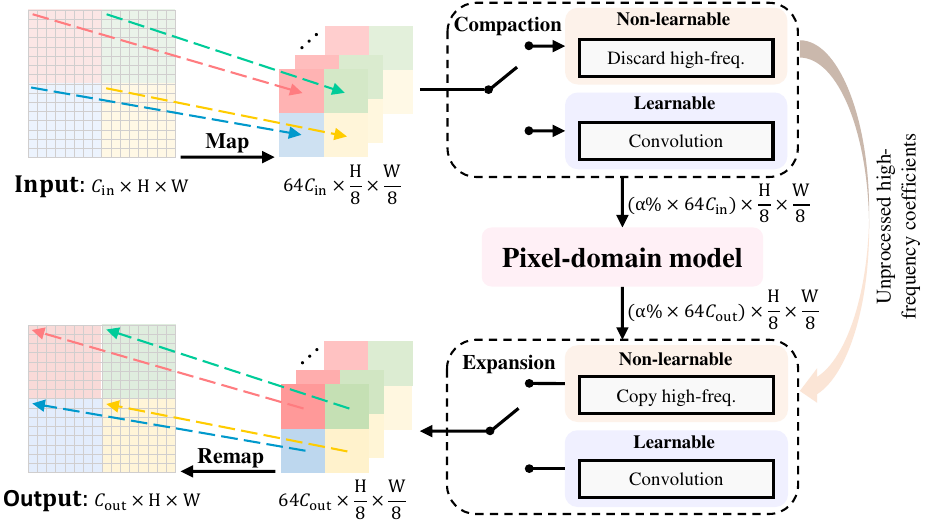}
    \caption{\textbf{Architecture of EMPC.}}
    \label{fig:EMPC}
\end{figure}

As illustrated in \cref{fig:EMPC}, we first generate coefficient maps for the input feature.
To mitigate the computational burden posed by the large number of coefficient maps (\eg, \(32 \times 64 = 2048\) maps for a 32-channel input feature), we apply one of two compaction strategies:
\begin{itemize}
    \item[1)] \textbf{Non-learnable:} Retain the top \( \alpha \)\% low-frequency coefficient maps and discard the rest.
    This is effective because low-frequency maps exhibit stronger spatial correlations than high-frequency ones, as observed in our findings.
    \item[2)] \textbf{Learnable:} Generate \( \lfloor \alpha\text{\%} \times 64 C_{\text{in}} \rfloor \) maps for a \(C_{\text{in}}\)-channel input through a few convolutional layers that are trained in an end-to-end manner.
    This approach may be more advantageous, as high-frequency maps also exhibit positive correlations that can be exploited, as noted in our findings.
\end{itemize}
Once compacted, we use a pixel-domain model to capture correlations and enhance the quality of these lossy coefficient maps.
Finally, we expand and remap these maps to restore the original number and spatial structure of coefficients, completing the inference of EMPC.

\subsection{Feature Fusion Module}
\label{sec:method:fusion}

As shown in \cref{fig:framework}, an EMPC and multiple EMBCs are employed to enhance the quality of DCT coefficients.
The outputs from these modules are integrated by a feature fusion module to produce the final coefficients: \(\mathbf{Y}_{\text{DCT}}^{\text{Enh.}}\), \(\mathbf{Cb}_{\text{DCT}}^{\text{Enh.}}\), and \(\mathbf{Cr}_{\text{DCT}}^{\text{Enh.}}\).
Specifically, the outputs from these modules are concatenated and sent to the fusion module \( f_{\theta} \), defined as:
\begin{align}
    f_{\theta }: \mathbb{R}^{N \times C_{\text{hid}} \times H \times W} \mapsto \mathbb{R}^{3 \times H \times W},
\end{align}
where \( N \) represents the number of fused outputs, \( C_{\text{hid}} \) denotes the channel number of each output, and \( H \) and \( W \) are the height and width of the DCT coefficients (and also the image).
Our fusion module can be implemented using either a residual block~\cite{He_2016_CVPR} or a series of convolution layers.

\subsection{DCT-domain Upsampling Module Equivalent to Pixel-domain (UMEP)}
\label{sec:method:upsampling}

\begin{table*}[t]
    \centering
    \begin{adjustbox}{max width=.95\linewidth}
        \begin{tabular}{c | c | c | c c c c c | c c }
            Dataset & QF & JPEG & AJQE w/ AR-CNN & AJQE w/ DCAD & AJQE w/ DnCNN & AJQE w/ FBCNN & AJQE w/ ARCRL & QGAC & JDEC \\
            \Xhline{1.2pt}
            \multicolumn{10}{c}{PSNR}\\
            \hline
            \multirow{4}{*}{BSDS500} & 10 & 25.47 & 27.14 \red{(+0.64)} & 27.22 \red{(+0.26)} & 27.29 \red{(+0.23)} & 27.69 \red{(+0.29)} & \textbf{27.78} \red{(+0.30)} & 26.83 & 27.11\\
            & 20 & 27.83 & 29.65 \red{(+0.76)} & 29.74 \red{(+0.31)} & 29.77 \red{(+0.23)} & 30.17 \red{(+0.30)} & \textbf{30.24} \red{(+0.32)} & 29.25 & 29.53\\
            & 30 & 29.19 & 31.02 \red{(+0.74)} & 31.12 \red{(+0.28)} & 31.25 \red{(+0.36)} & \textbf{31.29} \red{(+0.05)} & 31.28 \red{(+0.02)} & 30.54 & 30.95\\
            & 40 & 30.16 & 32.08 \red{(+0.88)} & 32.30 \red{(+0.46)} & 32.18 \red{(+0.22)} & 32.41 \red{(+0.22)} & \textbf{32.49} \red{(+0.32)} & 31.42 & 31.96\\
            \hline
            \multirow{4}{*}{LIVE-1} & 10 & 25.35 & 27.05 \red{(+0.60)} & 27.09 \red{(+0.14)} & 27.18 \red{(+0.20)} & 27.86 \red{(+0.29)} & \textbf{27.98} \red{(+0.39)} & 26.95 & 27.26\\
            & 20 & 27.76 & 29.65 \red{(+0.78)} & 29.74 \red{(+0.26)} & 29.86 \red{(+0.24)} & 30.22 \red{(+0.21)} & \textbf{30.36} \red{(+0.28)} & 29.28 & 29.64\\
            & 30 & 29.14 & 31.06 \red{(+0.79)} & 31.19 \red{(+0.27)} & 31.26 \red{(+0.42)} & 31.55 \red{(+0.16)} & \textbf{31.63} \red{(+0.21)} & 30.58 & 30.96\\
            & 40 & 30.10 & 32.08 \red{(+0.90)} & 32.32 \red{(+0.41)} & 32.30 \red{(+0.25)} & 32.46 \red{(+0.14)} & \textbf{32.61} \red{(+0.17)} & 31.42 & 32.05\\
            \hline
            \multirow{4}{*}{ICB} & 10 & 29.04 & 31.17 \red{(+0.90)} & 31.27 \red{(+0.42)} & 31.36 \red{(+0.30)} & 32.26 \red{(+0.24)} & \textbf{32.28} \red{(+0.32)} & 31.72 & 31.86\\
            & 20 & 32.09 & 33.94 \red{(+0.73)} & 34.04 \red{(+0.19)} & 34.26 \red{(+0.20)} & \textbf{34.82} \red{(+0.22)} & 34.80 \red{(+0.17)} & 34.11 & 34.39\\
            & 30 & 33.55 & 35.00 \red{(+0.34)} & 35.63 \red{(+0.39)} & 35.59 \red{(+0.44)} & \textbf{36.05} \red{(+0.22)} & 35.97 \red{(+0.12)} & 35.19 & 35.42\\
            & 40 & 34.59 & 36.42 \red{(+0.83)} & 36.56 \red{(+0.34)} & 36.47 \red{(+0.14)} & \textbf{36.82} \red{(+0.18)} & 36.80 \red{(+0.19)} & 35.98 & 36.22\\
            \hline
            \Xhline{1.2pt}
            \multicolumn{10}{c}{SSIM}\\
            \hline
            \multirow{4}{*}{BSDS500} & 10 & 0.738 & 0.793 \red{(+.023)} & 0.796 \red{(+.011)} & 0.798 \red{(+.010)} & 0.810 \red{(+.010)} & \textbf{0.813} \red{(+.015)} & 0.783 & 0.797\\
            & 20 & 0.826 & 0.870 \red{(+.020)} & 0.872 \red{(+.010)} & 0.873 \red{(+.009)} & 0.876 \red{(+.007)} & \textbf{0.879} \red{(+.006)} & 0.856 & 0.862\\
            & 30 & 0.864 & 0.900 \red{(+.016)} & 0.902 \red{(+.008)} & 0.904 \red{(+.007)} & \textbf{0.905} \red{(+.005)} & 0.904 \red{(+.002)} & 0.887 & 0.898\\
            & 40 & 0.886 & 0.919 \red{(+.017)} & 0.922 \red{(+.010)} & 0.919 \red{(+.005)} & \textbf{0.923} \red{(+.006)} & 0.922 \red{(+.003)} & 0.905 & 0.915\\
            \hline
            \multirow{4}{*}{LIVE-1} & 10 & 0.745 & 0.802 \red{(+.032)} & 0.804 \red{(+.009)} & 0.806 \red{(+.009)} & 0.813 \red{(+.003)} & \textbf{0.819} \red{(+.012)} & 0.790 & 0.801\\
            & 20 & 0.827 & 0.874 \red{(+.021)} & 0.876 \red{(+.010)} & 0.876 \red{(+.008)} & 0.881 \red{(+.006)} & \textbf{0.889} \red{(+.005)} & 0.858 & 0.872\\
            & 30 & 0.864 & 0.902 \red{(+.016)} & 0.904 \red{(+.008)} & 0.905 \red{(+.011)} & 0.920 \red{(+.002)} & \textbf{0.922} \red{(+.006)} & 0.888 & 0.901\\
            & 40 & 0.884 & 0.920 \red{(+.018)} & 0.923 \red{(+.011)} & 0.920 \red{(+.006)} & 0.923 \red{(+.005)} & \textbf{0.924} \red{(+.007)} & 0.904 & 0.915\\
            \hline
            \multirow{4}{*}{ICB} & 10 & 0.800 & 0.862 \red{(+.015)} & 0.864 \red{(+.009)} & 0.864 \red{(+.006)} & 0.875 \red{(+.007)} & \textbf{0.879} \red{(+.014)} & 0.859 & 0.862\\
            & 20 & 0.864 & 0.903 \red{(+.013)} & 0.904 \red{(+.008)} & 0.906 \red{(+.008)} & \textbf{0.912} \red{(+.010)} & 0.911 \red{(+.010)} & 0.895 & 0.900\\
            & 30 & 0.891 & 0.919 \red{(+.009)} & 0.924 \red{(+.010)} & 0.923 \red{(+.012)} & 0.924 \red{(+.006)} & \textbf{0.925} \red{(+.009)} & 0.912 & 0.916\\
            & 40 & 0.906 & 0.933 \red{(+.012)} & 0.934 \red{(+.009)} & 0.933 \red{(+.007)} & \textbf{0.935} \red{(+.007)} & 0.931 \red{(+.007)} & 0.922 & 0.925\\
            \hline
        \end{tabular}
    \end{adjustbox}
    \caption{\textbf{Quantitative performance of color JPEG image quality enhancement.}
    Performance improvements over the pixel-domain model are shown in parentheses.
    The best results are boldfaced.
    PSNR-B results are provided in the supplementary materials.}
    \label{tab:color_result}
\end{table*}

Our method is designed for quality enhancement of both grayscale and color JPEG images.
For grayscale images, only the first stage in \cref{fig:framework} is used, where EMBCs are parameterized only by \(\mathbf{Q}_{\text{Y}}\) to perform block-based convolution.
For color images, the second stage is also applied to enhance DCT-CbCr quality.
Since the chroma components are downsampled in JPEG, we need to upsample DCT-CbCr blocks to the same resolution as DCT-Y blocks.
However, DCT blocks consist of spatially-decorrelated coefficients, which are not suitable for spatial upsampling.
To address this, we propose a DCT-domain upsampling module that yields equivalent results to pixel-domain upsampling.

\begin{lemma}
    Let \( \hat{\mathbf{B}}_{N} \) represent an \( N \times N \) DCT block;
    \(\mathbf{U}_{N}\) denote the pixel-domain upsampling (\eg, nearest-neighbor) matrix of shape \( 2N \times N \);
    and \(\mathbf{T}_{N}\) represent the 1-D DCT basis~\cite{ahmedDiscreteCosineTransform1974} of shape \( N \times N \).
    Then, the upsampled DCT block \( \hat{\mathbf{B}}_{2N} \) can be obtained by:
    \begin{equation}
        \hat{\mathbf{B}}_{2N} = \left( \mathbf{T}_{2N} \mathbf{U}_{N} \mathbf{T}_{N}^{\top} \right) \hat{\mathbf{B}}_{N} \left( \mathbf{T}_{2N} \mathbf{U}_{N} \mathbf{T}_{N}^{\top} \right)^{\top}.
    \end{equation}
    Detailed derivations and proofs are provided in the supplementary material.
    \label{eq:DCT_upsampling}
\end{lemma}

Using \cref{eq:DCT_upsampling}, we can upsample each \( 4 \times 4 \) DCT block by a DCT-domain upsampling matrix \( \hat{\mathbf{U}}_{4} = \mathbf{T}_{8} \mathbf{U}_{4} \mathbf{T}_{4}^{\top} \), such that we can achieve equivalent results to pixel-domain upsampling by \( \mathbf{U}_{4} \).
However, in JPEG, there exist only \( 8 \times 8 \) chroma DCT blocks;
this is because four neighboring downsampled \( 4 \times 4 \) chroma pixel blocks are combined and transformed into a single \( 8 \times 8 \) chroma DCT block.
To address this, we adopt the sub-block conversion method~\cite{salazarComplexityScalableUniversal2007} to obtain four \( 4 \times 4 \) chroma DCT blocks from each \( 8 \times 8 \) chroma DCT block.
As a result, we can apply our DCT-domain upsampling as in \cref{eq:DCT_upsampling} to these converted \( 4 \times 4 \) blocks.
Detailed derivations and proofs are provided in the supplementary material.
Finally, we successfully upsample chroma DCT blocks with spatial upsampling operators in our UMEP.

\textbf{Discussion.}
Our AJQE method demonstrates broad applicability to existing pixel-domain models and also significantly decreasing computational burden.
Notably, the pixel-domain model in EMBC and EMPC is operating a reduced input size by 64\(\times\) compared to the original pixel-domain method.
To further enhance effectiveness, the adapted pixel-domain model is fine-tuned within our method.
We further investigate the effectiveness and efficiency of our method with pixel-domain models in the subsequent section.

\section{Experiments}
\label{sec:exp}

\subsection{Implementation Details}
\label{sec:exp:details}

\textbf{Datasets.}
Following prior works~\cite{ehrlichQuantizationGuidedJPEG2020a,wangJPEGArtifactsRemoval2022,ouyangJPEGQuantizedCoefficient2024}, we adopt the DIV2K dataset~\cite{Agustsson_2017_CVPR_Workshops} for training and four benchmark datasets for evaluation: the LIVE-1 dataset~\cite{sheikhStatisticalEvaluationRecent2006a}, the test set of BSDS500~\cite{arbelaezContourDetectionHierarchical2011a}, the ICB dataset~\cite{ImageCompressionBenchmark}, and the Classic5 dataset~\cite{zeydeSingleImageScaleUsing2012} (grayscale).
The raw images from these datasets are compressed into JPEG format using OpenCV~\cite{bradski2000opencv}, with quality factors (QF) of 10, 20, 30, and 40.
We employ the TorchJPEG library~\cite{ehrlichTorchJPEG} to extract the quantized DCT coefficients and the quantization matrix from JPEG files.
Min-max normalization is applied to rescale the dequantized DCT coefficients from their original range of \([-1024, 1024]\) to the standardized interval of \([-1, 1]\), which has proven beneficial in practice.

\textbf{Supervision.}
We supervise the training of our models directly in the DCT domain using only the L1 loss function:
\begin{equation}
    \mathcal{L}_{\text{DCT}} = \frac{1}{3HW} \sum_{\mathbf{C} \in \{\mathbf{Y}, \mathbf{Cb}, \mathbf{Cr}\}} \Vert \mathbf{C}^{\text{Enh.}}_{\text{DCT}} - \mathbf{C}^{\text{GT}}_{\text{DCT}} \Vert_{1},
\end{equation}
where the ground-truth DCT coefficients \( \mathbf{C}^{\text{GT}}_{\text{DCT}} \) are obtained by compressing raw images with a QF of 100.
For grayscale enhancement, this supervision is applied only to the Y channel.

\textbf{Optimization.}
Images are cropped into \(256 \times 256\) patches for training.
Inspired by prior work~\cite{Park_2023_CVPR}, we conduct data augmentation (\ie, rotation, flipping, and clipping) directly in the DCT domain.
Our models are trained for 500k iterations with a batch size of 32 using the Adam optimizer~\cite{kingmaAdamMethodStochastic2015}.
The learning rate is initialized at \(2 \times 10^{-4}\) and decayed by a factor of 0.5 at the iterations of 250k, 400k, 450k, and 475k.
The optimization is conducted within the PyTorch framework~\cite{paszkePyTorchImperativeStyle2019} with a maximum of two NVIDIA 4090 GPUs.

\textbf{Baselines.}
Our method can adapt numerous pixel-domain JPEG quality enhancement models.
Specifically, we select widely-adopted pixel-domain models, including AR-CNN~\cite{dongCompressionArtifactsReduction2015a}, DCAD~\cite{wangNovelDeepLearningBased2017a}, DnCNN~\cite{zhangGaussianDenoiserResidual2017}, FBCNN~\cite{Jiang_2021_ICCV}, and ARCRL~\cite{wangJPEGArtifactsRemoval2022}.
Additionally, we compare our method against state-of-the-art DCT-domain methods, specifically QGAC~\cite{ehrlichQuantizationGuidedJPEG2020a} and JDEC~\cite{Han_2024_CVPR}.
For a fair comparison, all methods are retrained on our dataset with consistent training settings.
Among these methods, FBCNN, ARCRL, QGAC, and JDEC are QF-blind methods, for which we obtain one model trained over a dataset mixed with four QFs.
In line with previous studies, we report Peak Signal-to-Noise Ratio (PSNR), the Structural SIMilarity (SSIM) index~\cite{wangImageQualityAssessment2004a}, and PSNR-Block (PSNR-B)~\cite{yimQualityAssessmentDeblocked2011a}.

\begin{table*}[t]
    \centering
    \begin{adjustbox}{max width=.95\linewidth}
        \begin{tabular}{c | c | c | c c c c c | c }
            Dataset & QF & JPEG & AJQE w/ AR-CNN & AJQE w/ DCAD & AJQE w/ DnCNN & AJQE w/ FBCNN & AJQE w/ ARCRL & QGAC \\
            \Xhline{1.2pt}
            \multirow{4}{*}{BSDS500} & 10 & 27.77 & 29.52 \red{(+0.56)} & 29.61 \red{(+0.38)} & 29.56 \red{(+0.22)} & 29.81 \red{(+0.24)} & \textbf{29.85} \red{(+0.10)} & 29.28\\
            & 20 & 29.94 & 31.95 \red{(+0.76)} & 31.99 \red{(+0.44)} & 31.96 \red{(+0.26)} & 32.09 \red{(+0.21)} & \textbf{32.13} \red{(+0.18)} & 31.50\\
            & 30 & 31.26 & 33.21 \red{(+1.02)} & 33.38 \red{(+0.43)} & 33.48 \red{(+0.41)} & \textbf{33.54} \red{(+0.28)} & 33.51 \red{(+0.08)} & 32.79\\
            & 40 & 32.21 & 34.42 \red{(+0.91)} & 34.45 \red{(+0.50)} & 34.44 \red{(+0.38)} & \textbf{34.54} \red{(+0.31)} & 34.50 \red{(+0.19)} & 33.70\\
            \hline
            \multirow{4}{*}{LIVE-1} & 10 & 27.67 & 29.56 \red{(+0.59)} & 29.68 \red{(+0.39)} & 29.61 \red{(+0.59)} & \textbf{29.97} \red{(+0.26)} & 29.92 \red{(+0.18)} & 29.37\\
            & 20 & 29.88 & 32.07 \red{(+0.86)} & 32.11 \red{(+0.48)} & 32.07 \red{(+0.29)} & 32.18 \red{(+0.12)} & \textbf{32.24} \red{(+0.08)} & 31.61\\
            & 30 & 31.19 & 33.26 \red{(+1.10)} & 33.46 \red{(+0.41)} & 33.45 \red{(+0.27)} & 33.53 \red{(+0.11)} & \textbf{33.60} \red{(+0.06)} & 32.89\\
            & 40 & 32.12 & 34.46 \red{(+0.96)} & 34.50 \red{(+0.49)} & 34.48 \red{(+0.32)} & 34.61 \red{(+0.25)} & \textbf{34.66} \red{(+0.27)} & 33.76\\
            \hline
            \multirow{4}{*}{ICB} & 10 & 32.67 & 35.17 \red{(+0.85)} & 35.24 \red{(+0.52)} & 35.28 \red{(+0.41)} & 35.95 \red{(+0.30)} & \textbf{36.06} \red{(+0.28)} & 35.28\\
            & 20 & 35.60 & 37.92 \red{(+0.84)} & 37.95 \red{(+0.44)} & 37.92 \red{(+0.20)} & \textbf{38.36} \red{(+0.21)} & 38.34 \red{(+0.13)} & 37.87\\
            & 30 & 37.18 & 39.24 \red{(+0.89)} & 39.44 \red{(+0.39)} & 39.44 \red{(+0.22)} & \textbf{39.67} \red{(+0.17)} & 39.59 \red{(+0.10)} & 39.17\\
            & 40 & 38.17 & 40.33 \red{(+0.75)} & 40.38 \red{(+0.40)} & 40.41 \red{(+0.28)} & 40.47 \red{(+0.13)} & \textbf{40.51} \red{(+0.12)} & 40.02\\
            \hline
            \multirow{4}{*}{Classic-5} & 10 & 28.84 & 30.86 \red{(+0.65)} & 31.04 \red{(+0.43)} & 30.91 \red{(+0.16)} & 31.37 \red{(+0.19)} & \textbf{31.38} \red{(+0.02)} & 30.85\\
            & 20 & 31.15 & 33.24 \red{(+0.80)} & 33.30 \red{(+0.42)} & 33.24 \red{(+0.25)} & \textbf{33.63} \red{(+0.32)} & 33.57 \red{(+0.15)} & 32.93\\
            & 30 & 32.50 & 34.41 \red{(+0.72)} & 34.63 \red{(+0.51)} & 34.63 \red{(+0.42)} & 34.72 \red{(+0.25)} & \textbf{34.75} \red{(+0.30)} & 34.08\\
            & 40 & 33.41 & 35.39 \red{(+0.85)} & 35.50 \red{(+0.53)} & 35.47 \red{(+0.51)} & \textbf{35.56} \red{(+0.32)} & 35.54 \red{(+0.16)} & 34.81\\
            \hline
        \end{tabular}
    \end{adjustbox}
    \caption{\textbf{PSNR results (dB) for grayscale JPEG image quality enhancement.}
    Performance improvements over the pixel-domain model are shown in parentheses.
    The best results are boldfaced.
    SSIM and PSNR-B results are provided in the supplementary materials. Note that JDEC does not support grayscale enhancement.}
    \label{tab:gray_result}
\end{table*}

\begin{figure*}[t]
    \centering
    \includegraphics[trim={15mm 15mm 15mm 15mm}, clip, width=\linewidth]{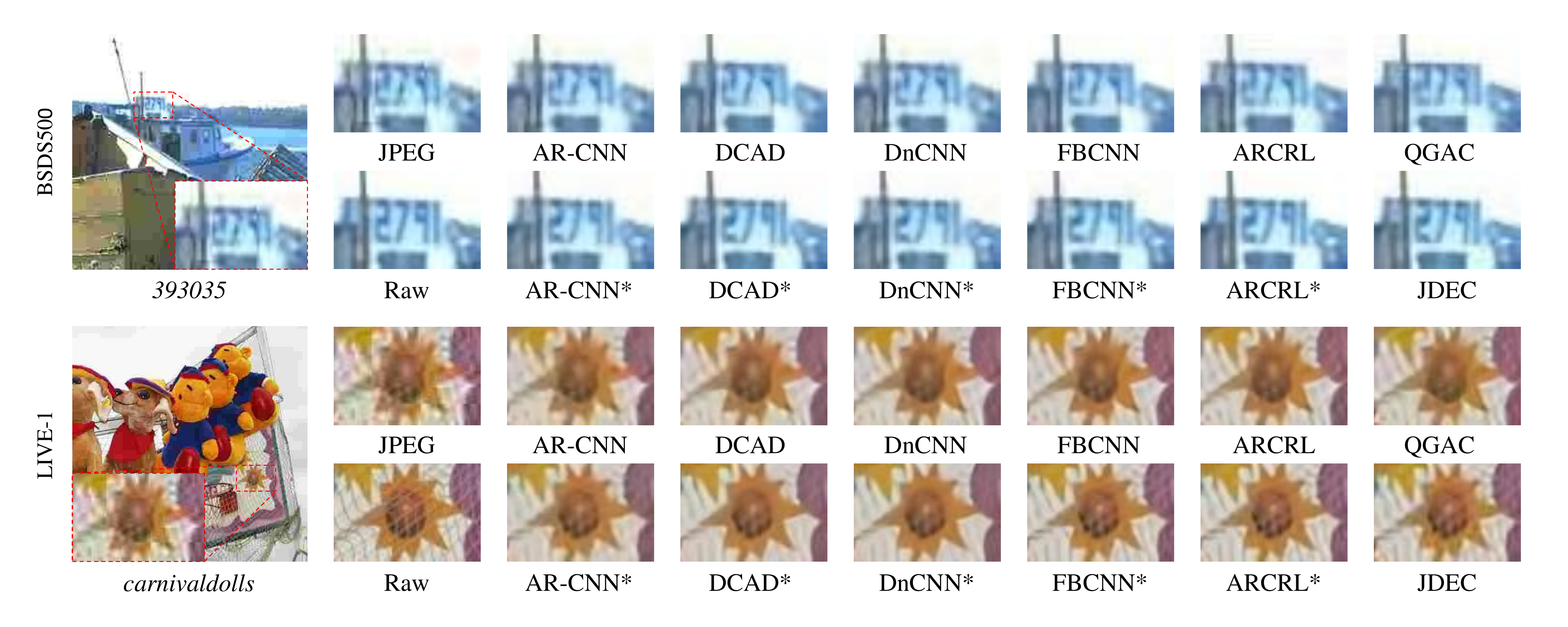}
    \caption{\textbf{Qualitative performance of color JPEG image quality enhancement.}
    AR-CNN\textsuperscript{*} indicates our AJQE method using the AR-CNN model.
    JPEG images are compressed with a QF of 10.}
    \label{fig:color_result}
\end{figure*}

\subsection{Evaluation}

\textbf{Effectiveness.}
We evaluate all re-trained models on three color image datasets.
The quantitative results are reported in \cref{tab:color_result}.
As shown, our DCT-domain models demonstrate significantly better image quality than their pixel-domain counterparts and other DCT-domain models across all QF.
On average, our models achieve a 0.35 dB improvement in PSNR over pixel-domain methods and a 0.41 dB improvement over state-of-the-art DCT-domain models.
Additionally, we evaluate all re-trained models on four grayscale image datasets, three of which are converted from the aforementioned color image datasets.
As presented in \cref{tab:gray_result}, our DCT-domain models also demonstrate superior image quality, resulting in an average 0.39 dB improvement in PSNR over pixel-domain methods and an average 0.48
dB improvement over state-of-the-art DCT-domain methods.
Moreover, we present qualitative results in \cref{fig:color_result}.
As shown, models using our method successfully suppress JPEG artifacts and restore more details compared to pixel-domain and other DCT-domain methods.
In conclusion, our method significantly improves the effectiveness in enhancing the quality of JPEG images.

\begin{figure}[t]
    \centering
    \includegraphics[trim={9mm 3mm 17mm 13mm}, clip, width=\linewidth]{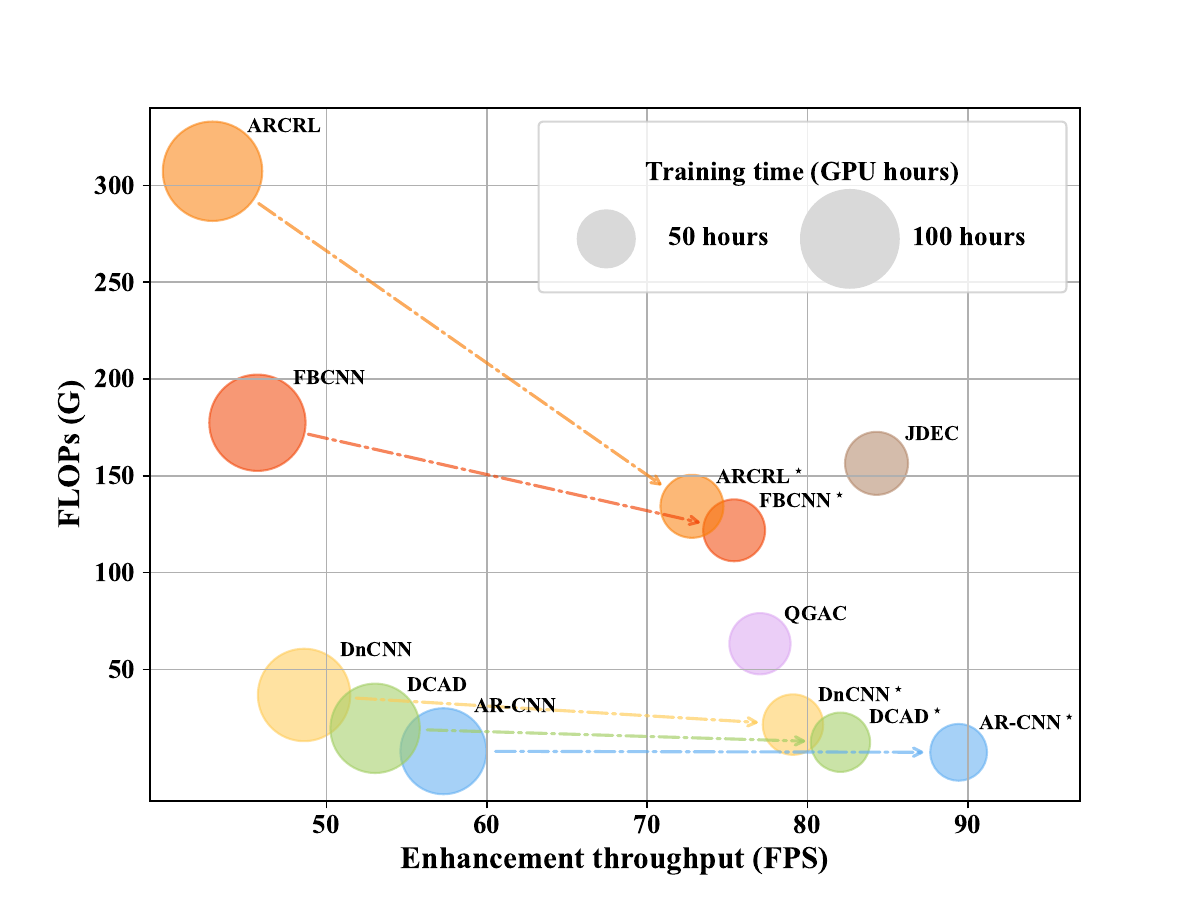}
    \caption{\textbf{Efficiency performance in terms of enhancement throughput, FLOPs, and training time.}
    ARCRL\textsuperscript{*} indicates our AJQE method using the ARCRL model.}
    \label{fig:efficiency}
\end{figure}

\textbf{Efficiency.}
We report the computational efficiency of all methods in terms of the total training time (GPU hours), the enhancement throughput (FPS), and the number of FLoating point OPerations (FLOPs).
Note that the enhancement throughput is calculated by averaging the enhancement speed over the BSDS500 test dataset.
All experiments are conducted on a single NVIDIA 4090 GPU.
As shown in \cref{fig:efficiency}, our DCT-domain models demonstrate significantly better efficiency than their pixel-domain counterparts, with 38.0\% lower training time, 60.5\% higher enhancement throughput, and 45.9\% fewer FLOPs.
Additionally, our DCT-domain models outperform all state-of-the-art DCT-domain methods by at least 45.9\% fewer FLOPs.
In conclusion, our method significantly improves the efficiency in enhancing the quality of JPEG images.

\subsection{Ablation Study}

In this section, we evaluate the effectiveness of the key components in our method: the Enhancement Module powered by Block-based Correlation (EMBC) and the Enhancement Module powered by Point-based Correlation (EMPC). We also assess the effectiveness of the block-based convolution in EMBC and the DCT-domain Upsampling Module Equivalent to Pixel-domain (UMEP) in the second stage of our method. Specifically, we use the FBCNN model as the pixel-domain model in our method.
We design the following experiments:
\begin{itemize}
    \item \textbf{EMBC-Res:} We replace all EMBCs with ResBlocks that have a similar number of parameters.
    \item \textbf{EMPC-Res:} We replace the EMPC with a ResBlock of similar parameter count.
    \item \textbf{All-Res:} We replace all EMBCs and the EMPC with ResBlocks of similar parameter count.
    \item \textbf{BConv-VConv:} We replace the block-based convolution in EMBC with a vanilla convolution.
    \item \textbf{Up-SpatialNN:} We replace the UMEP with spatial nearest-neighbor upsampling.
    \item \textbf{Up-TransConv:} We replace the UMEP with learnable upsampling using transposed convolution.
\end{itemize}
All models are re-trained and evaluated using the same settings described in \cref{sec:exp:details}.

The results are presented in \cref{tab:ablation}. As shown, there is a significant drop in enhancement performance for the models containing only EMPC and those containing only EMBC.
These results demonstrate that the proposed EMBC and EMPC are essential for achieving optimal enhancement performance due to the significant role of block-based and point-based correlations within DCT coefficients, as analyzed in \cref{sec:observation}.
Furthermore, the block-based convolution in EMBC is crucial for enhancing the quality of JPEG images, as evidenced by the 0.14 dB decrease in PSNR for BConv-VConv experiment.
Lastly, replacing our UMEP with nearest-neighbor interpolation and learnable upsampling (which uses transposed convolution) for DCT coefficients results in an average decrease in PSNR of 0.16 dB and 0.10 dB, respectively.
This indicates the importance of our proposed UMEP, which ensures that the upsampled results are equivalent to those obtained through pixel-domain upsampling.
In conclusion, each component in our method plays a crucial role in enhancing the quality of JPEG images in the DCT domain.

\begin{table}[t]
    \centering
    \renewcommand{\arraystretch}{1.2} % Adjust row spacing
    \begin{adjustbox}{max width=\linewidth}
        \begin{tabular}{l | c c c c}
            QF & 10 & 20 & 30 & 40 \\
            \hline
            \hline
            EMBC-Res  & 27.29/0.792 & 29.60/0.861 & 30.85/0.889 & 32.03/0.911  \\
            \hline
            EMPC-Res  & 27.59/0.808 & 30.04/0.874 & 31.20/0.903 & 32.26/0.920  \\
            \hline
            All-Res  & 26.92/0.789 & 29.42/0.856 & 30.63/0.884 & 31.87/0.908  \\
            \hline
            BConv-VConv  & 27.57/0.804 & 29.97/0.874 & 31.19/0.902 & 32.28/0.921  \\
            \hline
            Up-SpatialNN  & 27.54/0.804 & 29.99/0.871 & 31.15/0.903 & 32.24/0.918  \\
            \hline
            Up-TransConv  & 27.58/0.806 & 30.06/0.873 & 31.22/0.904 & 32.30/0.922  \\
            \hline
            Ours & \textbf{27.69/0.810} & \textbf{30.17/0.876} & \textbf{31.29/0.905} & \textbf{32.41/0.923}  \\
            \hline
        \end{tabular}
    \end{adjustbox}
    \caption{\textbf{Ablation results for key components.}
    Results for PSNR (dB) and SSIM are presented, with the best outcomes highlighted in bold.}
    \label{tab:ablation}
\end{table}

\section{Conclusion}

In this paper, we proposed an Advanced DCT-domain JPEG Quality Enhancement (AJQE) method that unravels the limitations of existing DCT-domain methods and achieves state-of-the-art enhancement performance. Specifically, our method exploits two critical correlations within DCT coefficients of JPEG images, namely block-based correlation and point-based correlation. Correspondingly, we designed two enhancement modules to capture these correlations in the DCT domain, enabling the adaptation of numerous well-established pixel-domain models to the DCT domain. We conducted extensive experiments to evaluate the performance of our method, comparing it with state-of-the-art pixel-domain and DCT-domain methods. Our method achieves superior enhancement performance with reduced computational complexity, demonstrating its effectiveness in enhancing JPEG images in the DCT domain.

{
    \small
    \bibliographystyle{ieeenat_fullname}
    \bibliography{arxiv}
}

\clearpage
\setcounter{page}{1}
\maketitlesupplementary

\section{Proof of DCT-domain Upsampling Module Equivalent to Pixel-domain (UMEP)}

\subsection{Block-based Chroma Upsampling}

In the main paper, we proposed a DCT-domain Upsampling Module Equivalent to Pixel-domain (UMEP) that achieves equivalent results to pixel-domain upsampling.
In this section, we provide detailed derivations and proofs for this method.

Assume we have an \( N \times N \) pixel block \( \mathbf{B}_{N} \).
To upsample this block by a factor of 2 into a \( 2N \times 2N \) pixel block \( \mathbf{B}_{2N} \), we can use nearest-neighbor upsampling, which can be efficiently implemented using NumPy broadcasting:
\begin{equation}
    \mathbf{B}_{2N} = \mathbf{U}_{N} \mathbf{B}_{N} \mathbf{U}_{N}^{\top},
    \label{eq:pixel_up}
\end{equation}
where \( \mathbf{U}_{N} \), with shape \( (2N, N) \), represents the upsampling matrix:
\begin{equation}
    \mathbf{U}_{N} = 
    \begin{cases} 
        \begin{array}{cc}
            \begin{bmatrix} 1 \\ 1 \end{bmatrix} & \text{if } N = 1, \\
            \begin{bmatrix}
                \mathbf{U}_{1} & \cdots & 0 \\
                \vdots & \ddots & \vdots \\
                0 & \cdots & \mathbf{U}_{1}
            \end{bmatrix} & \text{if } N > 1.
        \end{array}
    \end{cases}
\end{equation}

Similarly, we can upsample a DCT block \( \hat{\mathbf{B}}_{N} \) by a factor of 2 as follows:
\begin{equation}
    \hat{\mathbf{B}}_{2N} = \mathbf{U}_{N} \hat{\mathbf{B}}_{N} \mathbf{U}_{N}^{\top}.
    \label{eq:dct_up}
\end{equation}
However, DCT blocks are not suitable for spatial upsampling.
These blocks contain spatially decorrelated coefficients, while spatial upsampling methods assume that neighboring values are correlated and change continuously.
As a result, direct upsampling of DCT coefficients may result in aliasing and artifacts.\footnote{Our ablation study shows that directly upsampling chroma DCT blocks can lead to at least a 0.1 dB PSNR degradation in performance.}
The intuitive solution is to first inverse-transform the DCT block back to the pixel domain, upsample the pixel block, and then transform it back to the DCT domain.
Unfortunately, this process is computationally expensive.

Now, we aim to derive a DCT-domain upsampling matrix \( \hat{\mathbf{U}}_{N} \) that yields equivalent results to pixel-domain upsampling.
Our mathematical goal is to achieve:
\begin{align}
    \hat{\mathbf{B}}_{2N} &= \hat{\mathbf{U}}_{N} \hat{\mathbf{B}}_{N} \hat{\mathbf{U}}_{N}^{\top}, \label{eq:dct_up_goal1}\\
    &= \mathbf{T}_{2N} \mathbf{B}_{2N} \mathbf{T}_{2N}^{\top}, \label{eq:dct_up_goal2}
\end{align}
where \( \mathbf{T}_{2N} \) is the 1-D DCT basis matrix of size \( 2N \times 2N \)~\cite{ahmedDiscreteCosineTransform1974}.
Based on \cref{eq:pixel_up}, we replace \( \mathbf{B}_{2N} \) in \cref{eq:dct_up_goal2} and obtain:
\begin{align}
    \hat{\mathbf{B}}_{2N} &= \mathbf{T}_{2N} \left( \mathbf{U}_{N} \mathbf{B}_{N} \mathbf{U}_{N}^{\top} \right) \mathbf{T}_{2N}^{\top}, \label{eq:dct_up_goal3}\\
    &= \left( \mathbf{T}_{2N} \mathbf{U}_{N} \right) \mathbf{B}_{N} \left( \mathbf{T}_{2N} \mathbf{U}_{N} \right)^{\top}. \label{eq:dct_up_goal4}
\end{align}
Combining \cref{eq:dct_up_goal1} and \cref{eq:dct_up_goal4}, we have:
\begin{equation}
    \hat{\mathbf{U}}_{N} \hat{\mathbf{B}}_{N} \hat{\mathbf{U}}_{N}^{\top} = \left( \mathbf{T}_{2N} \mathbf{U}_{N} \right) \mathbf{B}_{N} \left( \mathbf{T}_{2N} \mathbf{U}_{N} \right)^{\top}.
    \label{eq:dct_up_goal5}
\end{equation}
Considering \( \hat{\mathbf{B}}_{N} = \mathbf{T}_{N} \mathbf{B}_{N} \mathbf{T}_{N}^{\top} \) in \cref{eq:dct_up_goal5}, we have:
\begin{align}
    \hat{\mathbf{U}}_{N} \left( \mathbf{T}_{N} \mathbf{B}_{N} \mathbf{T}_{N}^{\top} \right) \hat{\mathbf{U}}_{N}^{\top} &= \left(\hat{\mathbf{U}}_{N} \mathbf{T}_{N} \right) \mathbf{B}_{N} \left(\hat{\mathbf{U}}_{N} \mathbf{T}_{N} \right)^{\top},\\
    &= \left( \mathbf{T}_{2N} \mathbf{U}_{N} \right) \mathbf{B}_{N} \left( \mathbf{T}_{2N} \mathbf{U}_{N} \right)^{\top}.
\end{align}
Therefore, we derive \( \hat{\mathbf{U}}_{N} = \mathbf{T}_{2N} \mathbf{U}_{N} \mathbf{T}_{N}^{-1} = \mathbf{T}_{2N} \mathbf{U}_{N} \mathbf{T}_{N}^{\top} \). Note that the DCT basis matrix \( \mathbf{T} \) is orthogonal, meaning \( \mathbf{T}^{-1} = \mathbf{T}^{\top} \). As a result, we obtain the ideal DCT-domain upsampling by \( \hat{\mathbf{U}}_{N} \) that yields equivalent results to pixel-domain upsampling by \( \mathbf{U}_{N} \):
\begin{align}
    \hat{\mathbf{B}}_{2N} &= \hat{\mathbf{U}}_{N} \hat{\mathbf{B}}_{N} \hat{\mathbf{U}}_{N}^{\top}, \label{eq:dct_up_eq}\\
    &= \left( \mathbf{T}_{2N} \mathbf{U}_{N} \mathbf{T}_{N}^{\top} \right) \hat{\mathbf{B}}_{N} \left( \mathbf{T}_{2N} \mathbf{U}_{N} \mathbf{T}_{N}^{\top} \right)^{\top}.
    \label{eq:dct_up_equal}
\end{align}

\subsection{Application to JPEG Chroma Upsampling}

Ideally, we can upsample each \( 4 \times 4 \) DCT block by a DCT-domain upsampling matrix \( \hat{\mathbf{U}}_{4} = \mathbf{T}_{8} \mathbf{U}_{4} \mathbf{T}_{4}^{\top} \), such that we can achieve equivalent results to pixel-domain upsampling by \( \mathbf{U}_{4} \).
However, in JPEG, there exist only \( 8 \times 8 \) chroma DCT blocks, which are obtained by combining four neighboring downsampled \( 4 \times 4 \) chroma pixel blocks following DCT transformation.
To obtain four \( 4 \times 4 \) chroma DCT blocks from each \( 8 \times 8 \) chroma DCT block, we adopt the sub-block conversion method~\cite{salazarComplexityScalableUniversal2007,Park_2023_CVPR}:
\begin{equation}
    \mathbf{H}_{2}^{\top} \hat{\mathbf{B}}_{8} \mathbf{H}_{2} = \begin{bmatrix} \hat{\mathbf{B}}_{4}^{(0,0)} & \hat{\mathbf{B}}_{4}^{(0,1)} \\ \hat{\mathbf{B}}_{4}^{(1,0)} & \hat{\mathbf{B}}_{4}^{(1,1)} \end{bmatrix},
\end{equation}
where \( \mathbf{H}_{2} \) is the sub-block conversion matrix with shape \( 8 \times 8 \);
\( \hat{\mathbf{B}}_{4}^{(0,0)} \) to \( \hat{\mathbf{B}}_{4}^{(1,1)} \) are the converted \( 4 \times 4 \) chroma DCT blocks.
Finally, we can apply our DCT-domain upsampling as in \cref{eq:dct_up_eq} to these converted \( 4 \times 4 \) blocks:
\begin{equation}
    \hat{\mathbf{B}}_{8} = \hat{\mathbf{U}}_{4} \hat{\mathbf{B}}_{4} \hat{\mathbf{U}}_{4}^{\top}.
\end{equation}

\section{Additional Experimental Results}

\begin{table*}[htbp]
    \centering
    \begin{adjustbox}{max width=.95\linewidth}
        \begin{tabular}{c | c | c | c c c c c | c c }
            Dataset & QF & JPEG & AJQE w/ AR-CNN & AJQE w/ DCAD & AJQE w/ DnCNN & AJQE w/ FBCNN & AJQE w/ ARCRL & QGAC & JDEC \\
            \Xhline{1.2pt}
            \hline
            \multirow{4}{*}{BSDS500} & 10 & 23.67 & 26.83 \red{(+0.38)} & 26.80 \red{(+0.01)} & 26.96 \red{(+0.06)} & 27.45 \red{(+0.12)} & \textbf{27.58} \red{(+0.28)} & 26.64 & 27.04\\
            & 20 & 25.89 & 29.10 \red{(+0.30)} & 29.38 \red{(+0.07)} & 29.41 \red{(+0.09)} & 29.89 \red{(+0.14)} & \textbf{30.06} \red{(+0.32)} & 29.15 & 29.40\\
            & 30 & 27.28 & 30.31 \red{(+0.16)} & 30.70 \red{(+0.01)} & 30.51 \red{(+0.12)} & 31.10 \red{(+0.03)} & \textbf{31.12} (--0.03) & 30.41 & 30.78\\
            & 40 & 28.28 & 31.17 \red{(+0.13)} & 31.75 \red{(+0.10)} & 31.77 \red{(+0.01)} & \textbf{32.28} \red{(+0.30)} & 32.23 \red{(+0.24)} & 31.25 & 31.76\\
            \hline
            \multirow{4}{*}{LIVE-1} & 10 & 23.96 & 26.66 \red{(+0.27)} & 26.70 \red{(+0.02)} & 26.78 (--0.06) & 27.53 \red{(+0.04)} & \textbf{27.75} \red{(+0.43)} & 26.81 & 26.14\\
            & 20 & 26.27 & 28.98 \red{(+0.22)} & 29.25 \red{(+0.01)} & 29.49 \red{(+0.01)} & 29.91 \red{(+0.05)} & \textbf{30.16} \red{(+0.32)} & 29.05 & 29.48\\
            & 30 & 27.64 & 30.18 \red{(+0.06)} & 30.88 \red{(+0.14)} & 30.54 \red{(+0.12)} & 31.22 \red{(+0.03)} & \textbf{31.27} \red{(+0.07)} & 30.28 & 30.82\\
            & 40 & 28.63 & 31.05 \red{(+0.06)} & 31.73 \red{(+0.04)} & 31.96 \red{(+0.14)} & 32.19 \red{(+0.11)} & \textbf{32.26} \red{(+0.12)} & 31.07 & 31.65\\
            \hline
            \multirow{4}{*}{ICB} & 10 & 27.66 & 30.98 \red{(+0.75)} & 31.09 \red{(+0.28)} & 31.17 \red{(+0.15)} & 32.05 \red{(+0.17)} & \textbf{32.08} \red{(+0.17)} & 31.68 & 31.71\\
            & 20 & 30.65 & 33.58 \red{(+0.43)} & 33.69 \red{(+0.02)} & 33.88 \red{(+0.10)} & \textbf{34.42} \red{(+0.04)} & 34.41 \red{(+0.08)} & 34.02 & 34.21\\
            & 30 & 32.15 & 34.73 \red{(+0.16)} & 35.12 \red{(+0.02)} & 35.08 \red{(+0.29)} & \textbf{35.82} \red{(+0.10)} & 35.65 \red{(+0.07)} & 35.08 & 35.31\\
            & 40 & 33.23 & 35.82 \red{(+0.34)} & 36.13 \red{(+0.08)} & 36.16 \red{(+0.06)} & \textbf{36.68} \red{(+0.18)} & 36.61 \red{(+0.13)} & 35.85 & 36.08\\
            \hline
        \end{tabular}
    \end{adjustbox}
    \caption{\textbf{PSNR-B (dB) results for color JPEG image quality enhancement.}
    Performance improvements over the pixel-domain model are shown in parentheses.
    The best results are boldfaced.}
    \label{tab:color_result_supp}
\end{table*}

\begin{table*}[htbp]
    \centering
    \begin{adjustbox}{max width=.95\linewidth}
        \begin{tabular}{c | c | c | c c c c c | c }
            Dataset & QF & JPEG & AJQE w/ AR-CNN & AJQE w/ DCAD & AJQE w/ DnCNN & AJQE w/ FBCNN & AJQE w/ ARCRL & QGAC \\
            \Xhline{1.2pt}
            \multicolumn{9}{c}{SSIM}\\
            \hline
            \multirow{4}{*}{BSDS500} & 10 & 0.781 & 0.833 \red{(+.019)} & 0.835 \red{(+.013)} & 0.834 \red{(+.009)} & 0.837 \red{(+.007)} & \textbf{0.838} \red{(+.002)} & 0.822\\
            & 20 & 0.856 & 0.897 \red{(+.018)} & 0.898 \red{(+.012)} & 0.897 \red{(+.009)} & 0.899 \red{(+.008)} & \textbf{0.900} \red{(+.011)} & 0.885\\
            & 30 & 0.889 & 0.920 \red{(+.017)} & 0.923 \red{(+.009)} & 0.923 \red{(+.008)} & \textbf{0.924} \red{(+.007)} & \textbf{0.924} \red{(+.001)} & 0.911\\
            & 40 & 0.908 & 0.938 \red{(+.014)} & 0.938 \red{(+.009)} & 0.938 \red{(+.008)} & 0.939 \red{(+.007)} & \textbf{0.940} \red{(+.007)} & 0.926\\
            \hline
            \multirow{4}{*}{LIVE-1} & 10 & 0.784 & 0.839 \red{(+.019)} & 0.842 \red{(+.013)} & 0.840 \red{(+.008)} & 0.844 \red{(+.005)} & \textbf{0.845} \red{(+.004)} & 0.829\\
            & 20 & 0.856 & 0.900 \red{(+.019)} & 0.901 \red{(+.012)} & 0.900 \red{(+.008)} & 0.902 \red{(+.006)} & \textbf{0.903} \red{(+.005)} & 0.888\\
            & 30 & 0.888 & 0.921 \red{(+.018)} & 0.924 \red{(+.009)} & 0.924 \red{(+.007)} & 0.925 \red{(+.005)} & \textbf{0.929} \red{(+.002)} & 0.913\\
            & 40 & 0.905 & 0.938 \red{(+.015)} & 0.939 \red{(+.010)} & 0.939 \red{(+.008)} & 0.939 \red{(+.006)} & \textbf{0.940} \red{(+.009)} & 0.926\\
            \hline
            \multirow{4}{*}{ICB} & 10 & 0.863 & 0.913 \red{(+.012)} & 0.914 \red{(+.009)} & 0.914 \red{(+.007)} & 0.917 \red{(+.006)} & \textbf{0.918} \red{(+.001)} & 0.908\\
            & 20 & 0.912 & 0.943 \red{(+.011)} & 0.943 \red{(+.008)} & 0.943 \red{(+.007)} & 0.943 \red{(+.005)} & \textbf{0.944} \red{(+.005)} & 0.936\\
            & 30 & 0.932 & 0.953 \red{(+.009)} & 0.955 \red{(+.007)} & 0.955 \red{(+.006)} & \textbf{0.956} \red{(+.006)} & 0.955 \red{(+.001)} & 0.948\\
            & 40 & 0.943 & 0.961 \red{(+.008)} & 0.962 \red{(+.007)} & 0.962 \red{(+.006)} & 0.962 \red{(+.005)} & \textbf{0.963} \red{(+.007)} & 0.956\\
            \hline
            \multirow{4}{*}{Classic-5} & 10 & 0.783 & 0.848 \red{(+.026)} & 0.852 \red{(+.019)} & 0.851 \red{(+.011)} & 0.852 \red{(+.006)} & \textbf{0.856} \red{(+.008)} & 0.839\\
            & 20 & 0.850 & 0.896 \red{(+.024)} & 0.897 \red{(+.017)} & 0.896 \red{(+.014)} & 0.898 \red{(+.012)} & \textbf{0.899} \red{(+.007)} & 0.880\\
            & 30 & 0.879 & 0.914 \red{(+.018)} & 0.917 \red{(+.016)} & 0.917 \red{(+.015)} & 0.918 \red{(+.013)} & \textbf{0.919} \red{(+.017)} & 0.900\\
            & 40 & 0.895 & 0.927 \red{(+.019)} & 0.928 \red{(+.015)} & 0.928 \red{(+.007)} & 0.928 \red{(+.012)} & \textbf{0.930} \red{(+.006)} & 0.911\\
            \Xhline{1.2pt}
            \multicolumn{9}{c}{PSNR-B}\\
            \hline
            \multirow{4}{*}{BSDS500} & 10 & 25.03 & 28.96 \red{(+0.10)} & 29.14 \red{(+0.13)} & 29.20 \red{(+0.08)} & 29.52 \red{(+0.07)} & \textbf{29.56} \red{(+0.12)} & 29.17\\
            & 20 & 27.11 & 31.32 \red{(+0.30)} & 31.54 \red{(+0.17)} & 31.51 \red{(+0.03)} & 31.81 \red{(+0.13)} & \textbf{31.86} \red{(+0.19)} & 31.31\\
            & 30 & 28.48 & 32.24 \red{(+0.26)} & 32.97 \red{(+0.18)} & 32.87 \red{(+0.07)} & 33.22 \red{(+0.23)} & \textbf{33.27} \red{(+0.20)} & 32.54\\
            & 40 & 29.48 & 33.70 \red{(+0.47)} & 33.92 \red{(+0.28)} & 33.90 \red{(+0.16)} & 34.01 \red{(+0.11)} & \textbf{34.25} \red{(+0.16)} & 33.40\\
            \hline
            \multirow{4}{*}{LIVE-1} & 10 & 25.55 & 29.11 \red{(+0.25)} & 29.16 (--0.02) & 28.94 \red{(+0.06)} & \textbf{29.68} \red{(+0.12)} & 29.63 \red{(+0.13)} & 29.15\\
            & 20 & 27.70 & 31.48 \red{(+0.46)} & 31.51 \red{(+0.09)} & 31.59 \red{(+0.04)} & \textbf{31.97} \red{(+0.15)} & 31.96 \red{(+0.02)} & 31.26\\
            & 30 & 29.04 & 32.60 \red{(+0.68)} & 32.86 \red{(+0.03)} & 32.95 \red{(+0.08)} & 33.10 (+0.00) & \textbf{33.26} \red{(+0.02)} & 32.42\\
            & 40 & 30.02 & 33.70 \red{(+0.51)} & 33.73 \red{(+0.08)} & 33.91 \red{(+0.12)} & 34.13 \red{(+0.16)} & \textbf{34.17} \red{(+0.04)} & 33.21\\
            \hline
            \multirow{4}{*}{ICB} & 10 & 30.21 & 34.67 \red{(+0.43)} & 34.73 \red{(+0.10)} & 34.77 \red{(+0.09)} & \textbf{35.64} \red{(+0.09)} & \textbf{35.64} \red{(+0.15)} & 35.18\\
            & 20 & 33.08 & 37.50 \red{(+0.57)} & 37.43 \red{(+0.08)} & 37.49 \red{(+0.04)} & 38.01 \red{(+0.04)} & \textbf{38.08} \red{(+0.14)} & 37.69\\
            & 30 & 34.66 & 38.65 \red{(+0.41)} & 38.91 \red{(+0.09)} & 38.94 (--0.05) & \textbf{39.32} \red{(+0.06)} & 39.31 \red{(+0.15)} & 38.95\\
            & 40 & 35.75 & 39.83 \red{(+0.50)} & 39.87 \red{(+0.17)} & 39.88 \red{(+0.03)} & 40.16 \red{(+0.11)} & \textbf{40.25} \red{(+0.13)} & 39.75\\
            \hline
            \multirow{4}{*}{Classic-5} & 10 & 26.11 & 30.48 \red{(+0.31)} & 30.74 \red{(+0.17)} & 30.63 \red{(+0.13)} & \textbf{31.15} \red{(+0.03)} & 31.11 \red{(+0.11)} & 30.80\\
            & 20 & 28.50 & 32.71 \red{(+0.34)} & 32.86 \red{(+0.07)} & 32.71 \red{(+0.01)} & \textbf{33.37} \red{(+0.16)} & 33.29 \red{(+0.14)} & 32.84\\
            & 30 & 30.02 & 33.94 \red{(+0.36)} & 33.92 \red{(+0.03)} & 34.12 \red{(+0.05)} & \textbf{34.48} \red{(+0.16)} & 34.41 \red{(+0.19)} & 33.95\\
            & 40 & 31.08 & 34.85 \red{(+0.45)} & 34.92 \red{(+0.12)} & 34.60 \red{(+0.81)} & 35.19 \red{(+0.15)} & \textbf{35.20} \red{(+0.10)} & 34.66\\
            \hline
        \end{tabular}
    \end{adjustbox}
    \caption{\textbf{SSIM and PSNR-B (dB) results for grayscale JPEG image quality enhancement.}
    Performance improvements over the pixel-domain model are shown in parentheses.
    The best results are boldfaced.
    Note that JDEC does not support grayscale enhancement.}
    \label{tab:gray_result_supp}
\end{table*}

\begin{table*}[t]
    \centering
    \tiny
    \renewcommand{\arraystretch}{1.1} % Adjust row spacing
    \begin{adjustbox}{width=1.0\linewidth}
    \begin{tabular}{@{\extracolsep{0pt}} c | c | c c c c c c | c c c c c c@{}}
        \Xhline{0.7pt}
        \multirow{3}{*}{QF} & \multirow{3}{*}{Metric} & \multicolumn{6}{c|}{DIV2K} & \multicolumn{6}{c}{BSDS500} \\
        \cline{3-14}
        & & \multicolumn{2}{c}{Y} & \multicolumn{2}{c}{Cb} & \multicolumn{2}{c|}{Cr} & \multicolumn{2}{c}{Y} & \multicolumn{2}{c}{Cb} & \multicolumn{2}{c}{Cr} \\
        \cline{3-14}
        & & Pixel & DCT & Pixel & DCT & Pixel & DCT & Pixel & DCT & Pixel & DCT & Pixel & DCT \\
        \Xhline{0.7pt}
        \multirow{2}{*}{10} & \textit{MI}\(\uparrow\) & 0.947 & 0.007 & 0.978 & 0.009 & 0.978 & 0.008 & 0.935 & 0.006 & 0.989 & 0.010 & 0.990 & 0.008\\
        & \textit{GC}\(\downarrow\) & 0.053 & 1.003 & 0.022 & 0.998 & 0.022 & 0.998 & 0.065 & 0.999 & 0.011 & 0.999 & 0.010 & 0.995 \\
        \cline{1-14}
        \multirow{2}{*}{20} & \textit{MI}\(\uparrow\) & 0.946 & 0.007 & 0.977 & 0.008 & 0.977 & 0.008 &0.927 & 0.006 & 0.990 & 0.009 & 0.991 & 0.008\\
        & \textit{GC}\(\downarrow\) & 0.054 & 1.002 & 0.023 & 0.999 & 0.023 & 0.998 & 0.073 & 0.999 & 0.010 & 0.999 & 0.009 & 0.996 \\
        \cline{1-14}
        \multirow{2}{*}{30} & \textit{MI}\(\uparrow\) & 0.946 & 0.007 & 0.977 & 0.008 & 0.977 & 0.009 & 0.922 & 0.006 & 0.990 & 0.009 & 0.991 & 0.008 \\
        & \textit{GC}\(\downarrow\) & 0.054 & 1.003 & 0.023 & 0.999 & 0.023 & 1.000 & 0.078 & 0.999 & 0.010 & 0.999 & 0.009 & 0.996  \\
        \cline{1-14}
        \multirow{2}{*}{40} & \textit{MI}\(\uparrow\) & 0.946 & 0.007 & 0.976 & 0.008 & 0.977 & 0.008 & 0.919 & 0.006 & 0.990 & 0.009 & 0.991 & 0.008 \\
        & \textit{GC}\(\downarrow\) & 0.054 & 1.002 & 0.024 & 1.000 & 0.023 & 0.999 & 0.081 & 0.999 & 0.010 & 0.999 & 0.009 & 0.996 \\
        \cline{1-14}
        \multirow{2}{*}{50} & \textit{MI}\(\uparrow\) & 0.945 & 0.007 & 0.976 & 0.008 & 0.976 & 0.008 & 0.915 & 0.006 & 0.990 & 0.009 & 0.990 & 0.008 \\
        & \textit{GC}\(\downarrow\) & 0.055 & 1.003 & 0.024 & 0.999 & 0.024 & 0.999 & 0.085 & 0.999 & 0.010 & 0.999 & 0.010 & 0.996 \\
        \cline{1-14}
        \multirow{2}{*}{60} & \textit{MI}\(\uparrow\) & 0.946 & 0.007 & 0.976 & 0.009 & 0.975 & 0.009 &0.913 & 0.006 & 0.990 & 0.009 & 0.990 & 0.008 \\
        & \textit{GC}\(\downarrow\) & 0.054 & 1.003 & 0.024 & 0.999 & 0.025 & 0.999 & 0.087 & 0.999 & 0.010 & 0.999 & 0.010 & 0.996 \\
        \cline{1-14}
        \multirow{2}{*}{70} & \textit{MI}\(\uparrow\) & 0.946 & 0.007 & 0.975 & 0.009 & 0.975 & 0.008 & 0.910 & 0.006 & 0.990 & 0.009 & 0.990 & 0.008 \\
        & \textit{GC}\(\downarrow\) & 0.054 & 1.002 & 0.025 & 0.999 & 0.025 & 0.999 & 0.090 & 0.999 & 0.010 & 0.999 & 0.010 & 0.996 \\
        \cline{1-14}
        \multirow{2}{*}{80} & \textit{MI}\(\uparrow\) & 0.946 & 0.007 & 0.974 & 0.008 & 0.975 & 0.009 & 0.907 & 0.006 & 0.989 & 0.009 & 0.990 & 0.008 \\
        & \textit{GC}\(\downarrow\) & 0.054 & 1.002 & 0.026 & 0.998 & 0.025 & 0.999 & 0.093 & 0.999 & 0.010 & 0.999 & 0.010 & 0.996 \\
        \cline{1-14}
        \multirow{2}{*}{90} & \textit{MI}\(\uparrow\) & 0.947 & 0.007 & 0.973 & 0.009 & 0.973 & 0.008 & 0.905 & 0.006 & 0.989 & 0.009 & 0.989 & 0.008 \\
        & \textit{GC}\(\downarrow\) & 0.053 & 1.003 & 0.027 & 1.000 & 0.027 & 0.999 & 0.095 & 0.999 & 0.011 & 0.999 & 0.011 & 0.996 \\
        \cline{1-14}
        \multirow{2}{*}{100} & \textit{MI}\(\uparrow\) & 0.946 & 0.007 & 0.971 & 0.008 & 0.972 & 0.008 & 0.911 & 0.006 & 0.988 & 0.009 & 0.987 & 0.008 \\
        & \textit{GC}\(\downarrow\) & 0.054 & 1.003 & 0.029 & 0.999 & 0.028 & 0.999 & 0.089 & 0.999 & 0.012 & 0.999 & 0.013 & 0.996 \\
        \Xhline{0.7pt}
    \end{tabular}
    \end{adjustbox}
    \caption{\textbf{Correlations within pixel values and DCT coefficients with QF ranging 10 to 100 on the DIV2K and BSDS500 datasets.}
    Note that \textit{MI} ranges from \(-1\) to \(1\), where values close to \(1\) and \(0\) represent strong positive and negligible correlations, respectively.
    Conversely, \textit{GC} ranges from \(0\) to \(2\), where values close to \(0\) and \(1\) represent strong positive and negligible correlations, respectively.}
    \label{tab:corre_supp}
\end{table*}

The main paper presents the PSNR and SSIM results for enhanced color JPEG images, as well as the PSNR results for grayscale JPEG images, with QF values ranging from 10 to 40 for both.
In this section, we provide the PSNR-B results for color JPEG image quality enhancement in \cref{tab:color_result_supp}, and the SSIM and PSNR-B results for grayscale JPEG image quality enhancement in \cref{tab:gray_result_supp}.

Consistently, our results show that our DCT-domain models outperform both pixel-domain models and other DCT-domain models in terms of quantitative performance for both color and grayscale JPEG image quality enhancement.
Specifically, our DCT-domain models demonstrate an average PSNR-B improvement of 0.14 dB over pixel-domain methods and an average improvement of 0.23 dB over other DCT-domain methods for color JPEG image quality enhancement.
Moreover, for grayscale JPEG images, our DCT-domain models achieve a 0.17 dB improvement over pixel-domain methods and a 0.26 dB improvement over other DCT-domain models.
In conclusion, our method significantly enhances the effectiveness of JPEG image quality enhancement, achieving substantial improvements in all metrics.

\section{Additional Finding Results}

\begin{figure*}[htbp]
    \centering
    \begin{subfigure}{0.33\linewidth}
        \centering
        \includegraphics[trim={9mm 102mm 22mm 109mm}, clip, width=\linewidth]{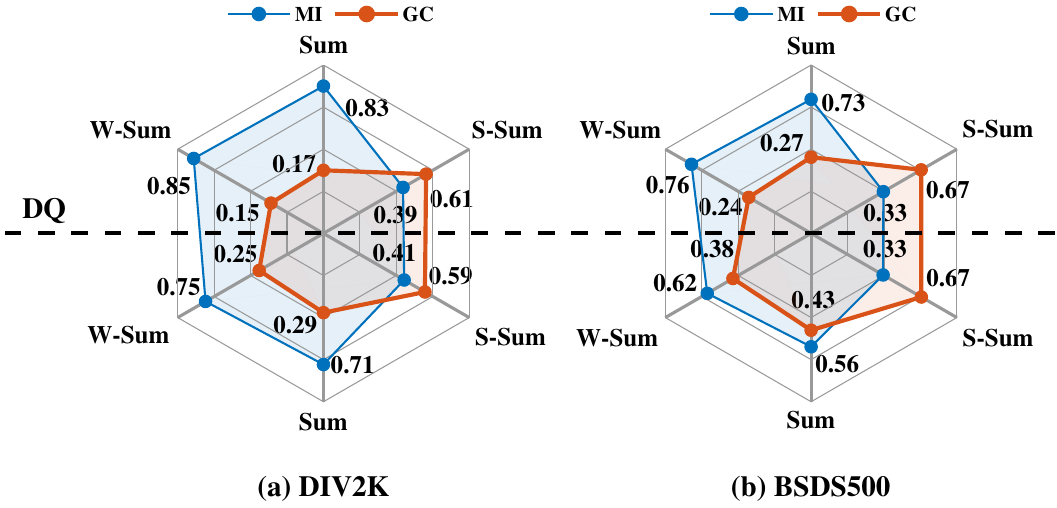}
        \caption{Y}
    \end{subfigure}
    \begin{subfigure}{.33\linewidth}
        \centering
        \includegraphics[trim={9mm 102mm 22mm 109mm}, clip, width=\linewidth]{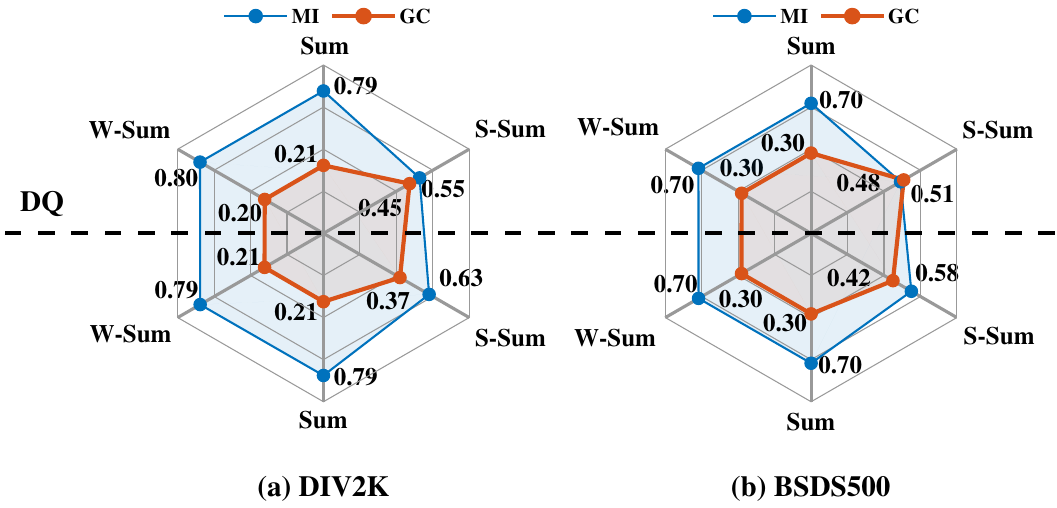}
        \caption{Cb}
    \end{subfigure}
    \begin{subfigure}{.33\linewidth}
        \centering
        \includegraphics[trim={9mm 102mm 22mm 109mm}, clip, width=\linewidth]{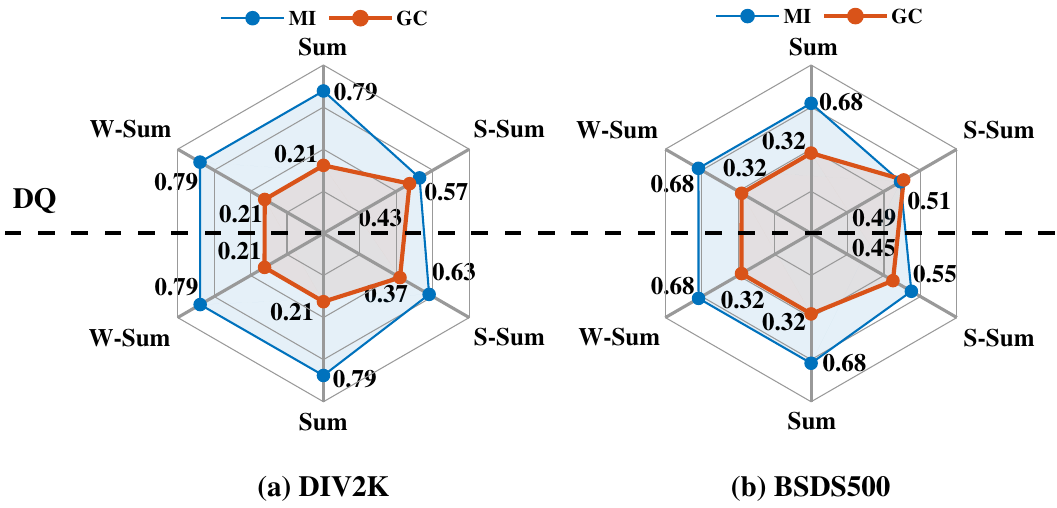}
        \caption{Cr}
    \end{subfigure}
    \vspace{\vspacelength}
    \caption{\textbf{Block-based correlations using different block-based features on the DIV2K and BSDS500 datasets with QF set to 10.}
    Upper: DCT blocks are dequantized before calculating feature values.
    Lower: DCT blocks remain quantized.}
    \label{fig:block_based_corr_supp_10}
\end{figure*}

\begin{figure*}[htbp]
    \centering
    \begin{subfigure}{0.33\linewidth}
        \centering
        \includegraphics[trim={0mm 0mm 0mm 0mm}, clip, width=\linewidth]{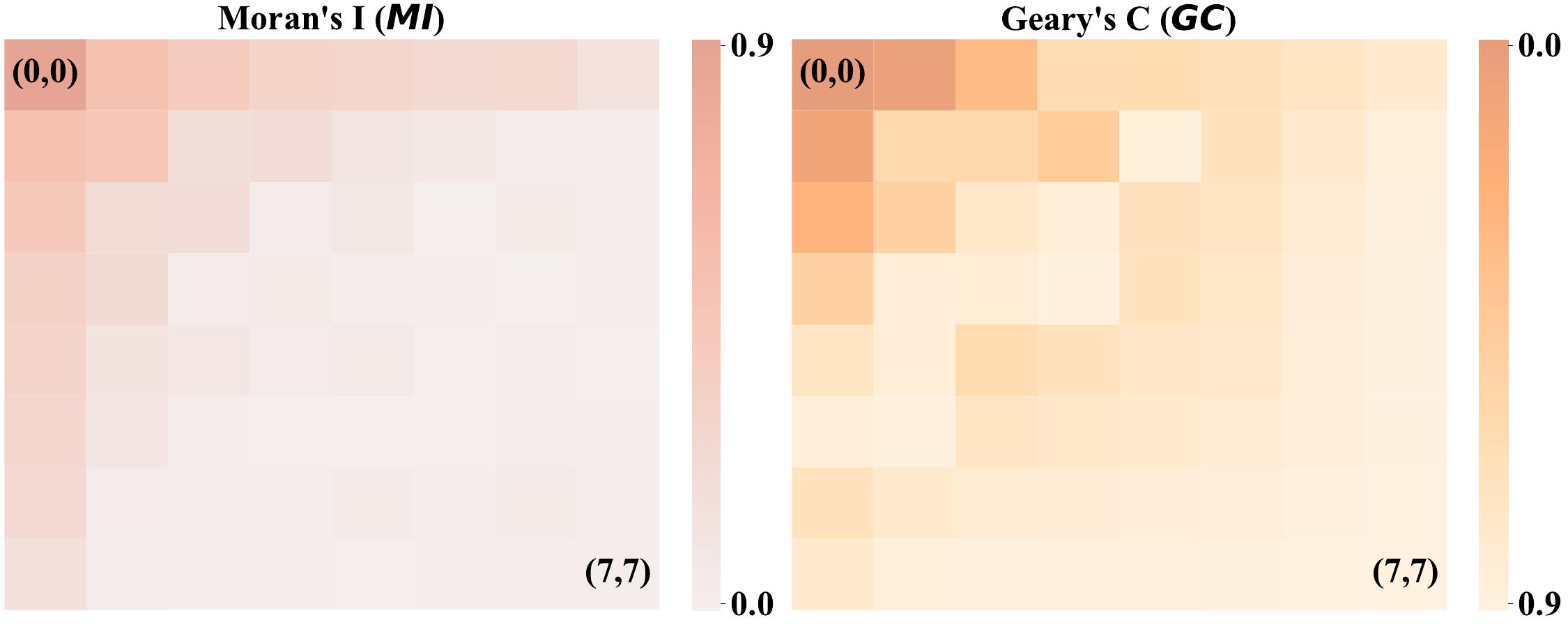}
        \caption{Y of DIV2K}
    \end{subfigure}
    \begin{subfigure}{0.33\linewidth}
        \centering
        \includegraphics[trim={0mm 0mm 0mm 0mm}, clip, width=\linewidth]{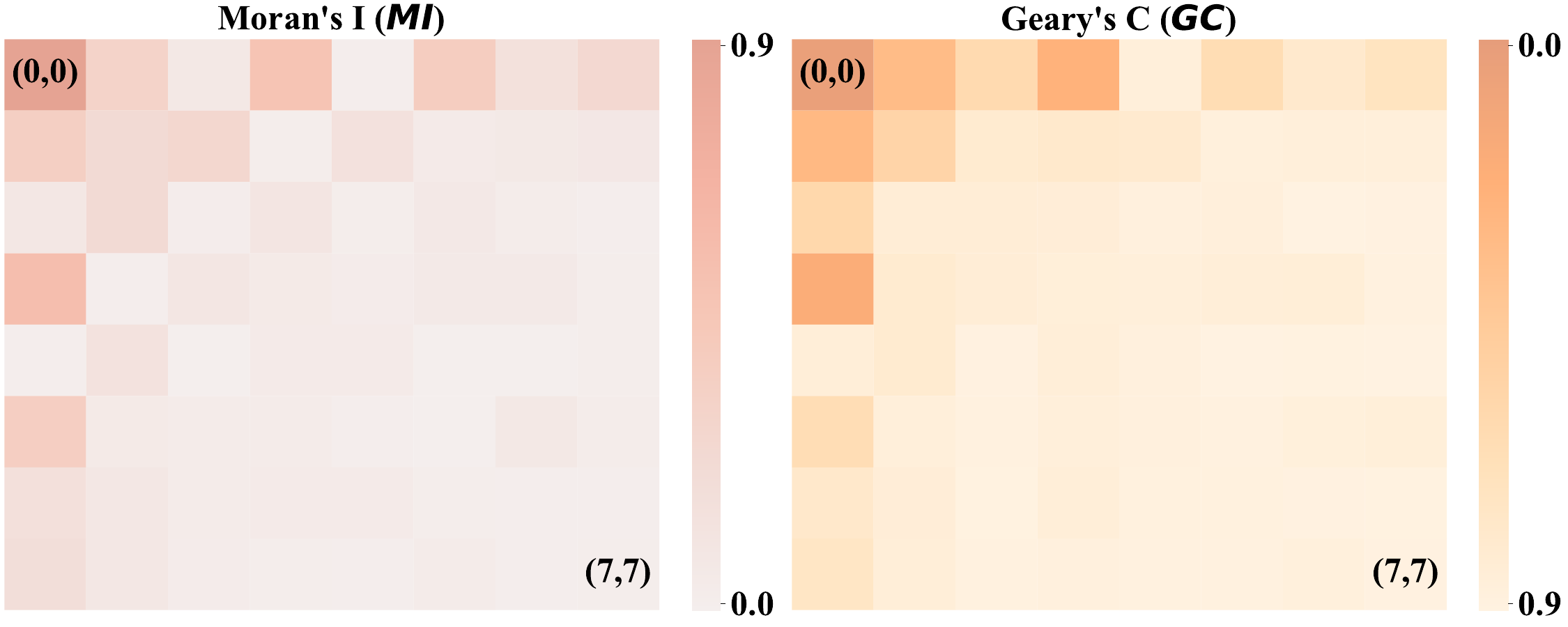}
        \caption{Cb of DIV2K}
    \end{subfigure}
    \begin{subfigure}{0.33\linewidth}
        \centering
        \includegraphics[trim={0mm 0mm 0mm 0mm}, clip, width=\linewidth]{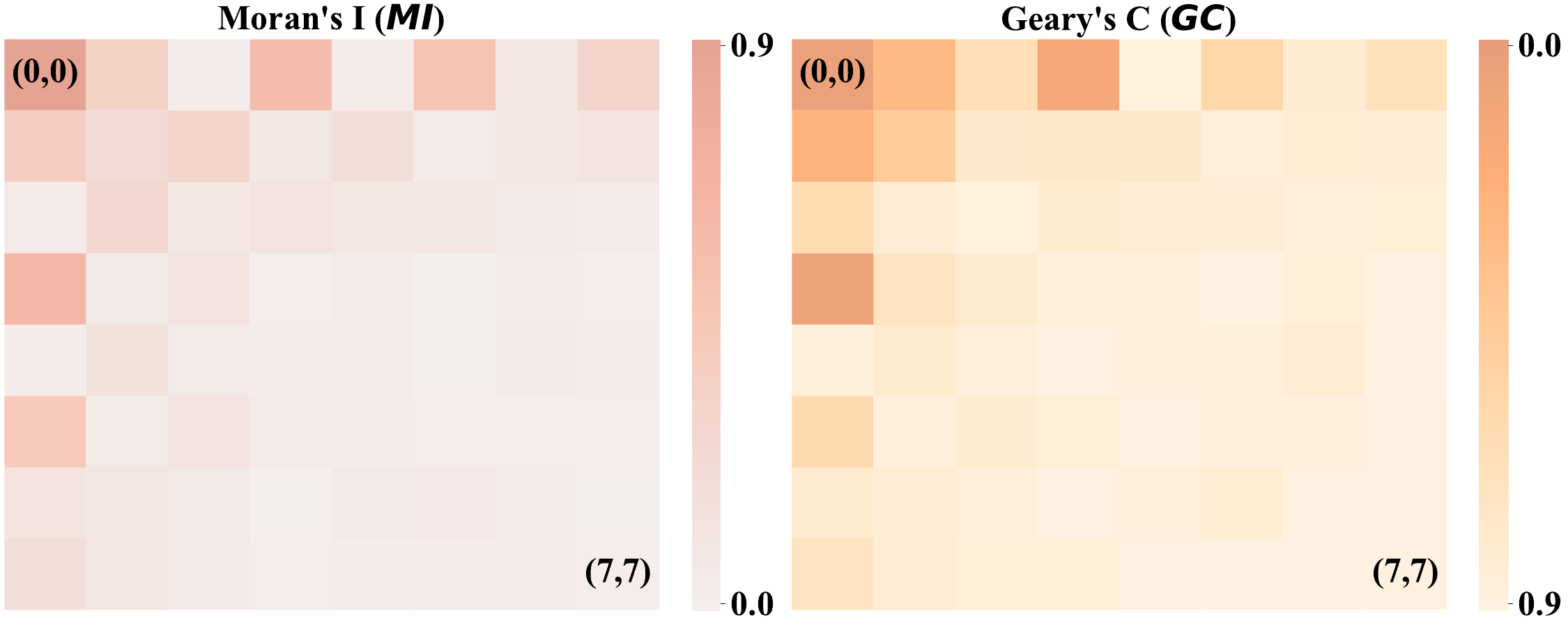}
        \caption{Cr of DIV2K}
    \end{subfigure}
    \begin{subfigure}{0.33\linewidth}
        \centering
        \includegraphics[trim={0mm 0mm 0mm 0mm}, clip, width=\linewidth]{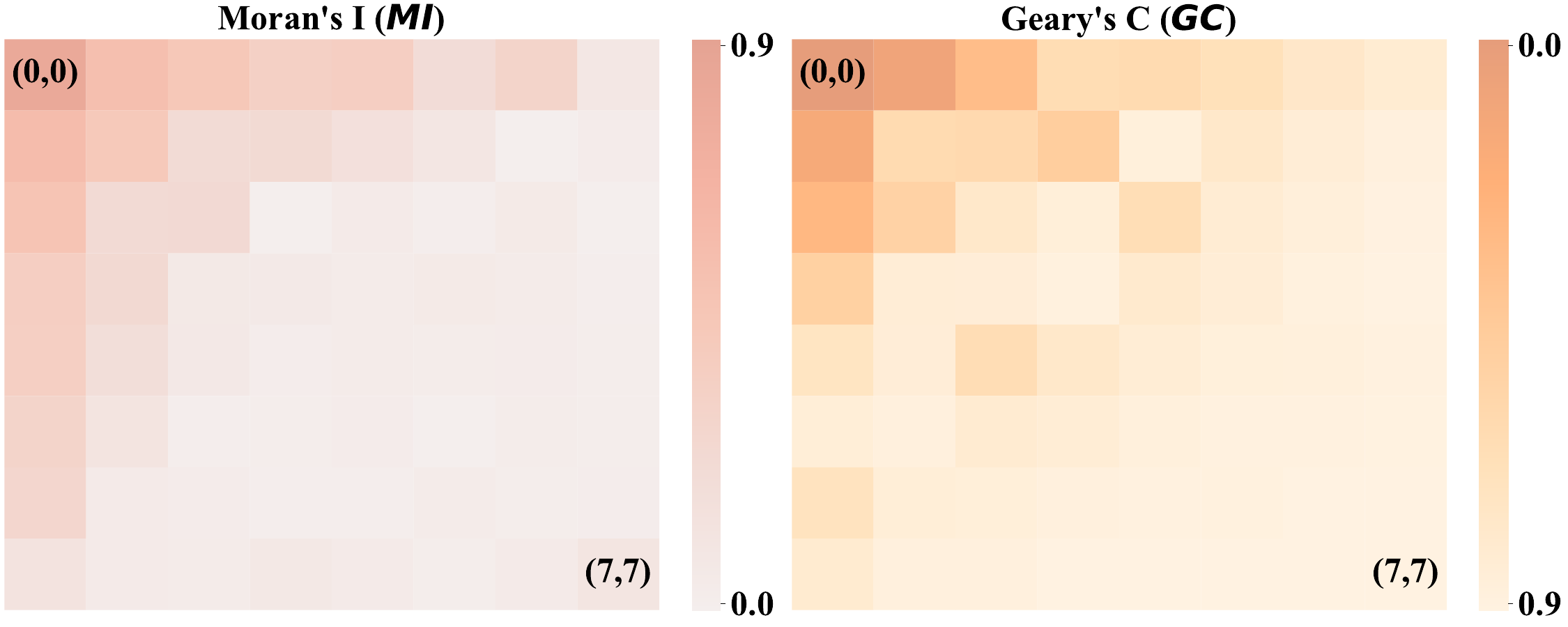}
        \caption{Y of BSDS500}
    \end{subfigure}
    \begin{subfigure}{0.33\linewidth}
        \centering
        \includegraphics[trim={0mm 0mm 0mm 0mm}, clip, width=\linewidth]{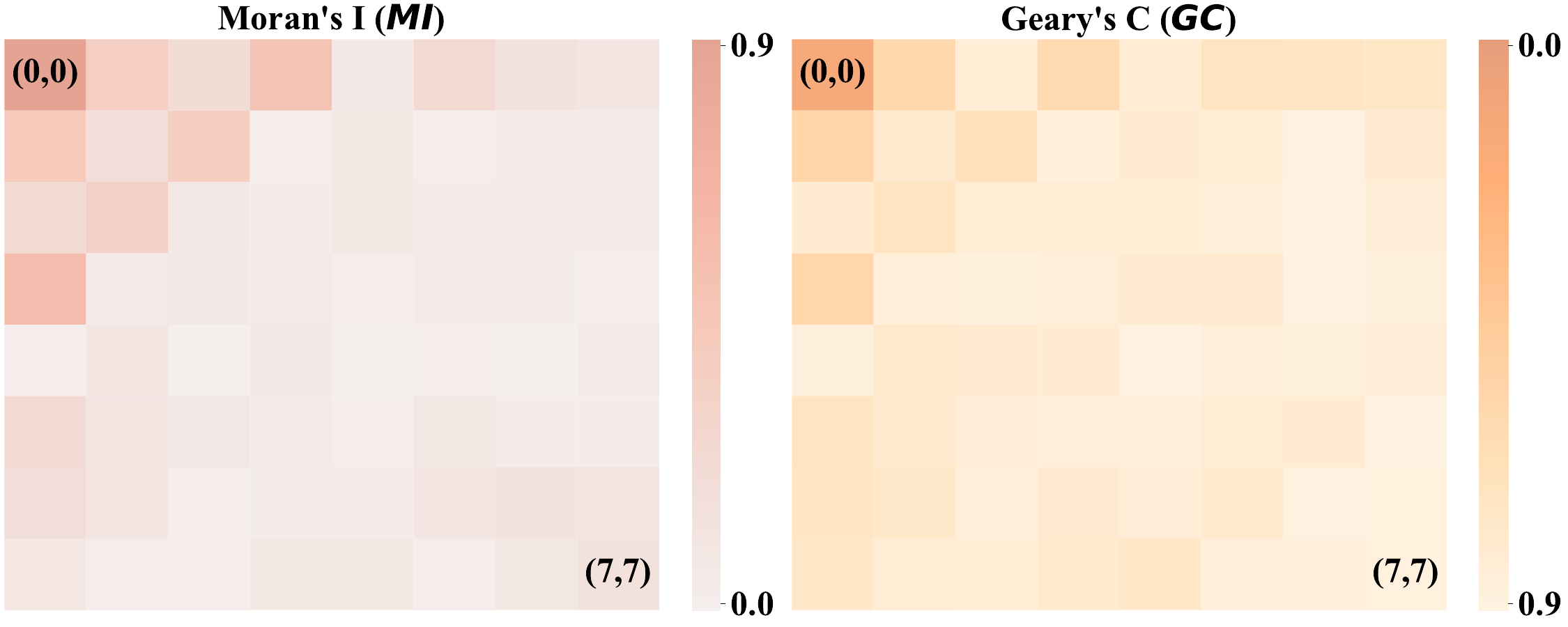}
        \caption{Cb of BSDS500}
    \end{subfigure}
    \begin{subfigure}{0.33\linewidth}
        \centering
        \includegraphics[trim={0mm 0mm 0mm 0mm}, clip, width=\linewidth]{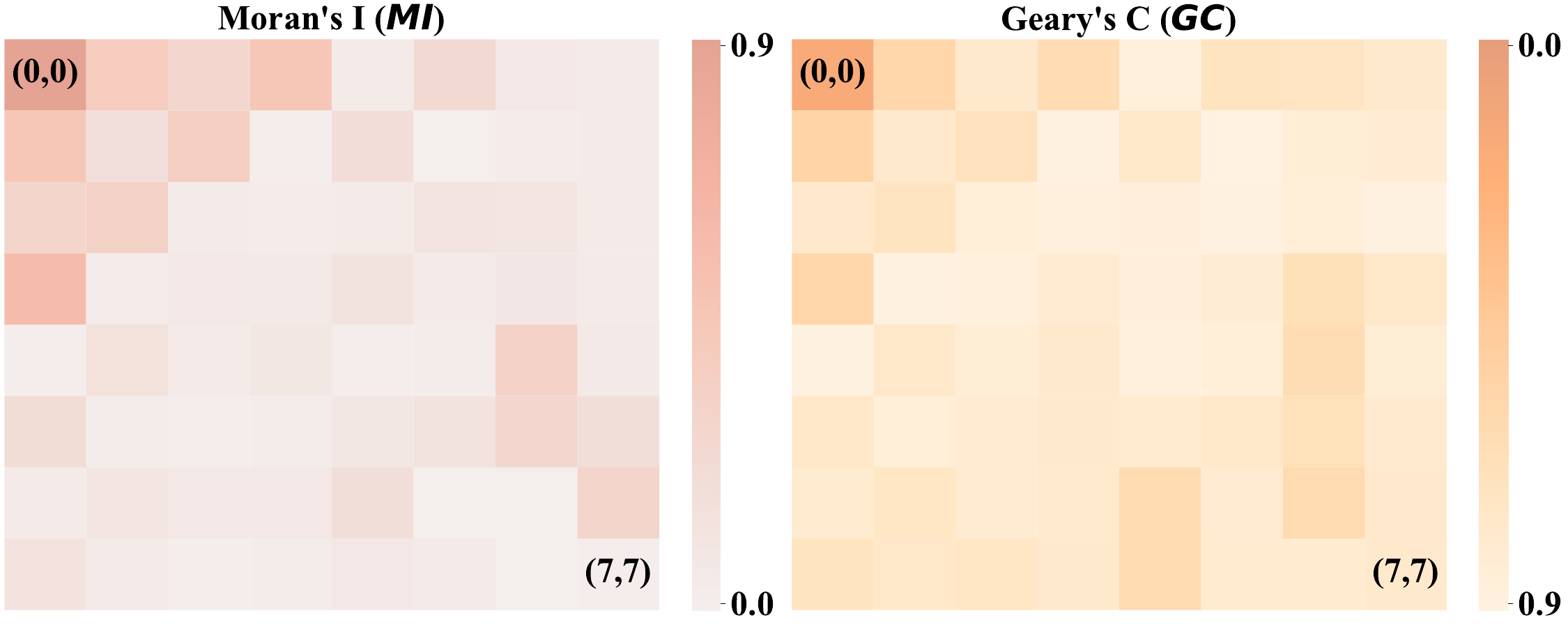}
        \caption{Cr of BSDS500}
    \end{subfigure}
    \vspace{\vspacelength}
    \caption{\textbf{Point-based correlations using coefficient maps on the DIV2K and BSDS500 datasets with QF set to 10.}
    Note that the intensity of heat maps indicate the strength of the correlations.}
    \label{fig:point_based_corr_supp_10}
\end{figure*}

\begin{figure*}[htbp]
    \centering
    \begin{subfigure}{0.33\linewidth}
        \centering
        \includegraphics[trim={9mm 102mm 22mm 109mm}, clip, width=\linewidth]{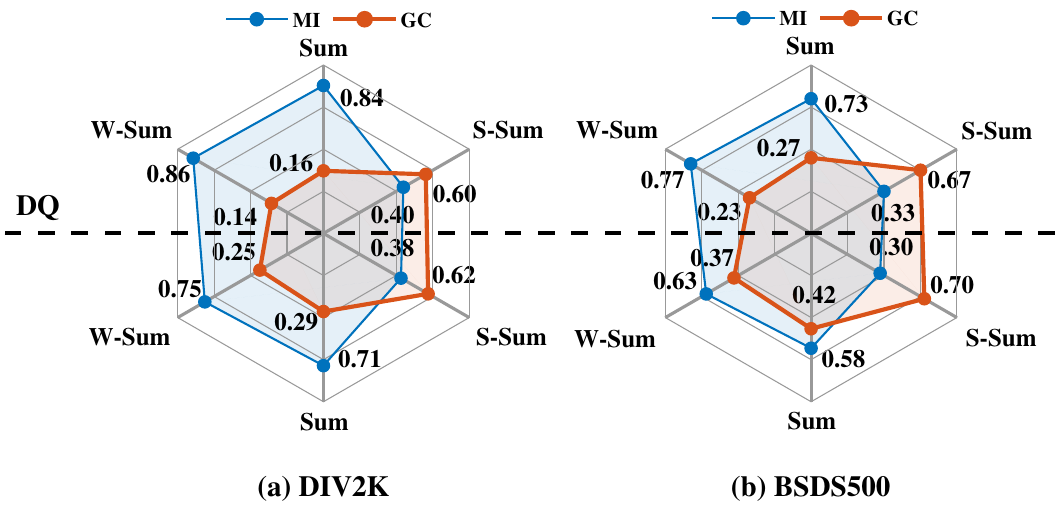}
        \caption{Y}
    \end{subfigure}
    \begin{subfigure}{.33\linewidth}
        \centering
        \includegraphics[trim={9mm 102mm 22mm 109mm}, clip, width=\linewidth]{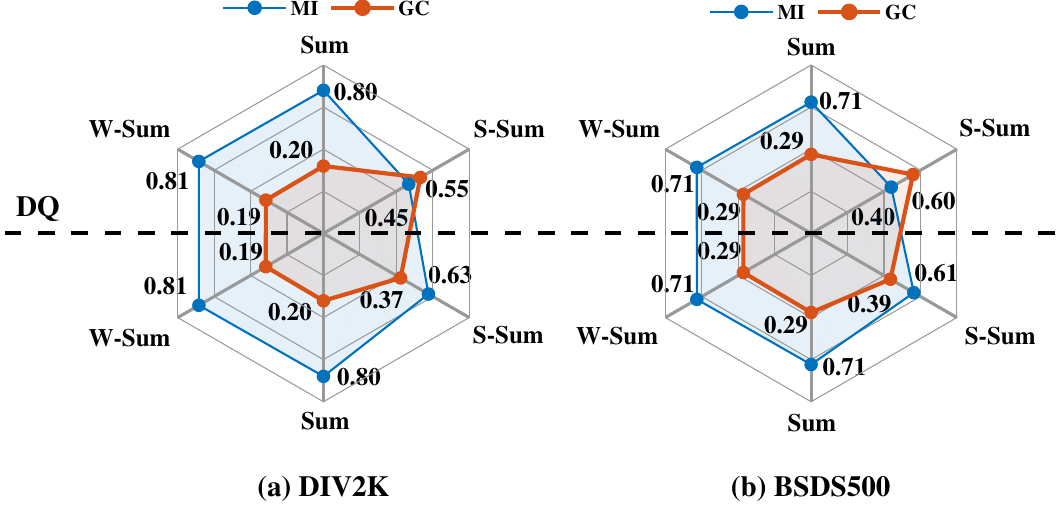}
        \caption{Cb}
    \end{subfigure}
    \begin{subfigure}{.33\linewidth}
        \centering
        \includegraphics[trim={9mm 102mm 22mm 109mm}, clip, width=\linewidth]{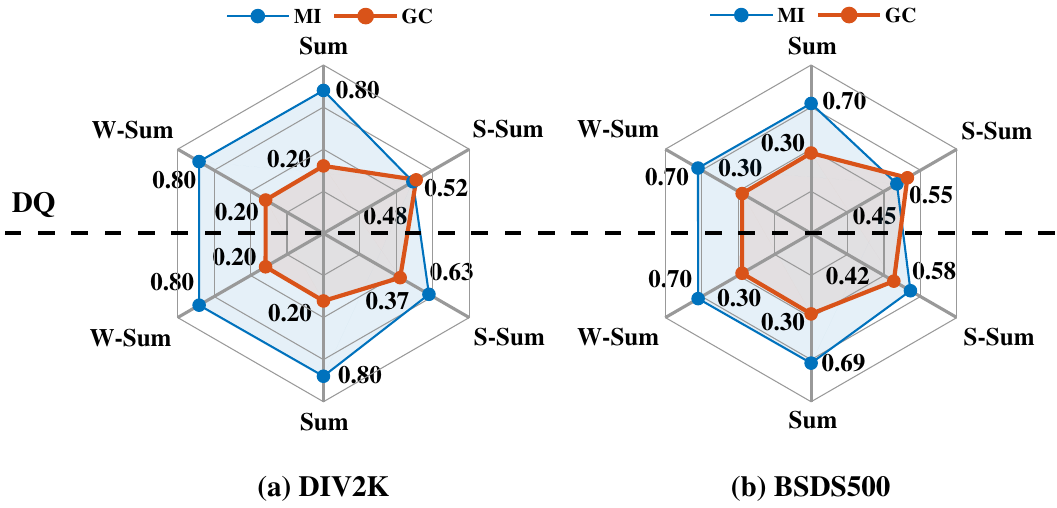}
        \caption{Cr}
    \end{subfigure}
    \vspace{\vspacelength}
    \caption{\textbf{Block-based correlations using different block-based features on the DIV2K and BSDS500 datasets with QF set to 20.}
    Upper: DCT blocks are dequantized before calculating feature values.
    Lower: DCT blocks remain quantized.}
    \label{fig:block_based_corr_supp_20}
\end{figure*}

\begin{figure*}[htbp]
    \centering
    \begin{subfigure}{0.33\linewidth}
        \centering
        \includegraphics[trim={0mm 0mm 0mm 0mm}, clip, width=\linewidth]{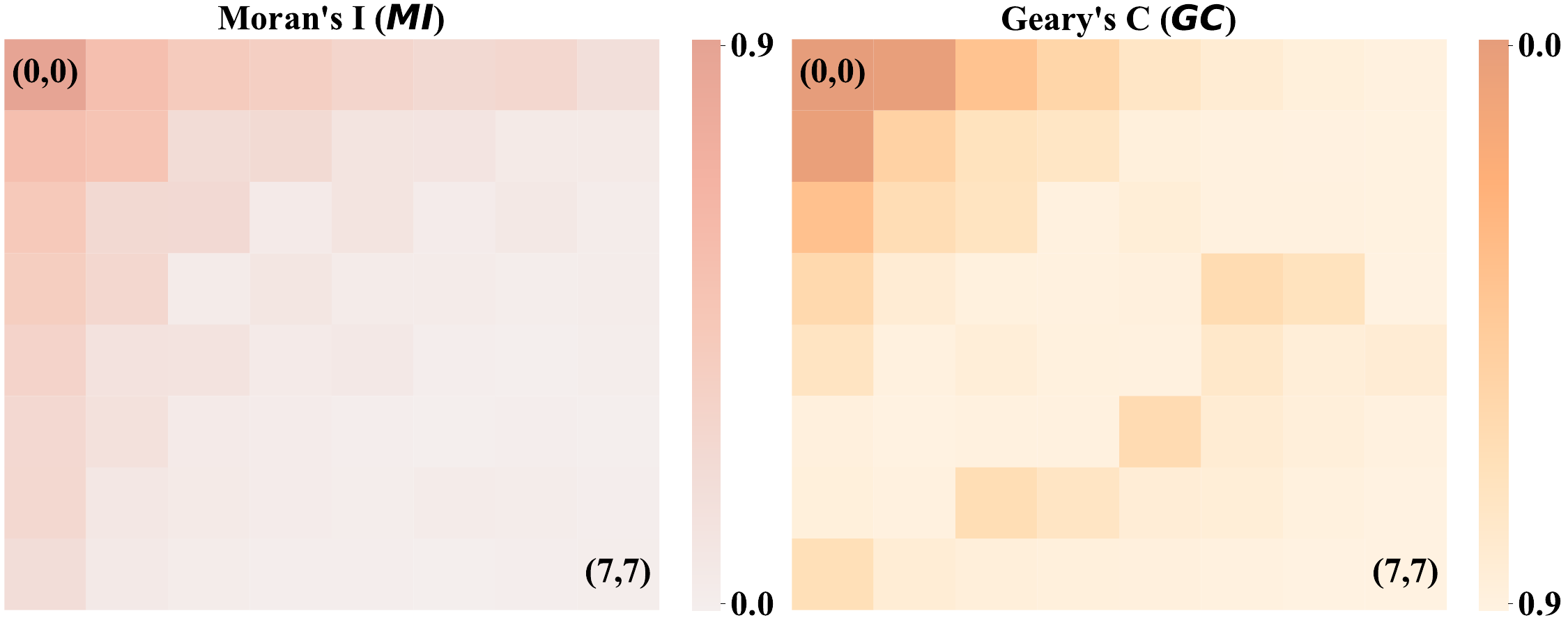}
        \caption{Y of DIV2K}
    \end{subfigure}
    \begin{subfigure}{0.33\linewidth}
        \centering
        \includegraphics[trim={0mm 0mm 0mm 0mm}, clip, width=\linewidth]{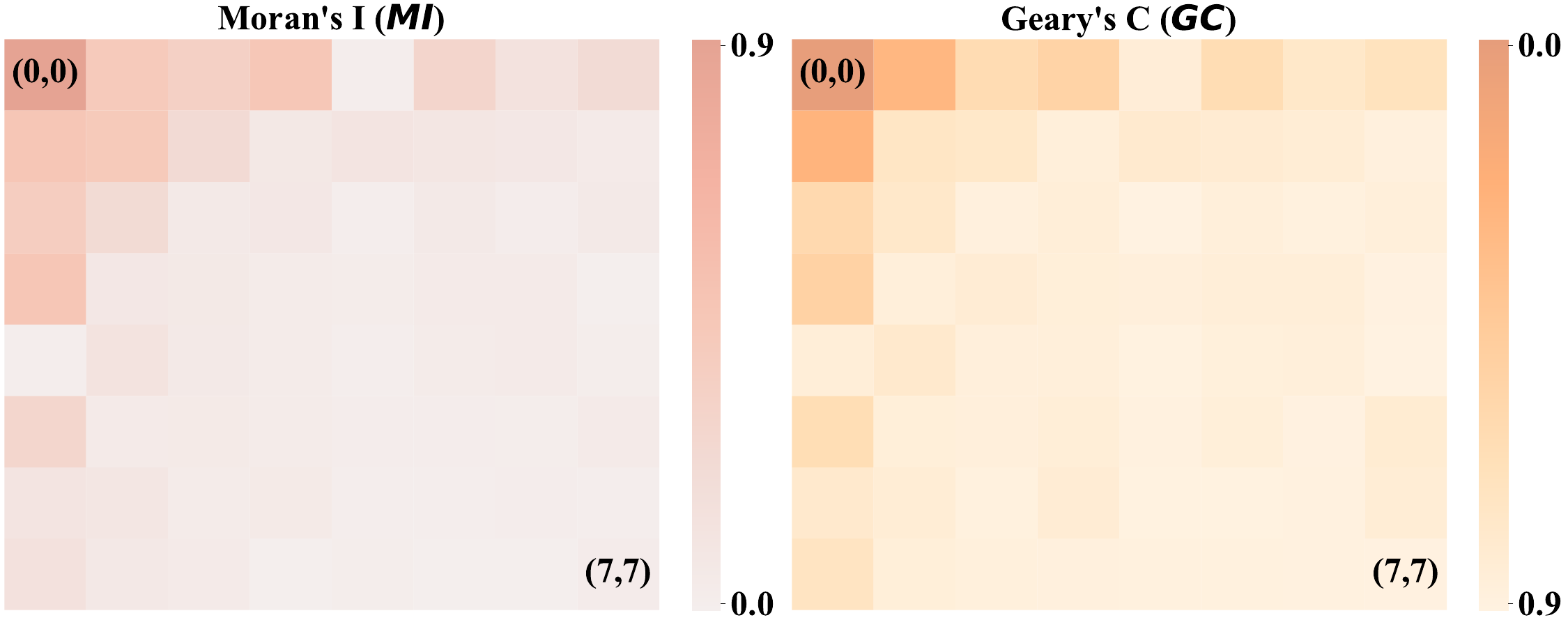}
        \caption{Cb of DIV2K}
    \end{subfigure}
    \begin{subfigure}{0.33\linewidth}
        \centering
        \includegraphics[trim={0mm 0mm 0mm 0mm}, clip, width=\linewidth]{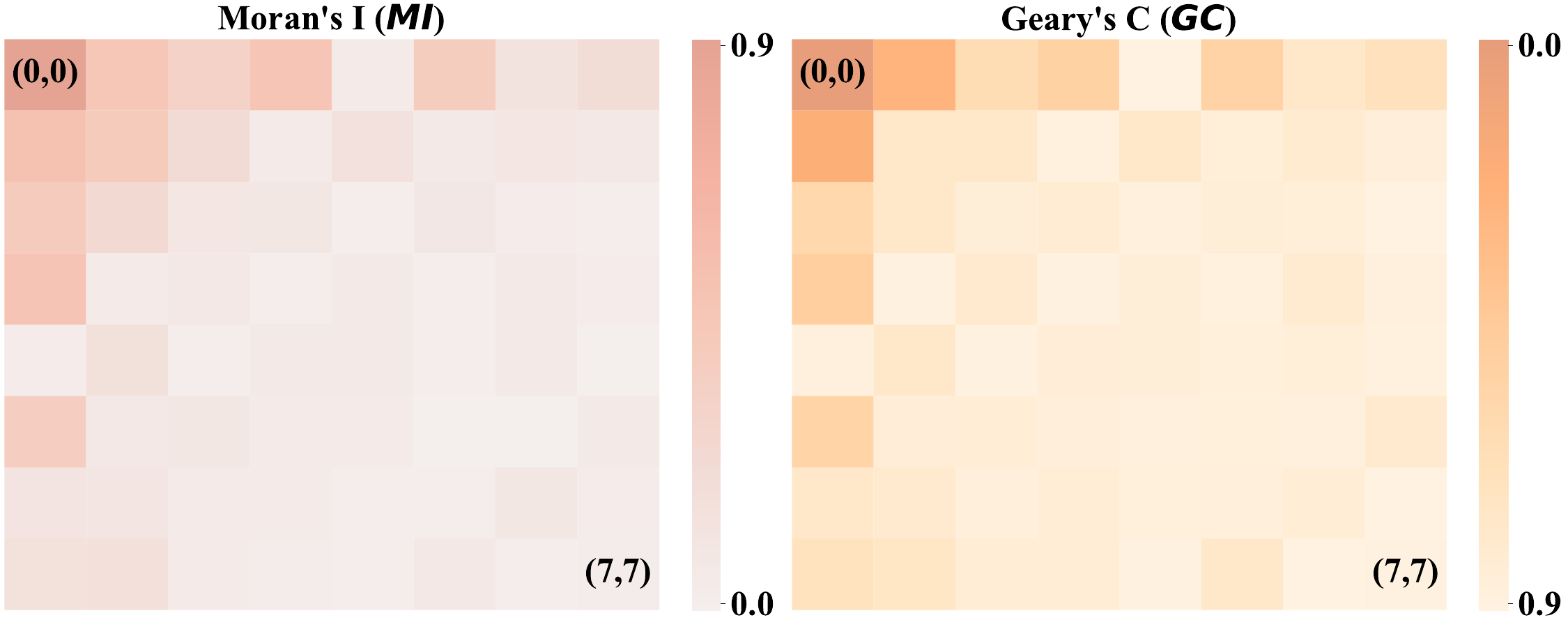}
        \caption{Cr of DIV2K}
    \end{subfigure}
    \begin{subfigure}{0.33\linewidth}
        \centering
        \includegraphics[trim={0mm 0mm 0mm 0mm}, clip, width=\linewidth]{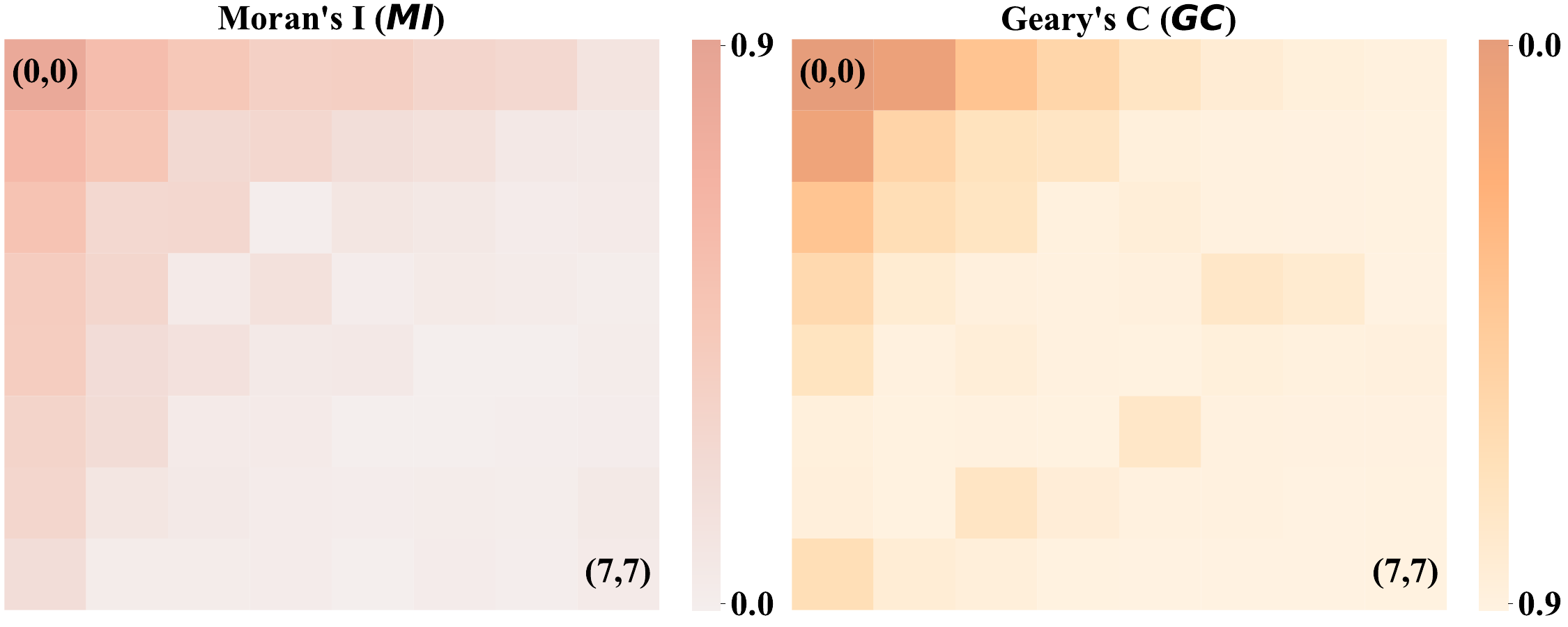}
        \caption{Y of BSDS500}
    \end{subfigure}
    \begin{subfigure}{0.33\linewidth}
        \centering
        \includegraphics[trim={0mm 0mm 0mm 0mm}, clip, width=\linewidth]{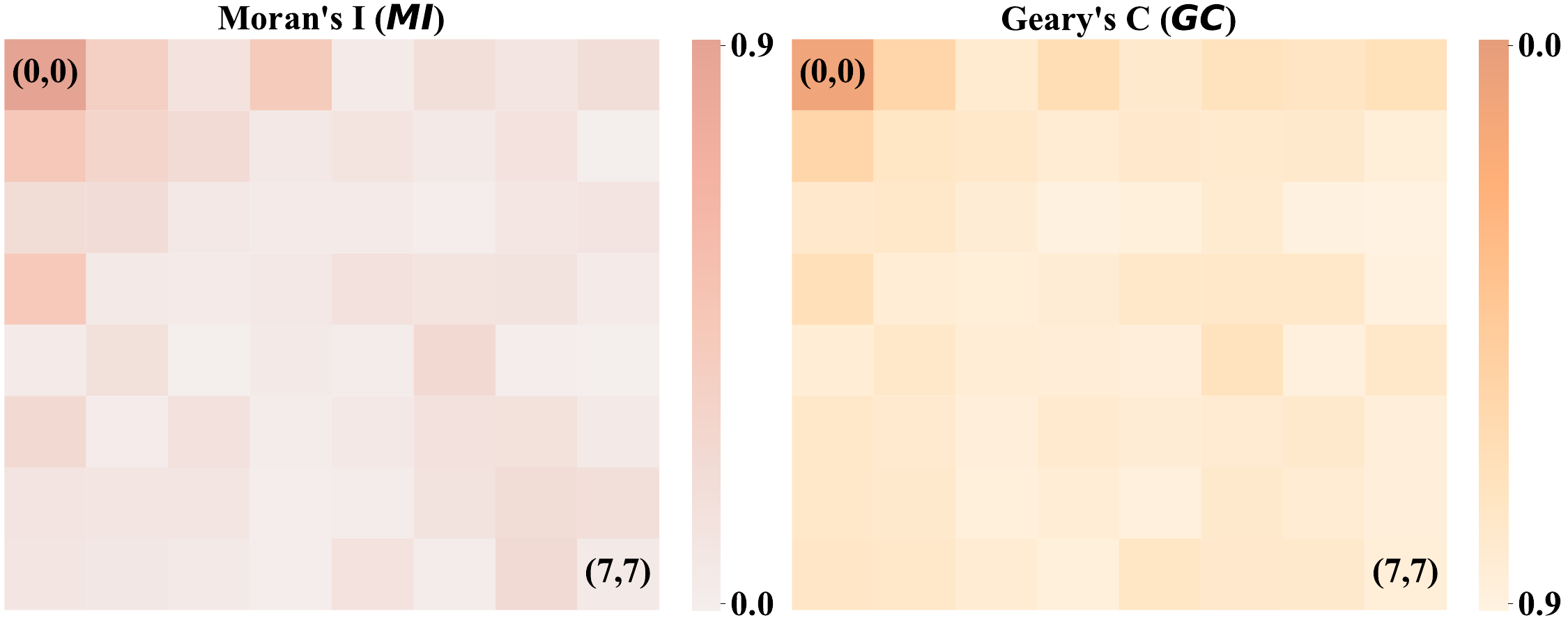}
        \caption{Cb of BSDS500}
    \end{subfigure}
    \begin{subfigure}{0.33\linewidth}
        \centering
        \includegraphics[trim={0mm 0mm 0mm 0mm}, clip, width=\linewidth]{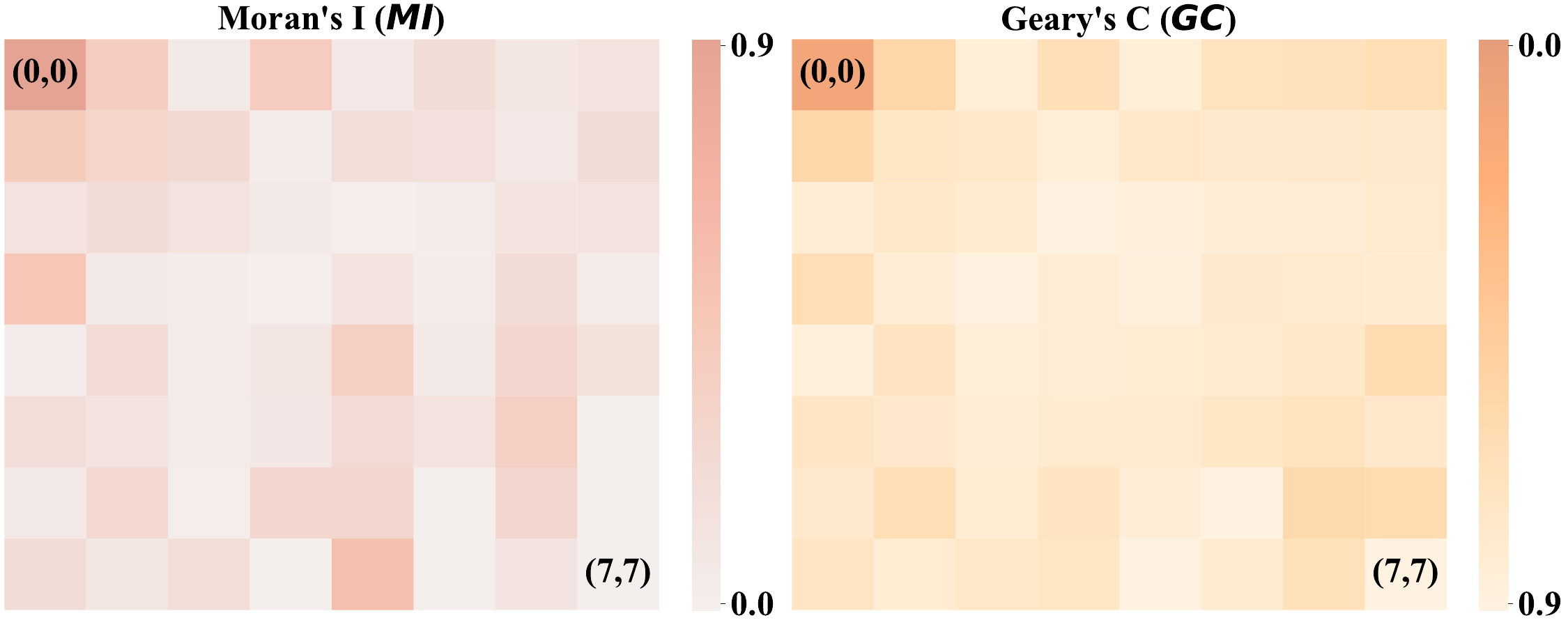}
        \caption{Cr of BSDS500}
    \end{subfigure}
    \vspace{\vspacelength}
    \caption{\textbf{Point-based correlations using coefficient maps on the DIV2K and BSDS500 datasets with QF set to 20.}
    Note that the intensity of heat maps indicate the strength of the correlations.}
    \label{fig:point_based_corr_supp_20}
\end{figure*}

\begin{figure*}[htbp]
    \centering
    \begin{subfigure}{0.33\linewidth}
        \centering
        \includegraphics[trim={9mm 102mm 22mm 109mm}, clip, width=\linewidth]{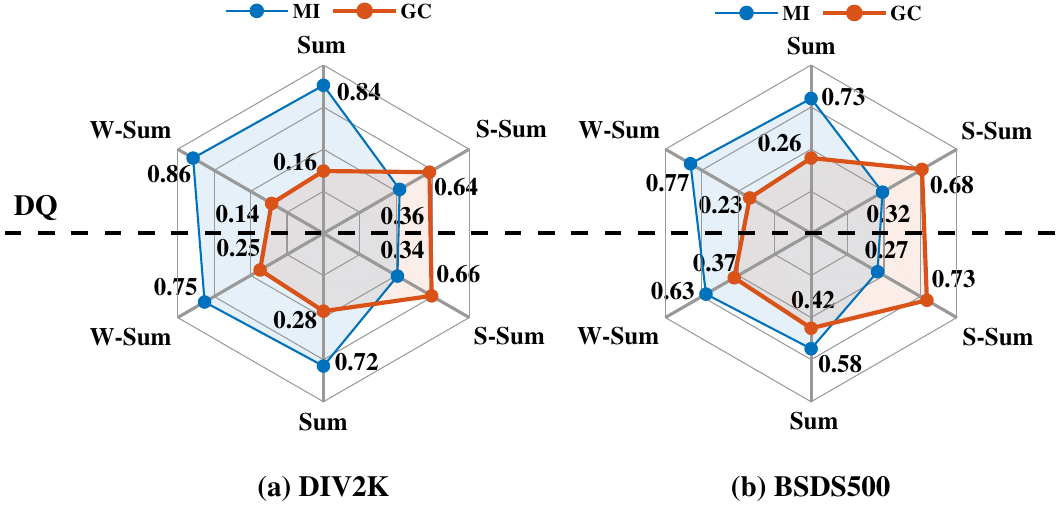}
        \caption{Y}
    \end{subfigure}
    \begin{subfigure}{.33\linewidth}
        \centering
        \includegraphics[trim={9mm 102mm 22mm 109mm}, clip, width=\linewidth]{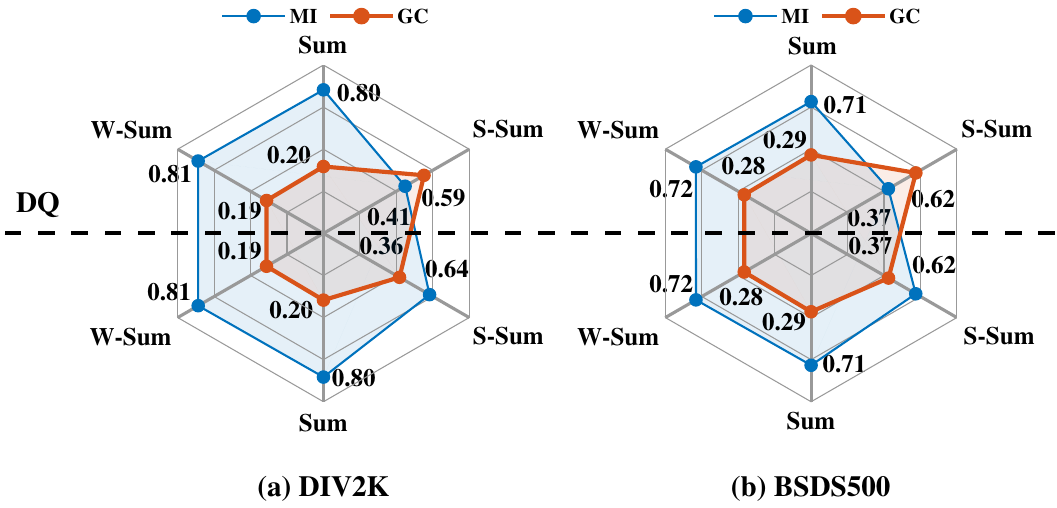}
        \caption{Cb}
    \end{subfigure}
    \begin{subfigure}{.33\linewidth}
        \centering
        \includegraphics[trim={9mm 102mm 22mm 109mm}, clip, width=\linewidth]{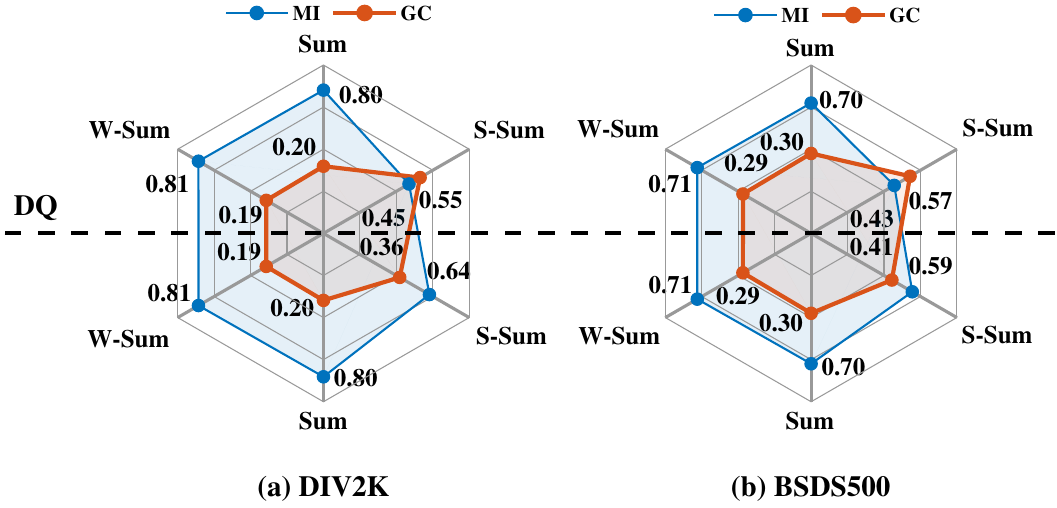}
        \caption{Cr}
    \end{subfigure}
    \vspace{\vspacelength}
    \caption{\textbf{Block-based correlations using different block-based features on the DIV2K and BSDS500 datasets with QF set to 30.}
    Upper: DCT blocks are dequantized before calculating feature values.
    Lower: DCT blocks remain quantized.}
    \label{fig:block_based_corr_supp_30}
\end{figure*}

\begin{figure*}[htbp]
    \centering
    \begin{subfigure}{0.33\linewidth}
        \centering
        \includegraphics[trim={0mm 0mm 0mm 0mm}, clip, width=\linewidth]{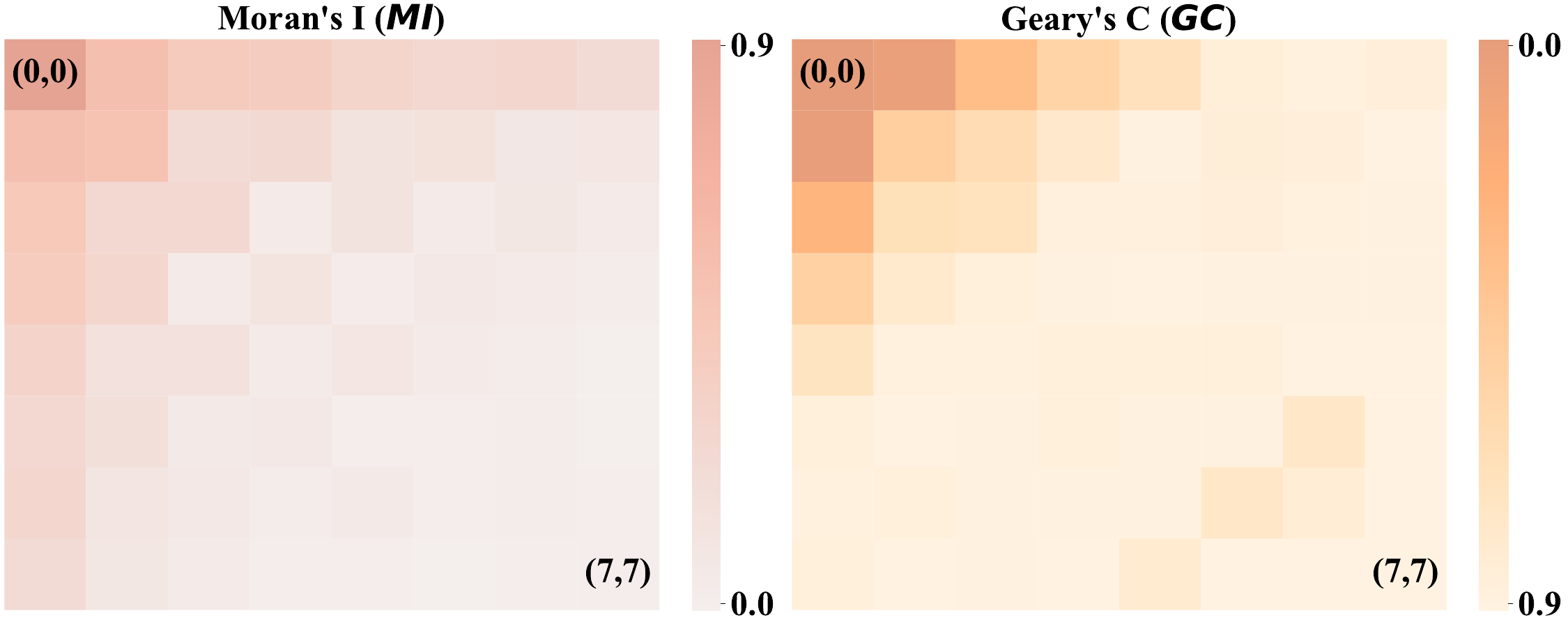}
        \caption{Y of DIV2K}
    \end{subfigure}
    \begin{subfigure}{0.33\linewidth}
        \centering
        \includegraphics[trim={0mm 0mm 0mm 0mm}, clip, width=\linewidth]{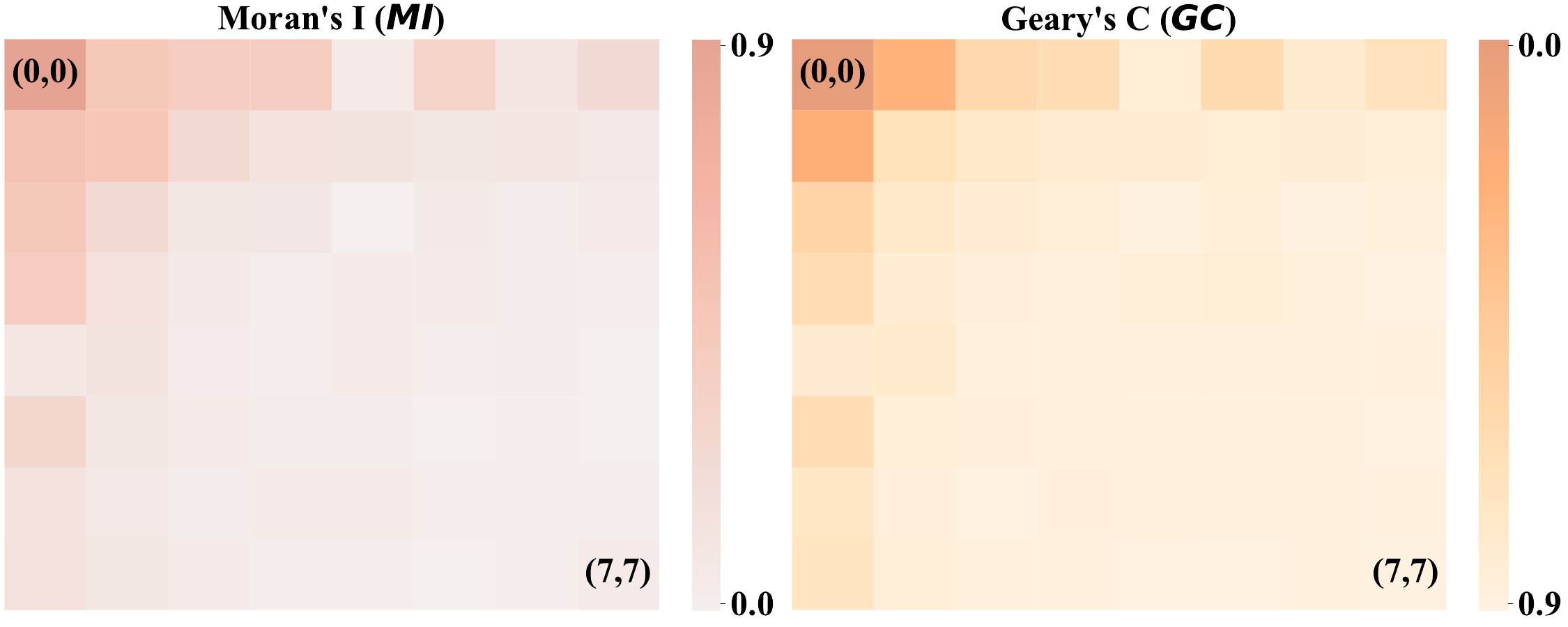}
        \caption{Cb of DIV2K}
    \end{subfigure}
    \begin{subfigure}{0.33\linewidth}
        \centering
        \includegraphics[trim={0mm 0mm 0mm 0mm}, clip, width=\linewidth]{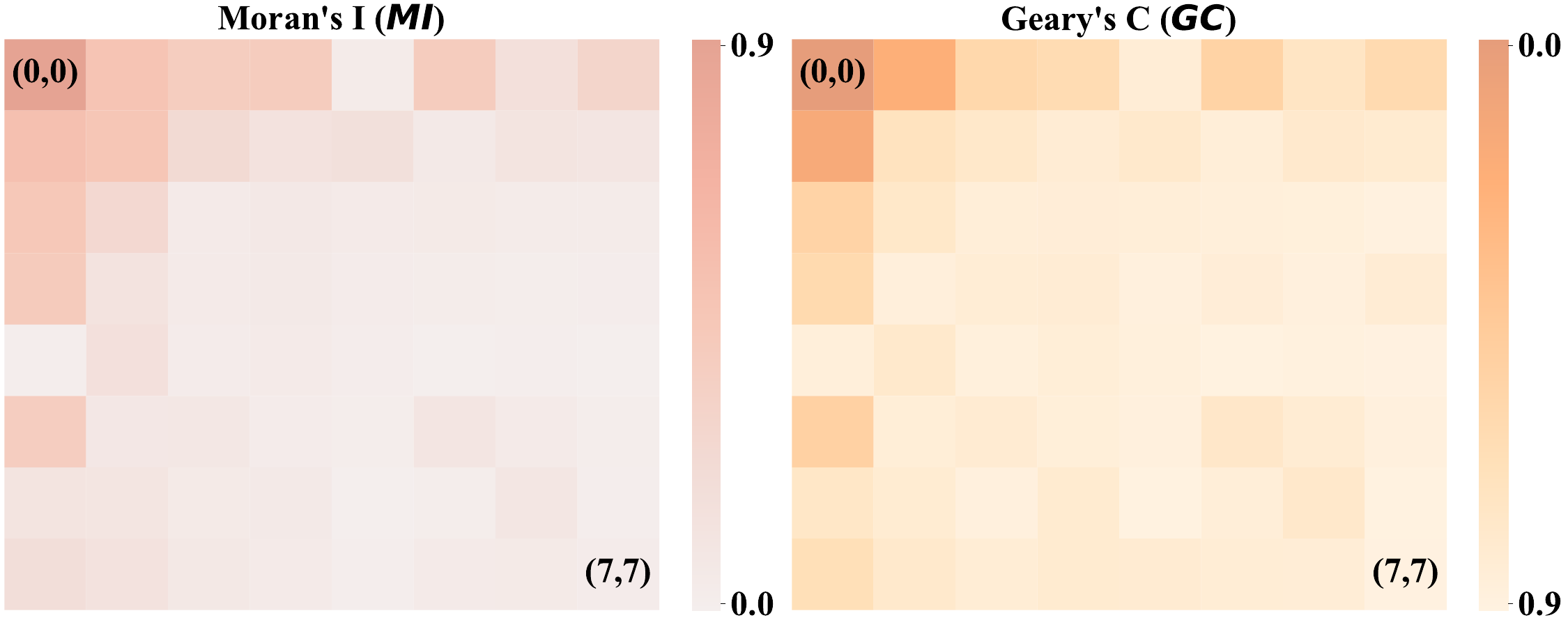}
        \caption{Cr of DIV2K}
    \end{subfigure}
    \begin{subfigure}{0.33\linewidth}
        \centering
        \includegraphics[trim={0mm 0mm 0mm 0mm}, clip, width=\linewidth]{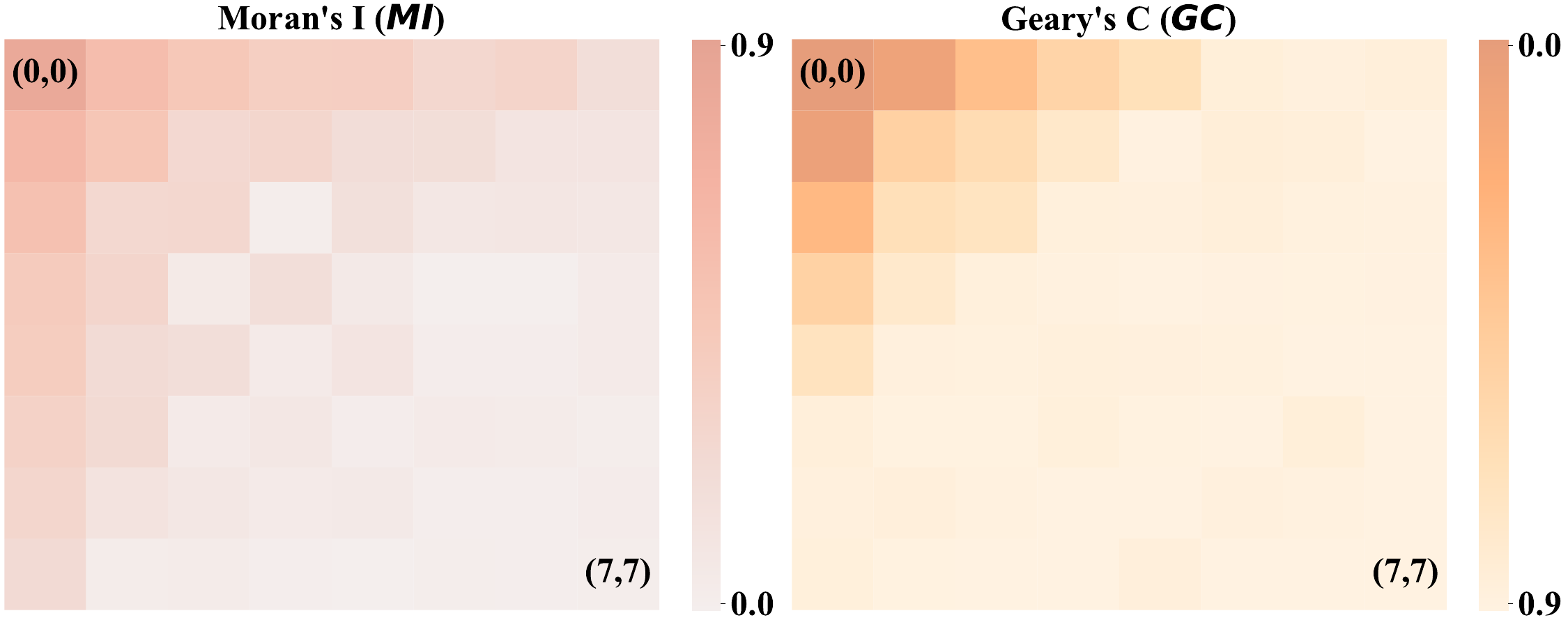}
        \caption{Y of BSDS500}
    \end{subfigure}
    \begin{subfigure}{0.33\linewidth}
        \centering
        \includegraphics[trim={0mm 0mm 0mm 0mm}, clip, width=\linewidth]{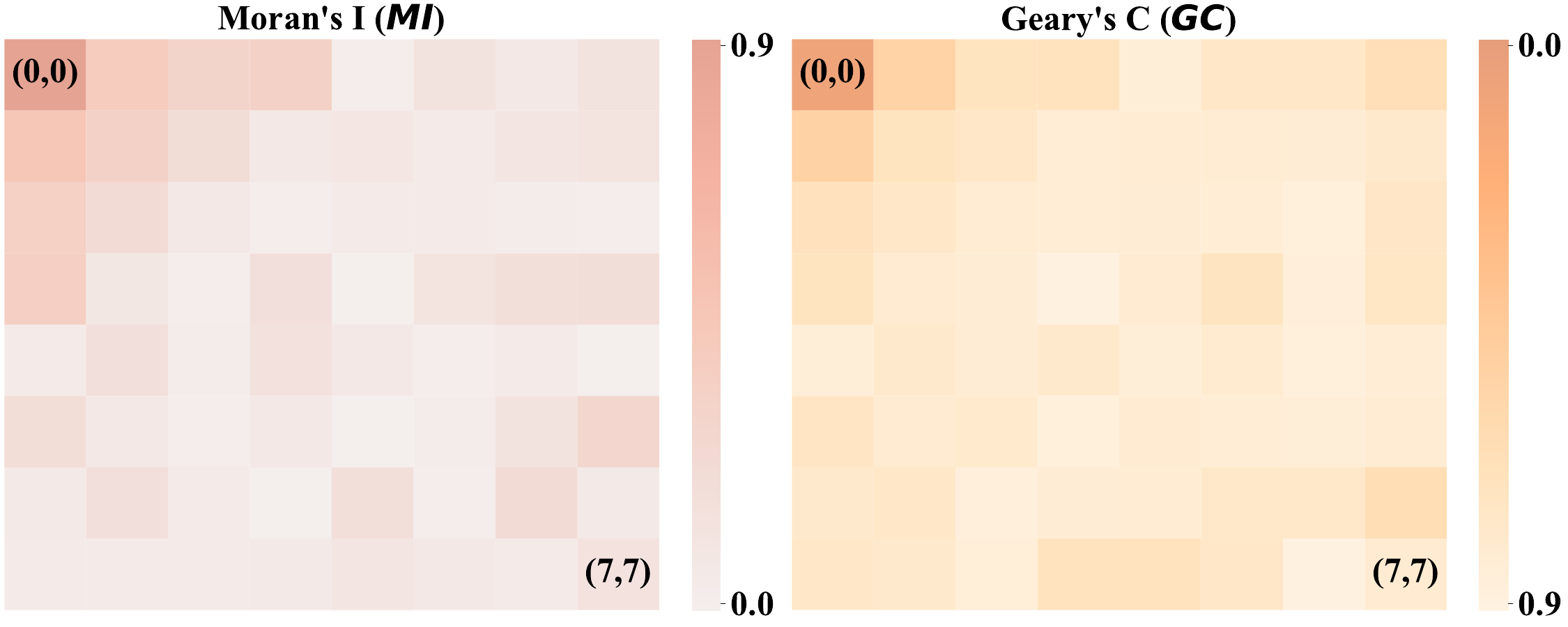}
        \caption{Cb of BSDS500}
    \end{subfigure}
    \begin{subfigure}{0.33\linewidth}
        \centering
        \includegraphics[trim={0mm 0mm 0mm 0mm}, clip, width=\linewidth]{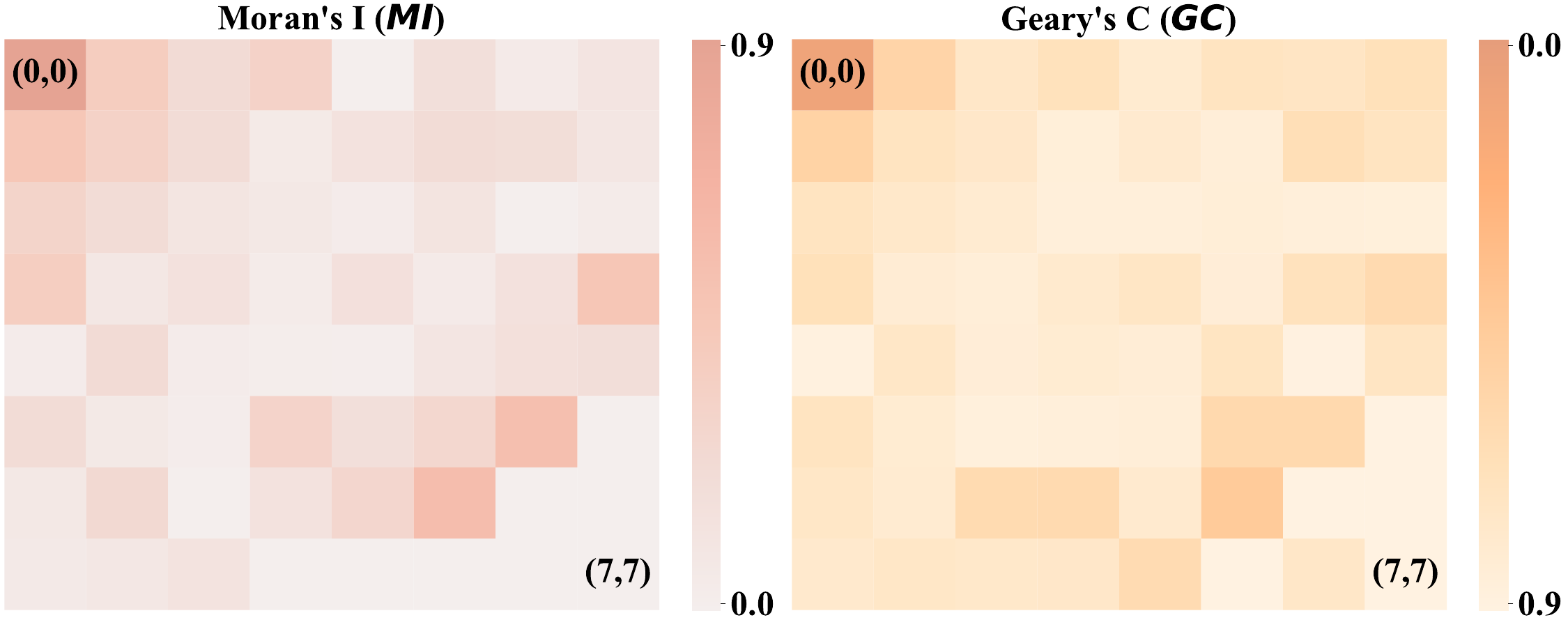}
        \caption{Cr of BSDS500}
    \end{subfigure}
    \vspace{\vspacelength}
    \caption{\textbf{Point-based correlations using coefficient maps on the DIV2K and BSDS500 datasets with QF set to 30.}
    Note that the intensity of heat maps indicate the strength of the correlations.}
    \label{fig:point_based_corr_supp_30}
\end{figure*}

\begin{figure*}[htbp]
    \centering
    \begin{subfigure}{0.33\linewidth}
        \centering
        \includegraphics[trim={9mm 102mm 22mm 109mm}, clip, width=\linewidth]{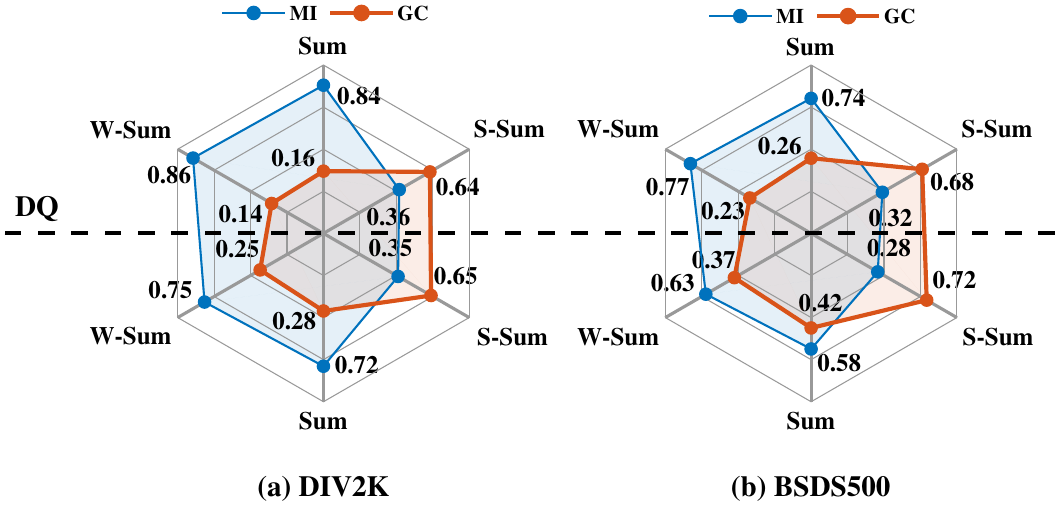}
        \caption{Y}
    \end{subfigure}
    \begin{subfigure}{.33\linewidth}
        \centering
        \includegraphics[trim={9mm 102mm 22mm 109mm}, clip, width=\linewidth]{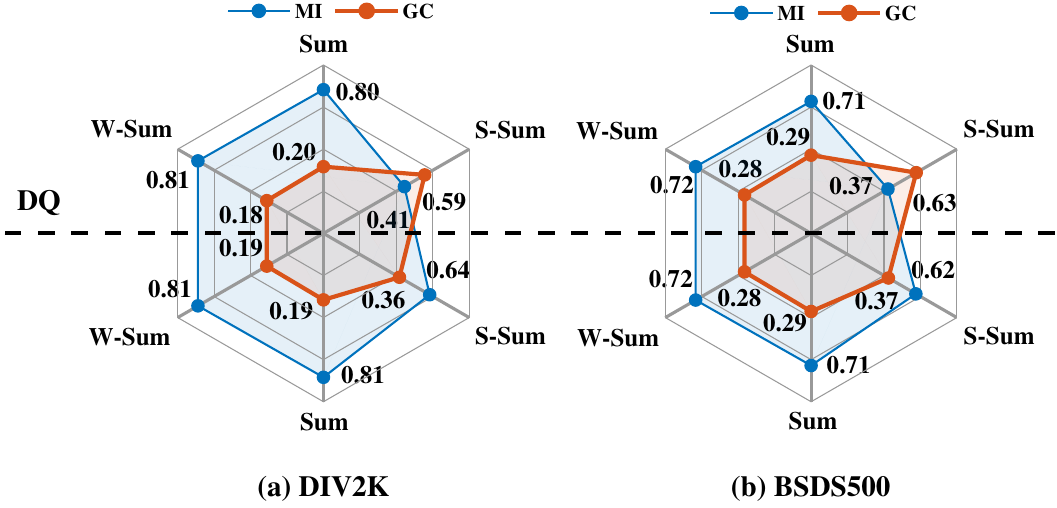}
        \caption{Cb}
    \end{subfigure}
    \begin{subfigure}{.33\linewidth}
        \centering
        \includegraphics[trim={9mm 102mm 22mm 109mm}, clip, width=\linewidth]{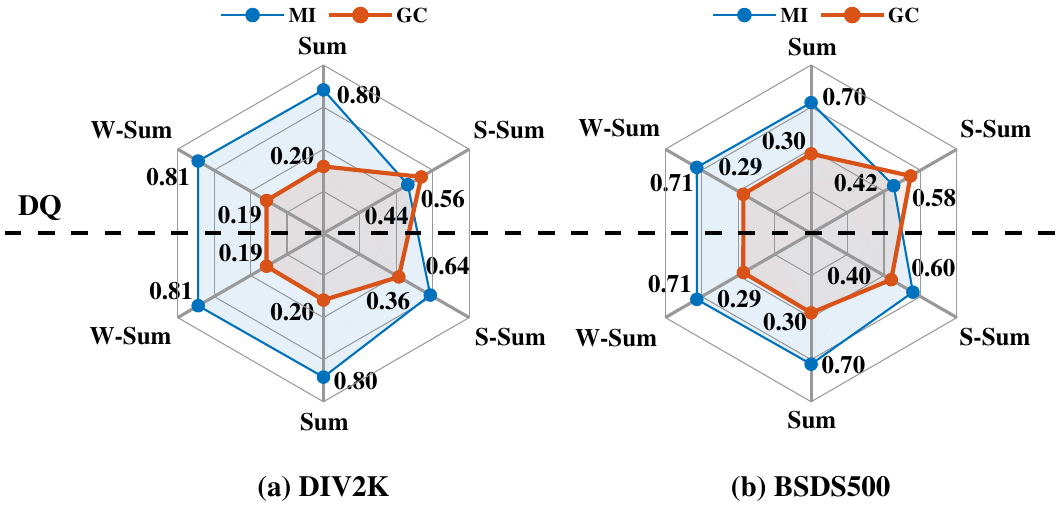}
        \caption{Cr}
    \end{subfigure}
    \vspace{\vspacelength}
    \caption{\textbf{Block-based correlations using different block-based features on the DIV2K and BSDS500 datasets with QF set to 40.}
    Upper: DCT blocks are dequantized before calculating feature values.
    Lower: DCT blocks remain quantized.}
    \label{fig:block_based_corr_supp_40}
\end{figure*}

\begin{figure*}[htbp]
    \centering
    \begin{subfigure}{0.33\linewidth}
        \centering
        \includegraphics[trim={0mm 0mm 0mm 0mm}, clip, width=\linewidth]{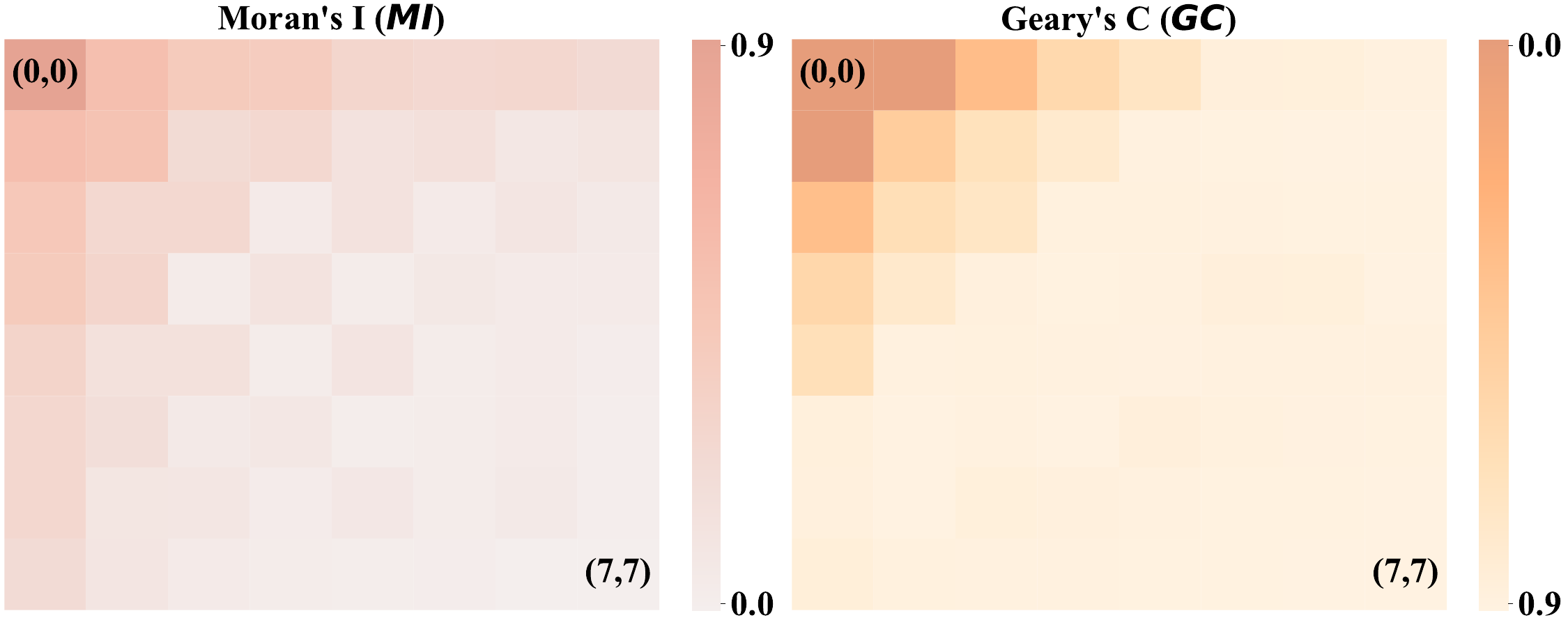}
        \caption{Y of DIV2K}
    \end{subfigure}
    \begin{subfigure}{0.33\linewidth}
        \centering
        \includegraphics[trim={0mm 0mm 0mm 0mm}, clip, width=\linewidth]{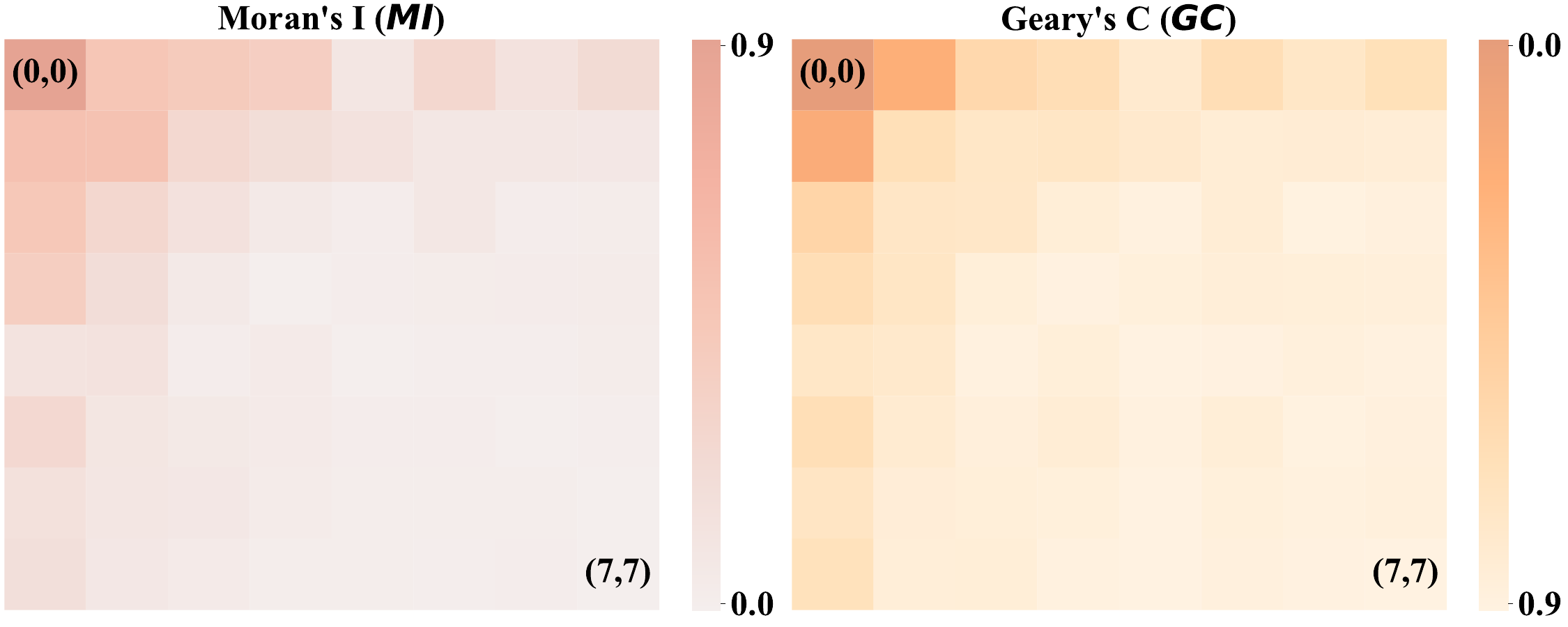}
        \caption{Cb of DIV2K}
    \end{subfigure}
    \begin{subfigure}{0.33\linewidth}
        \centering
        \includegraphics[trim={0mm 0mm 0mm 0mm}, clip, width=\linewidth]{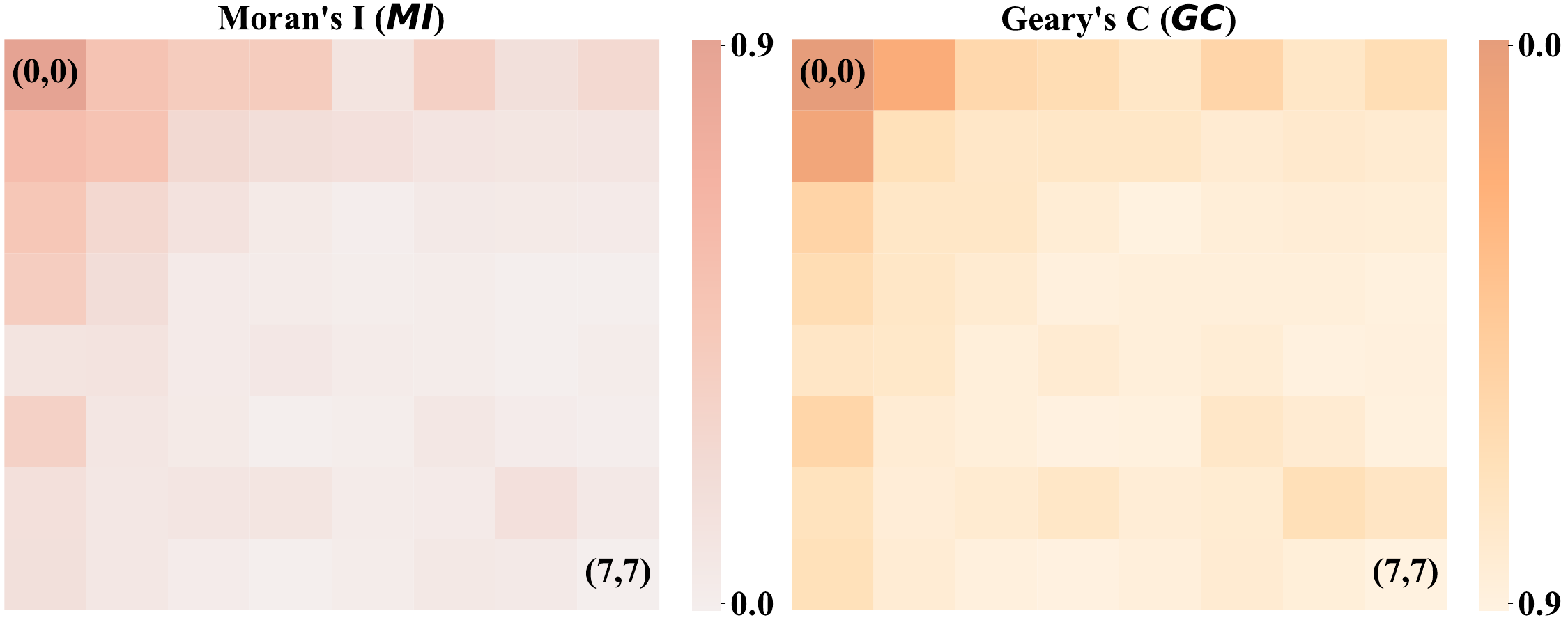}
        \caption{Cr of DIV2K}
    \end{subfigure}
    \begin{subfigure}{0.33\linewidth}
        \centering
        \includegraphics[trim={0mm 0mm 0mm 0mm}, clip, width=\linewidth]{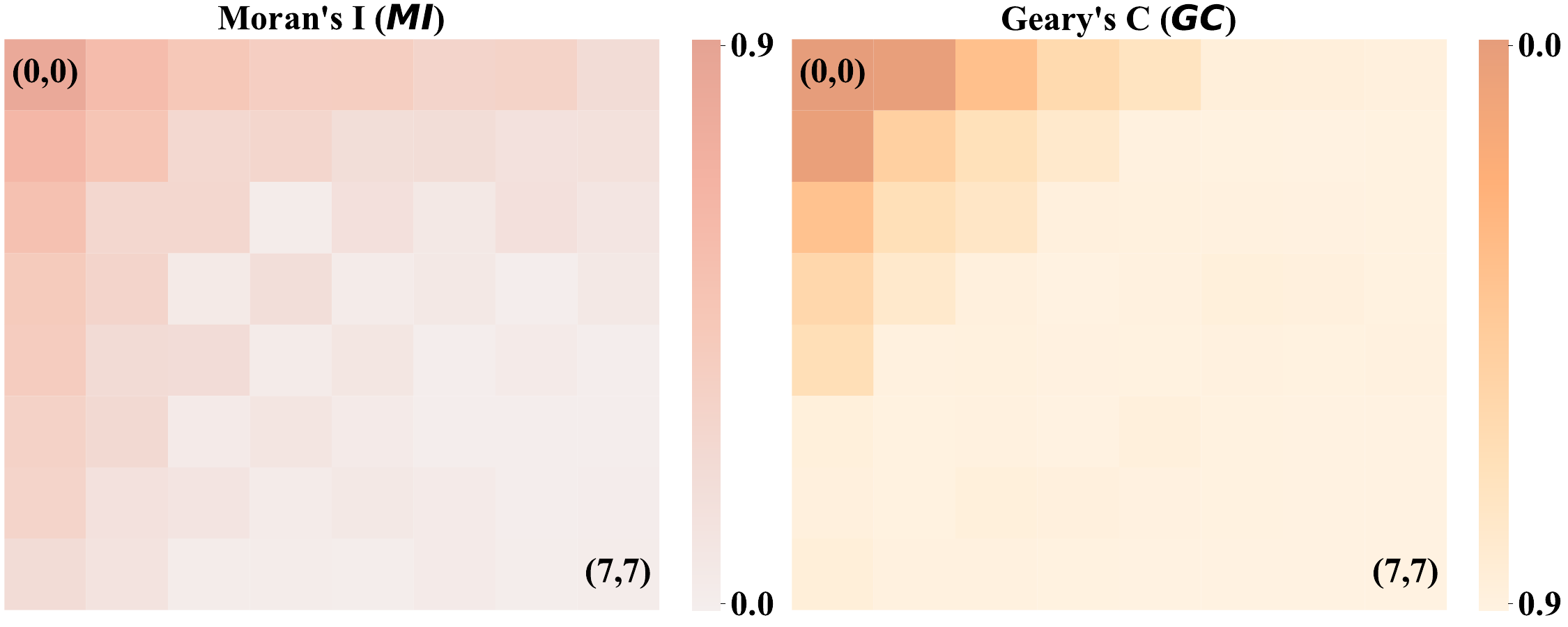}
        \caption{Y of BSDS500}
    \end{subfigure}
    \begin{subfigure}{0.33\linewidth}
        \centering
        \includegraphics[trim={0mm 0mm 0mm 0mm}, clip, width=\linewidth]{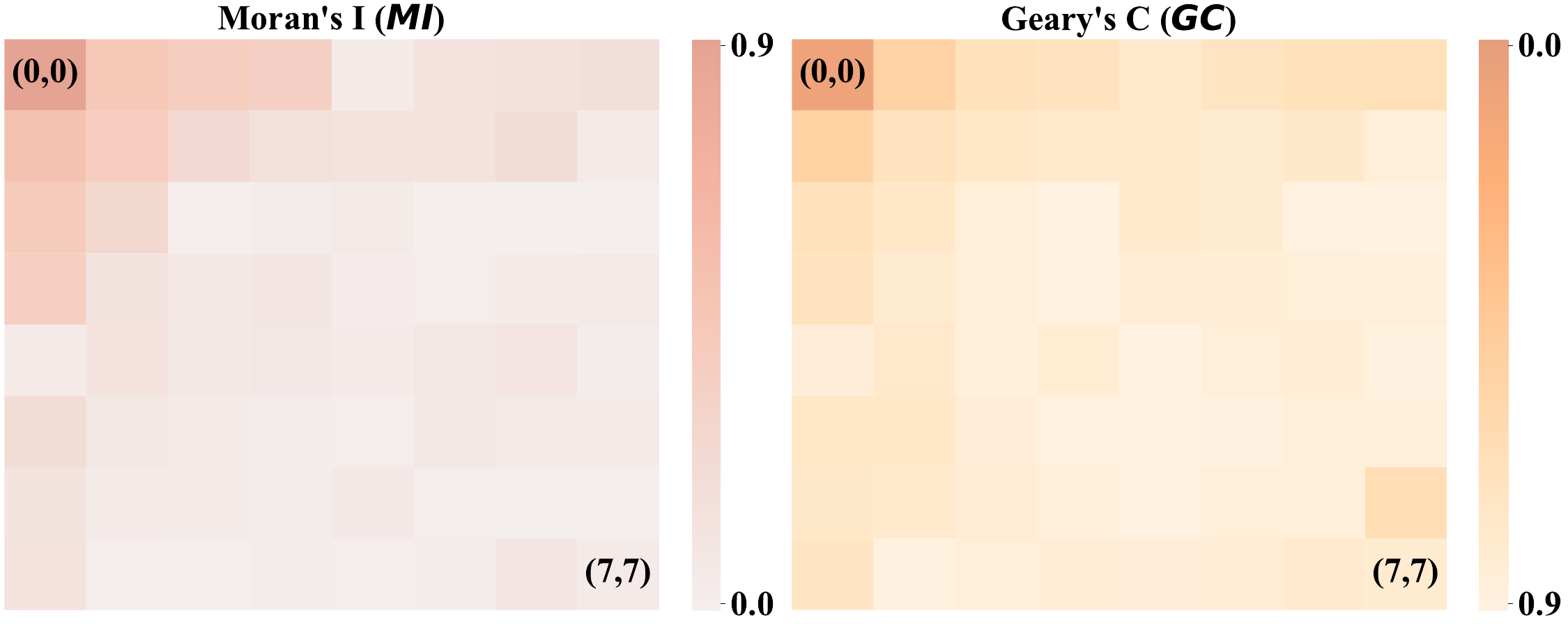}
        \caption{Cb of BSDS500}
    \end{subfigure}
    \begin{subfigure}{0.33\linewidth}
        \centering
        \includegraphics[trim={0mm 0mm 0mm 0mm}, clip, width=\linewidth]{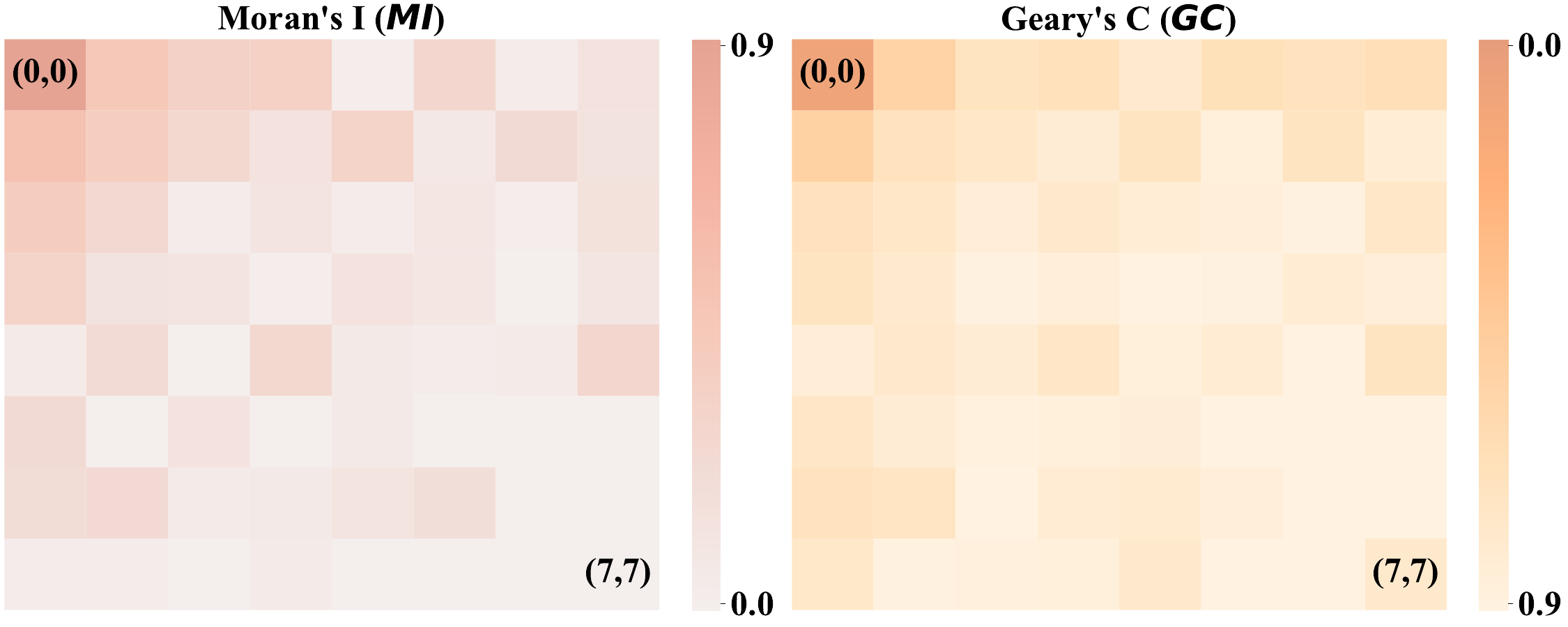}
        \caption{Cr of BSDS500}
    \end{subfigure}
    \vspace{\vspacelength}
    \caption{\textbf{Point-based correlations using coefficient maps on the DIV2K and BSDS500 datasets with QF set to 40.}
    Note that the intensity of heat maps indicate the strength of the correlations.}
    \label{fig:point_based_corr_supp_40}
\end{figure*}

\begin{figure*}[htbp]
    \centering
    \begin{subfigure}{0.33\linewidth}
        \centering
        \includegraphics[trim={9mm 102mm 22mm 109mm}, clip, width=\linewidth]{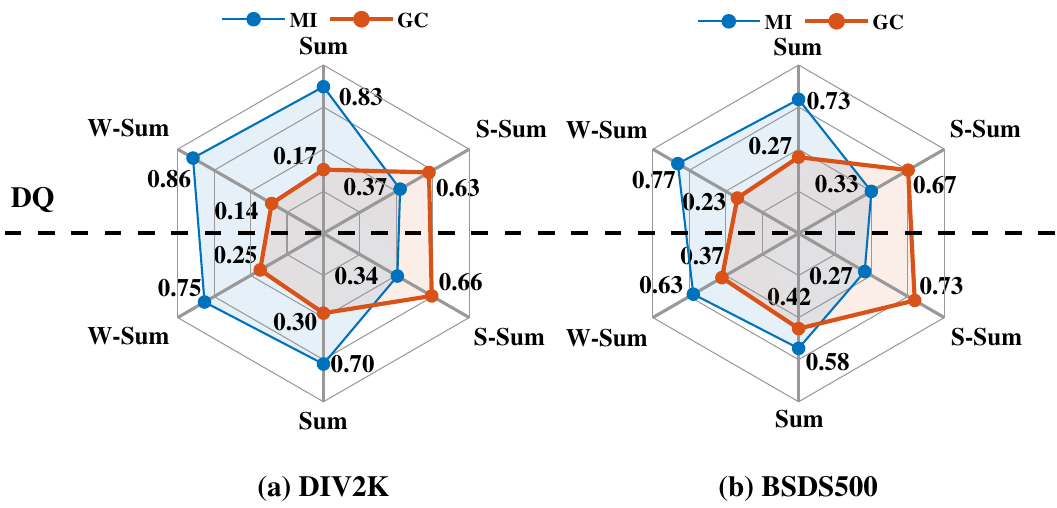}
        \caption{Y}
    \end{subfigure}
    \begin{subfigure}{.33\linewidth}
        \centering
        \includegraphics[trim={9mm 102mm 22mm 109mm}, clip, width=\linewidth]{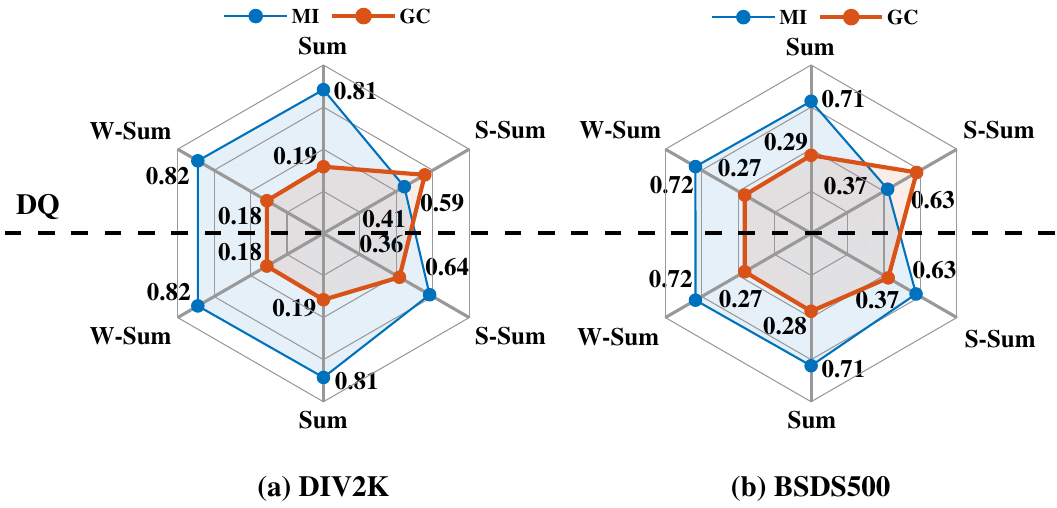}
        \caption{Cb}
    \end{subfigure}
    \begin{subfigure}{.33\linewidth}
        \centering
        \includegraphics[trim={9mm 102mm 22mm 109mm}, clip, width=\linewidth]{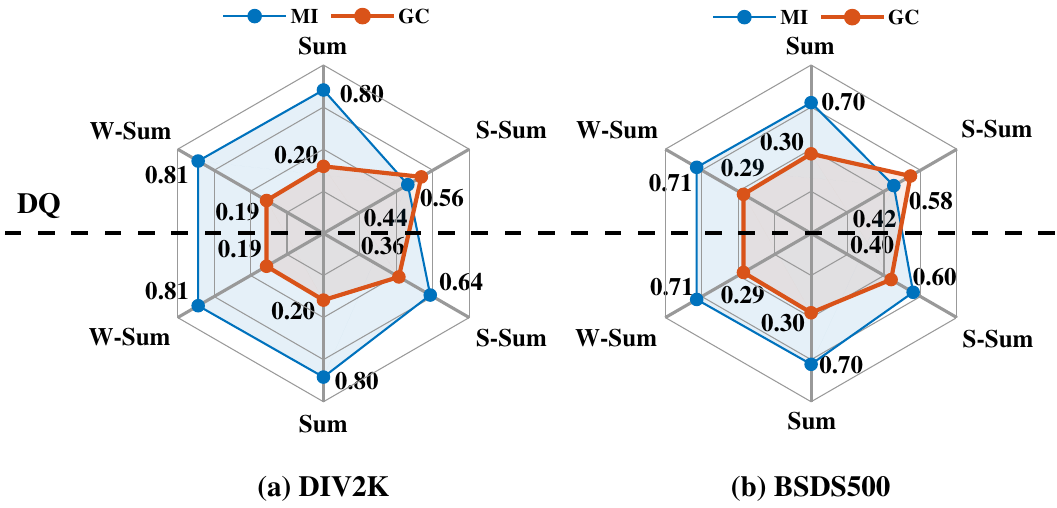}
        \caption{Cr}
    \end{subfigure}
    \vspace{\vspacelength}
    \caption{\textbf{Block-based correlations using different block-based features on the DIV2K and BSDS500 datasets with QF set to 50.}
    Upper: DCT blocks are dequantized before calculating feature values.
    Lower: DCT blocks remain quantized.}
    \label{fig:block_based_corr_supp_50}
\end{figure*}

\begin{figure*}[htbp]
    \centering
    \begin{subfigure}{0.33\linewidth}
        \centering
        \includegraphics[trim={0mm 0mm 0mm 0mm}, clip, width=\linewidth]{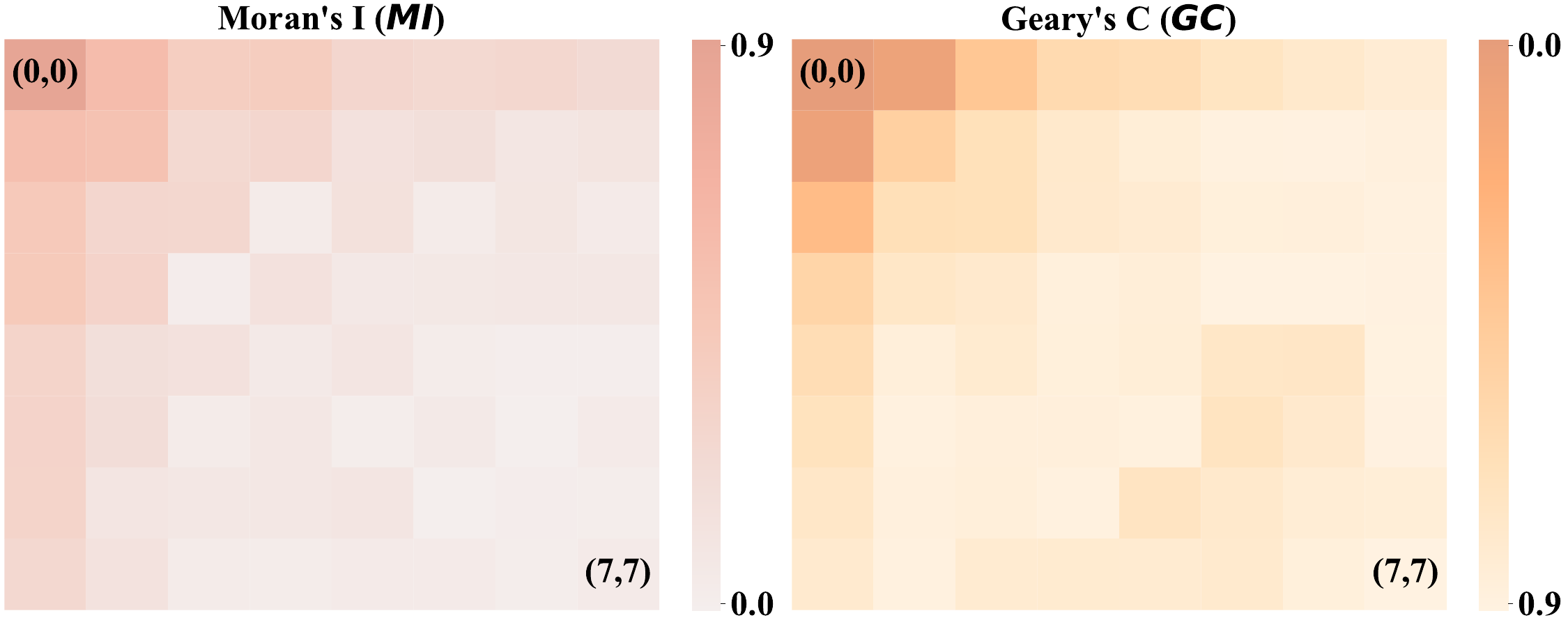}
        \caption{Y of DIV2K}
    \end{subfigure}
    \begin{subfigure}{0.33\linewidth}
        \centering
        \includegraphics[trim={0mm 0mm 0mm 0mm}, clip, width=\linewidth]{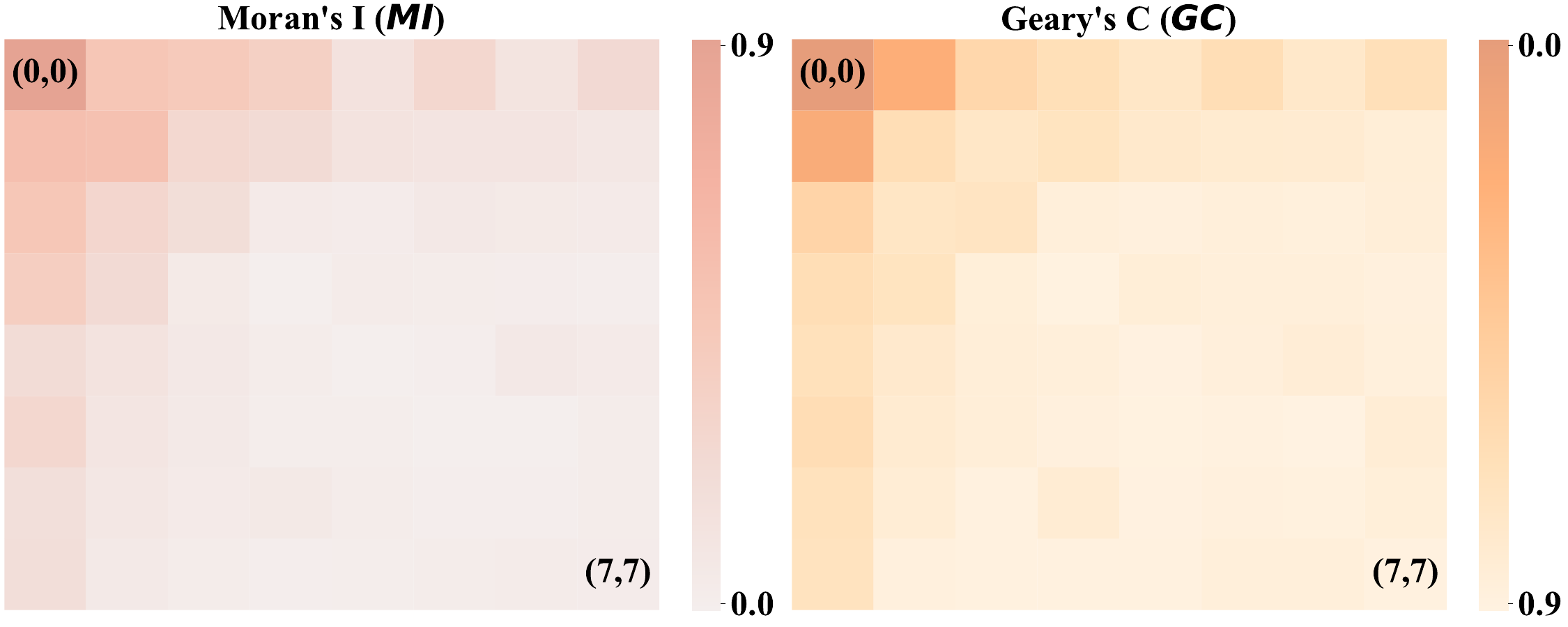}
        \caption{Cb of DIV2K}
    \end{subfigure}
    \begin{subfigure}{0.33\linewidth}
        \centering
        \includegraphics[trim={0mm 0mm 0mm 0mm}, clip, width=\linewidth]{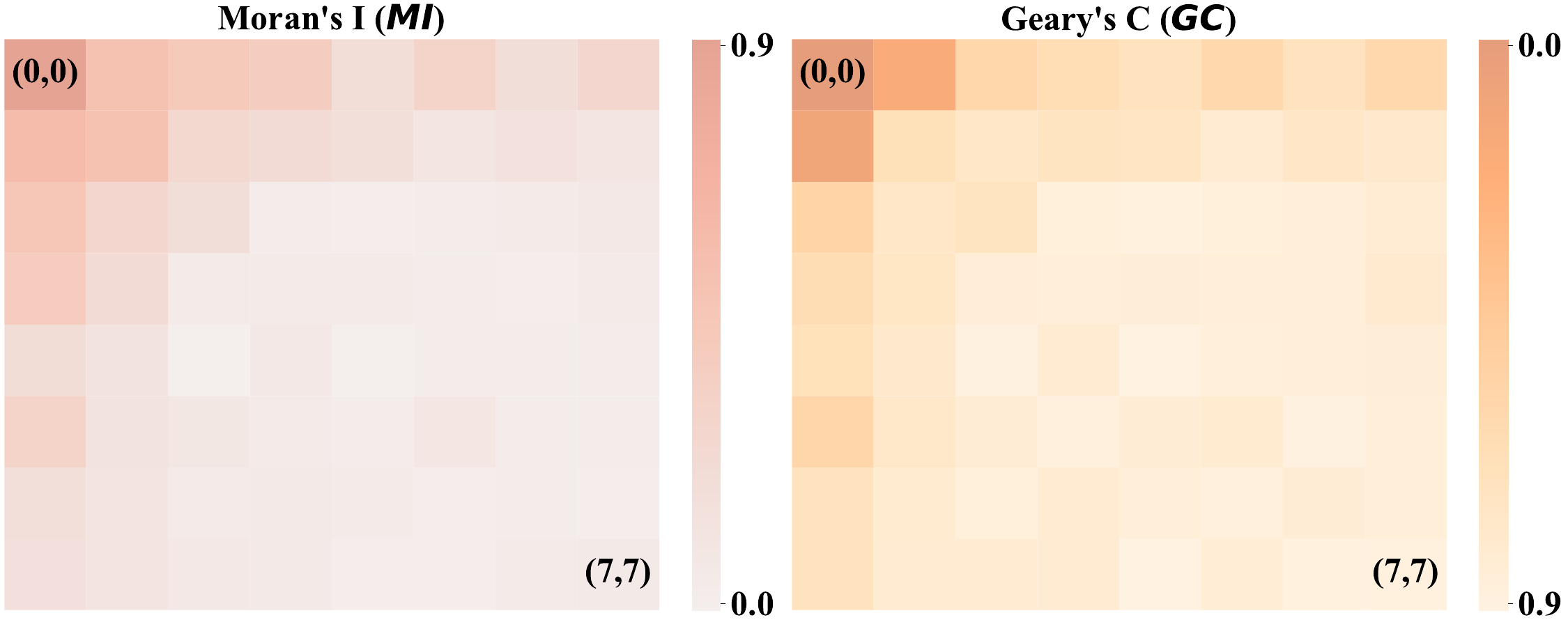}
        \caption{Cr of DIV2K}
    \end{subfigure}
    \begin{subfigure}{0.33\linewidth}
        \centering
        \includegraphics[trim={0mm 0mm 0mm 0mm}, clip, width=\linewidth]{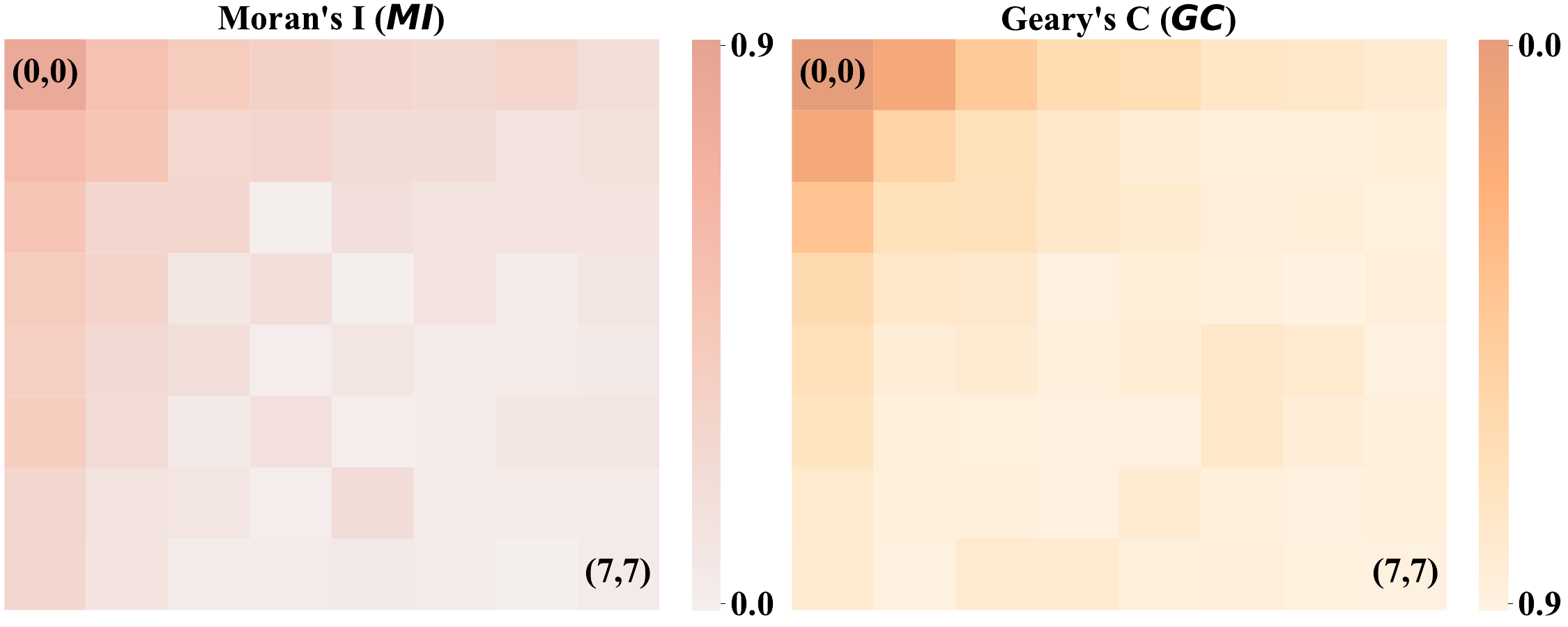}
        \caption{Y of BSDS500}
    \end{subfigure}
    \begin{subfigure}{0.33\linewidth}
        \centering
        \includegraphics[trim={0mm 0mm 0mm 0mm}, clip, width=\linewidth]{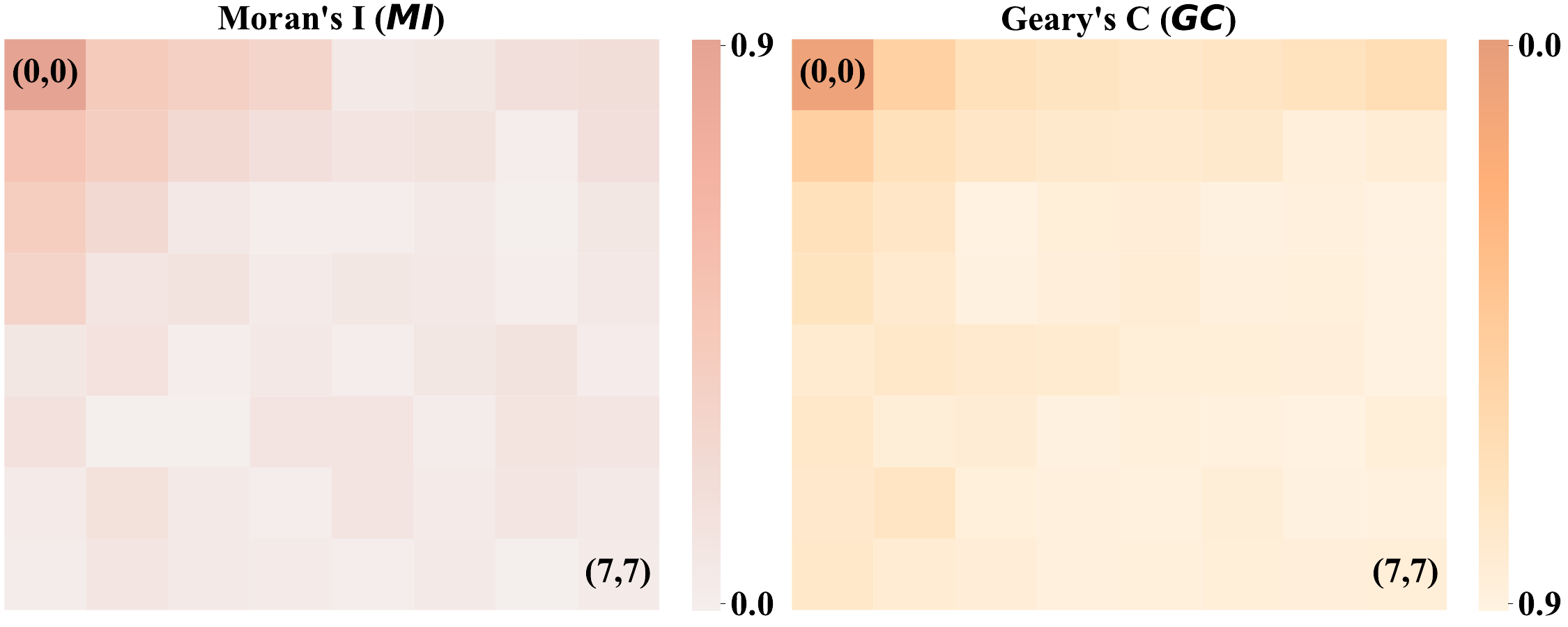}
        \caption{Cb of BSDS500}
    \end{subfigure}
    \begin{subfigure}{0.33\linewidth}
        \centering
        \includegraphics[trim={0mm 0mm 0mm 0mm}, clip, width=\linewidth]{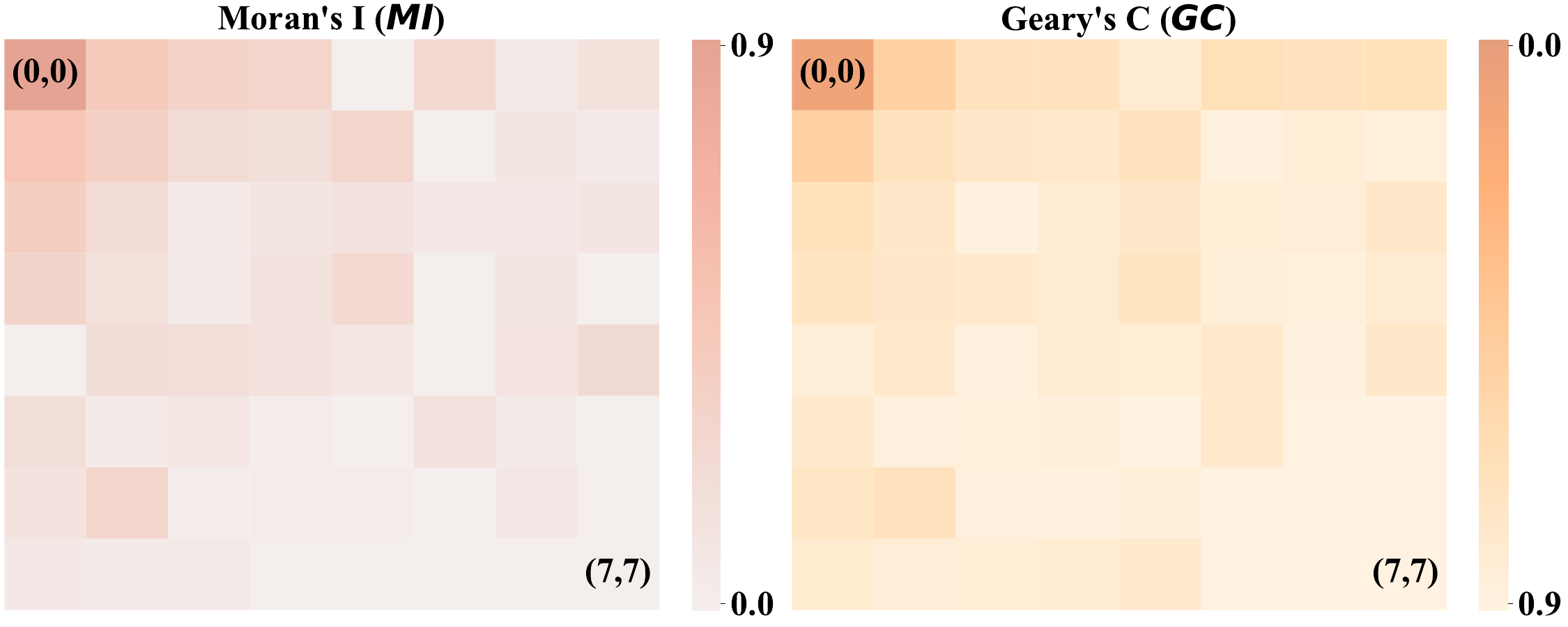}
        \caption{Cr of BSDS500}
    \end{subfigure}
    \vspace{\vspacelength}
    \caption{\textbf{Point-based correlations using coefficient maps on the DIV2K and BSDS500 datasets with QF set to 50.}
    Note that the intensity of heat maps indicate the strength of the correlations.}
    \label{fig:point_based_corr_supp_50}
\end{figure*}

\begin{figure*}[htbp]
    \centering
    \begin{subfigure}{0.33\linewidth}
        \centering
        \includegraphics[trim={9mm 102mm 22mm 109mm}, clip, width=\linewidth]{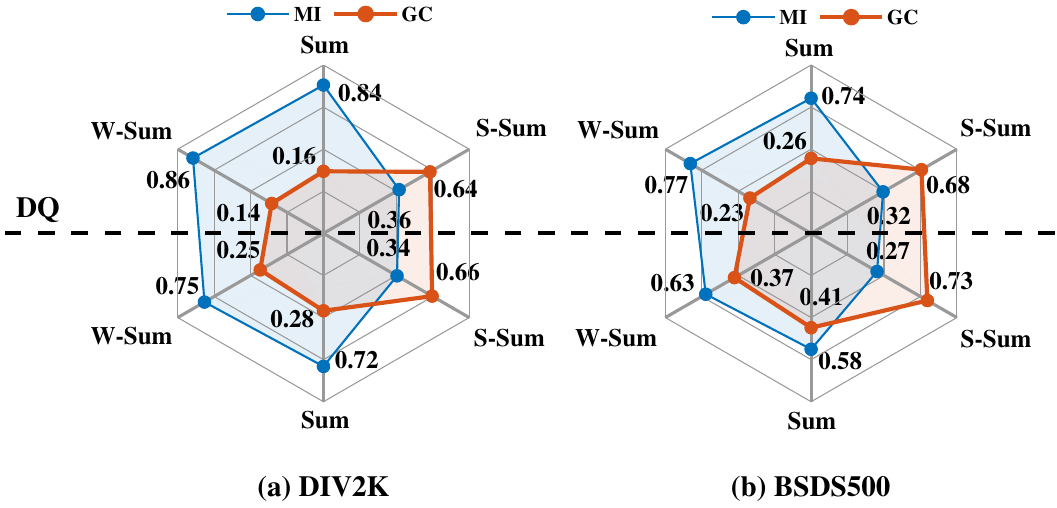}
        \caption{Y}
    \end{subfigure}
    \begin{subfigure}{.33\linewidth}
        \centering
        \includegraphics[trim={9mm 102mm 22mm 109mm}, clip, width=\linewidth]{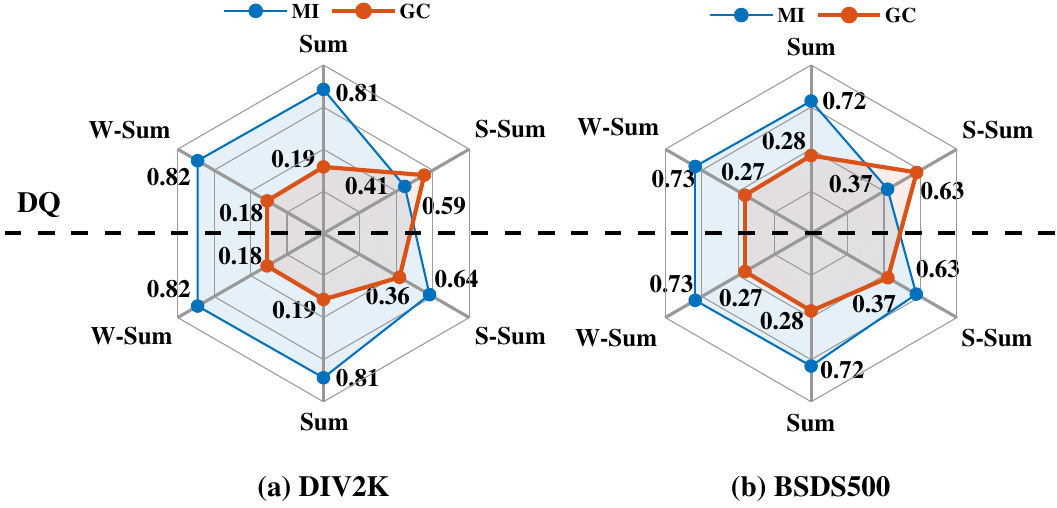}
        \caption{Cb}
    \end{subfigure}
    \begin{subfigure}{.33\linewidth}
        \centering
        \includegraphics[trim={9mm 102mm 22mm 109mm}, clip, width=\linewidth]{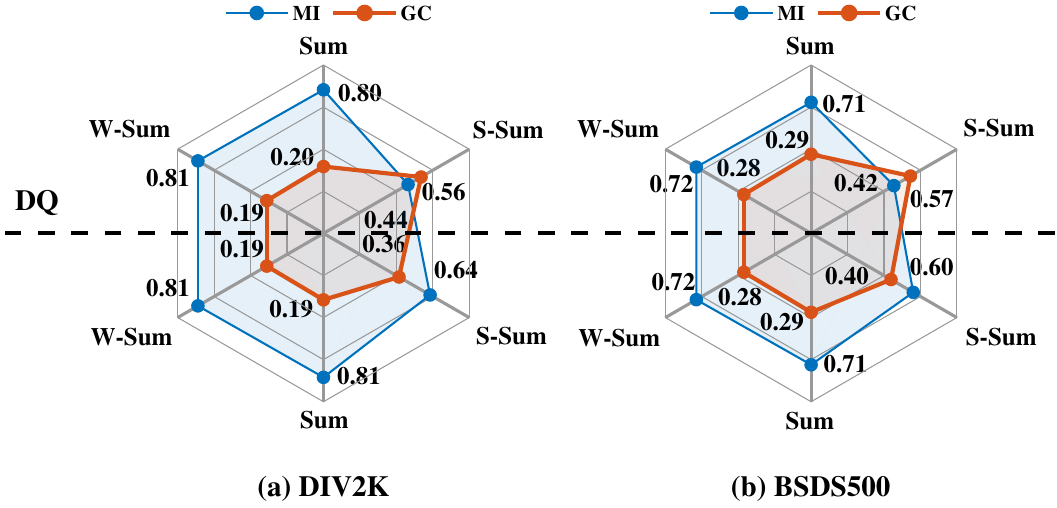}
        \caption{Cr}
    \end{subfigure}
    \vspace{\vspacelength}
    \caption{\textbf{Block-based correlations using different block-based features on the DIV2K and BSDS500 datasets with QF set to 60.}
    Upper: DCT blocks are dequantized before calculating feature values.
    Lower: DCT blocks remain quantized.}
    \label{fig:block_based_corr_supp_60}
\end{figure*}

\begin{figure*}[htbp]
    \centering
    \begin{subfigure}{0.33\linewidth}
        \centering
        \includegraphics[trim={0mm 0mm 0mm 0mm}, clip, width=\linewidth]{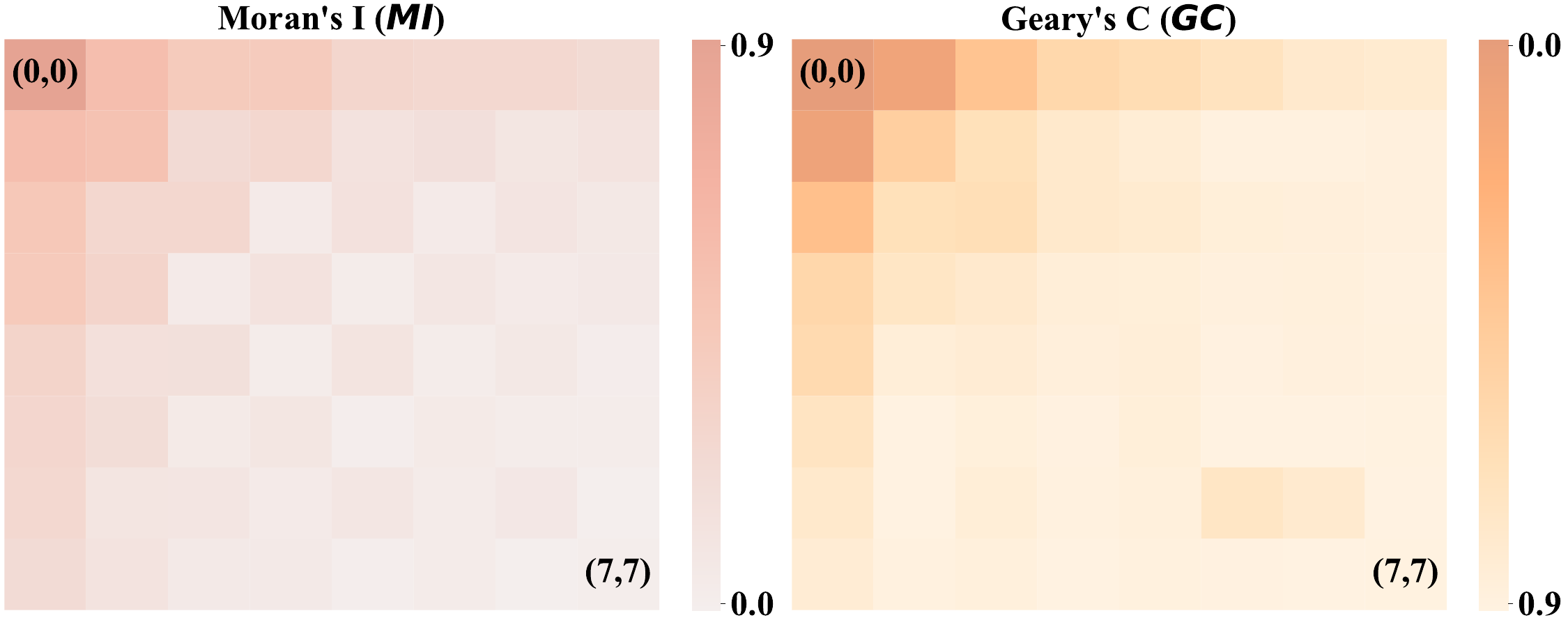}
        \caption{Y of DIV2K}
    \end{subfigure}
    \begin{subfigure}{0.33\linewidth}
        \centering
        \includegraphics[trim={0mm 0mm 0mm 0mm}, clip, width=\linewidth]{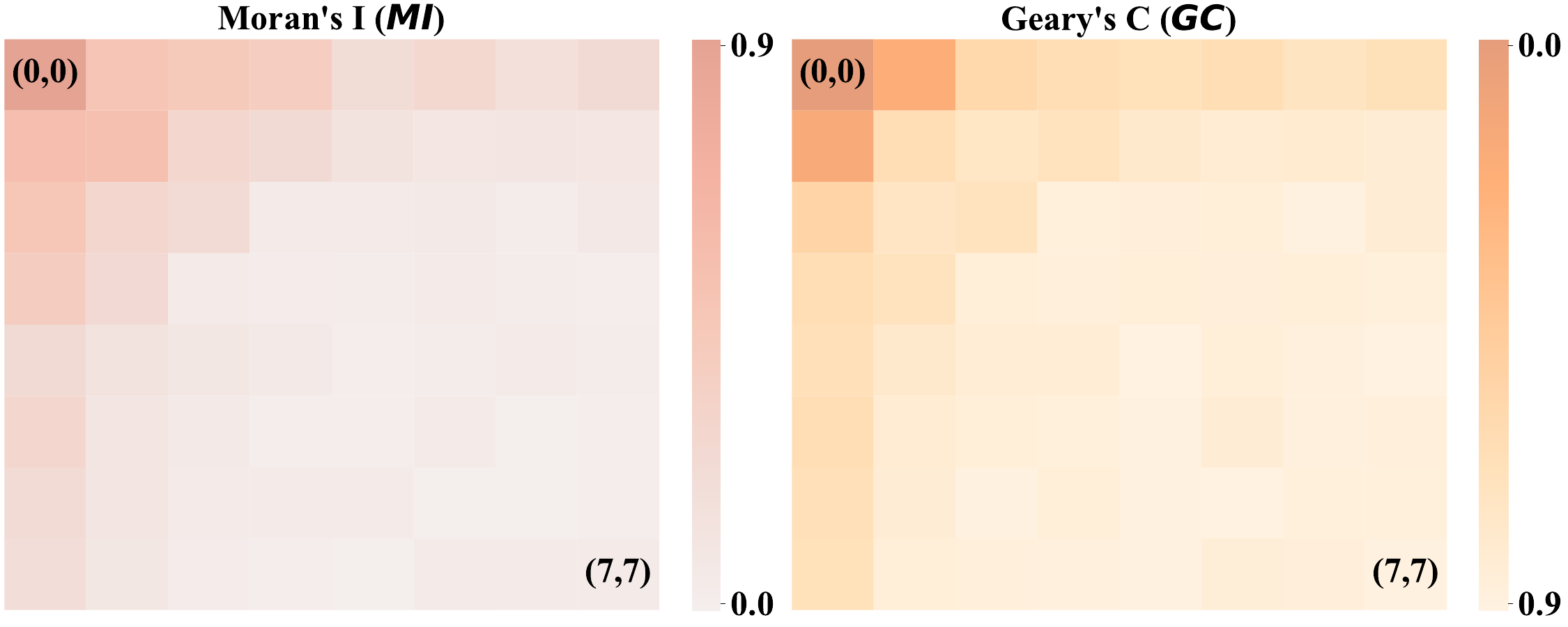}
        \caption{Cb of DIV2K}
    \end{subfigure}
    \begin{subfigure}{0.33\linewidth}
        \centering
        \includegraphics[trim={0mm 0mm 0mm 0mm}, clip, width=\linewidth]{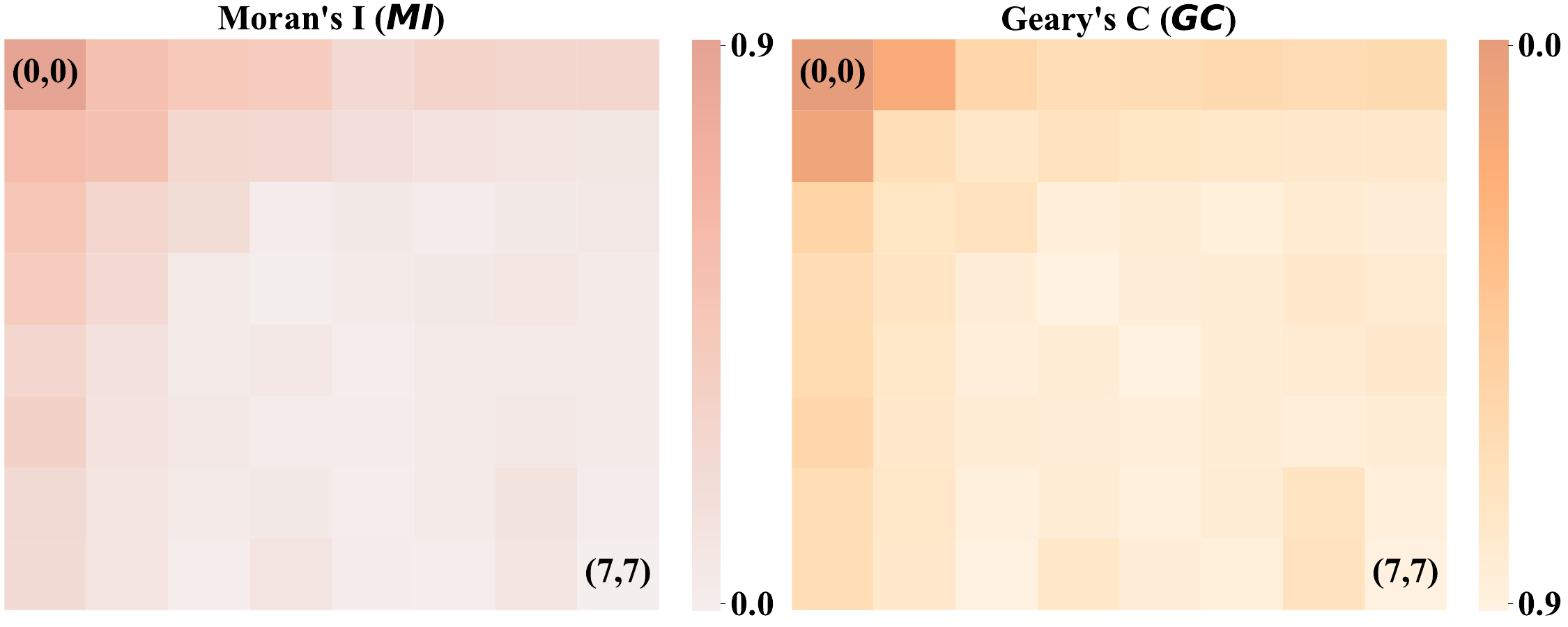}
        \caption{Cr of DIV2K}
    \end{subfigure}
    \begin{subfigure}{0.33\linewidth}
        \centering
        \includegraphics[trim={0mm 0mm 0mm 0mm}, clip, width=\linewidth]{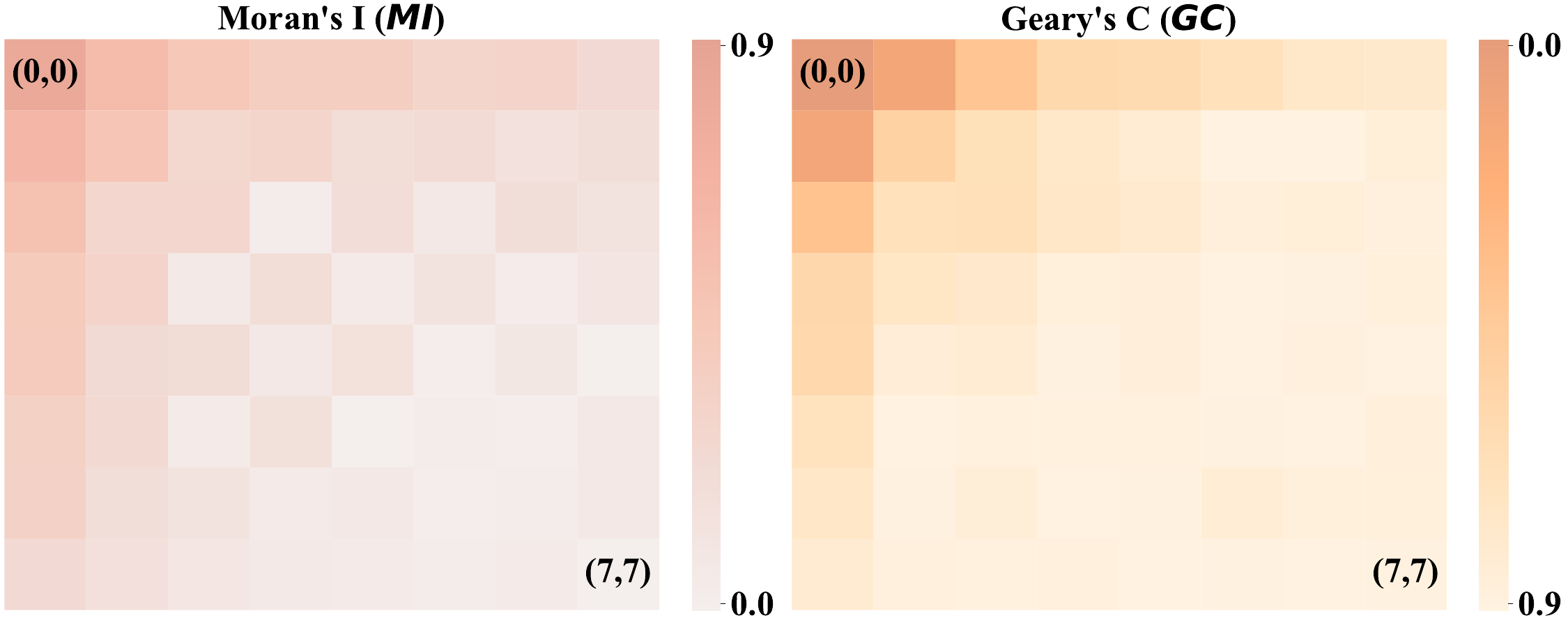}
        \caption{Y of BSDS500}
    \end{subfigure}
    \begin{subfigure}{0.33\linewidth}
        \centering
        \includegraphics[trim={0mm 0mm 0mm 0mm}, clip, width=\linewidth]{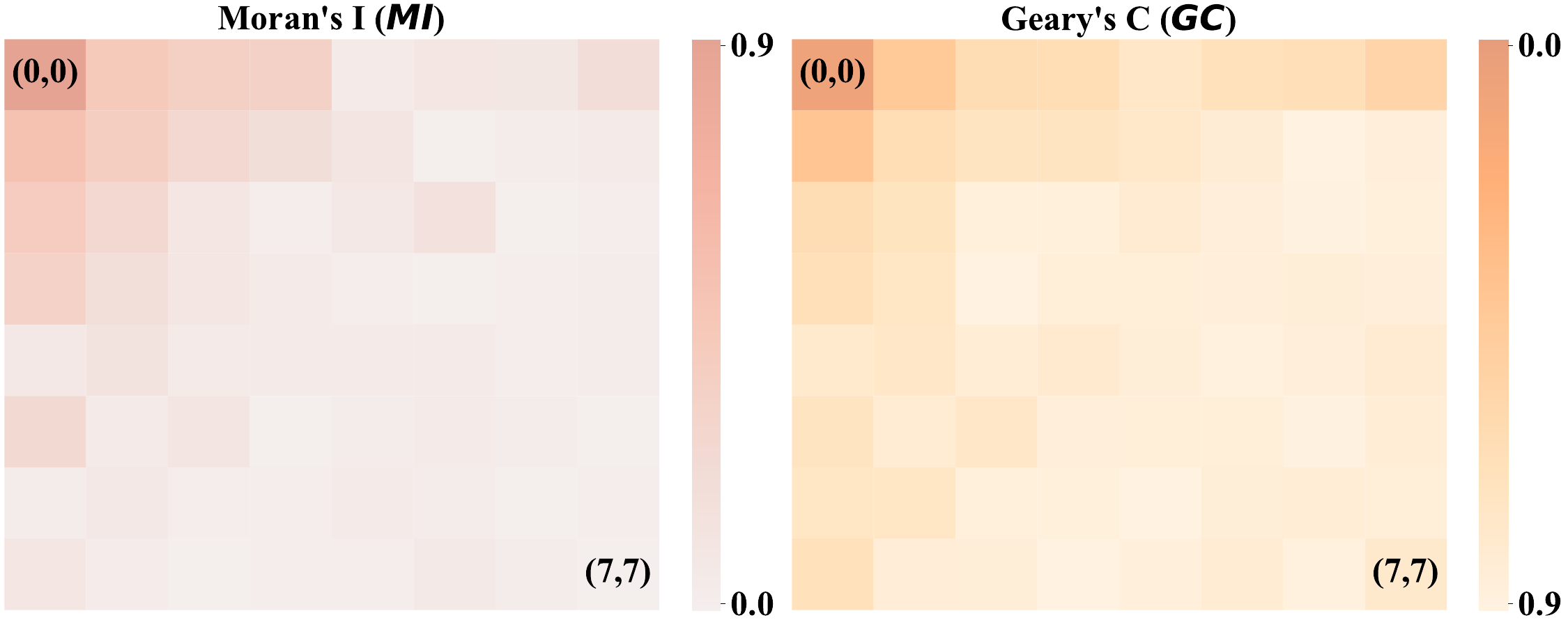}
        \caption{Cb of BSDS500}
    \end{subfigure}
    \begin{subfigure}{0.33\linewidth}
        \centering
        \includegraphics[trim={0mm 0mm 0mm 0mm}, clip, width=\linewidth]{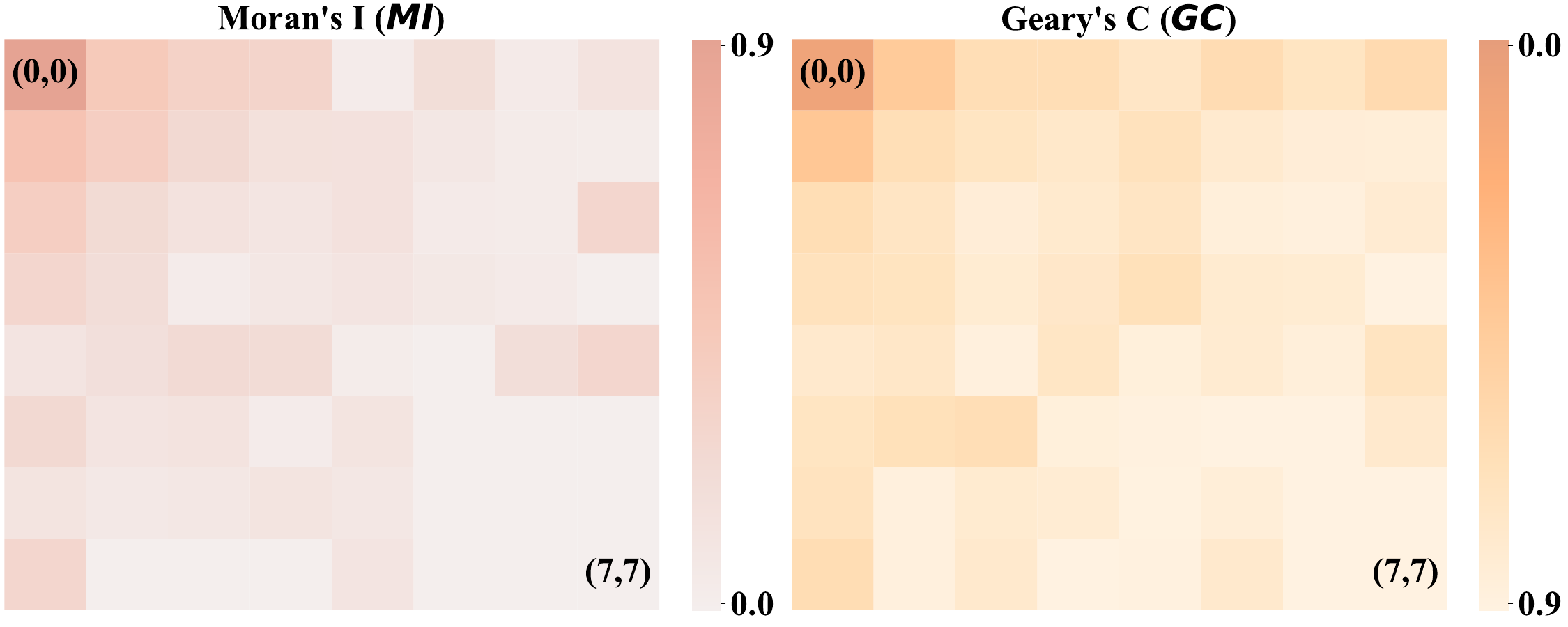}
        \caption{Cr of BSDS500}
    \end{subfigure}
    \vspace{\vspacelength}
    \caption{\textbf{Point-based correlations using coefficient maps on the DIV2K and BSDS500 datasets with QF set to 60.}
    Note that the intensity of heat maps indicate the strength of the correlations.}
    \label{fig:point_based_corr_supp_60}
\end{figure*}

\begin{figure*}[htbp]
    \centering
    \begin{subfigure}{0.33\linewidth}
        \centering
        \includegraphics[trim={9mm 102mm 22mm 109mm}, clip, width=\linewidth]{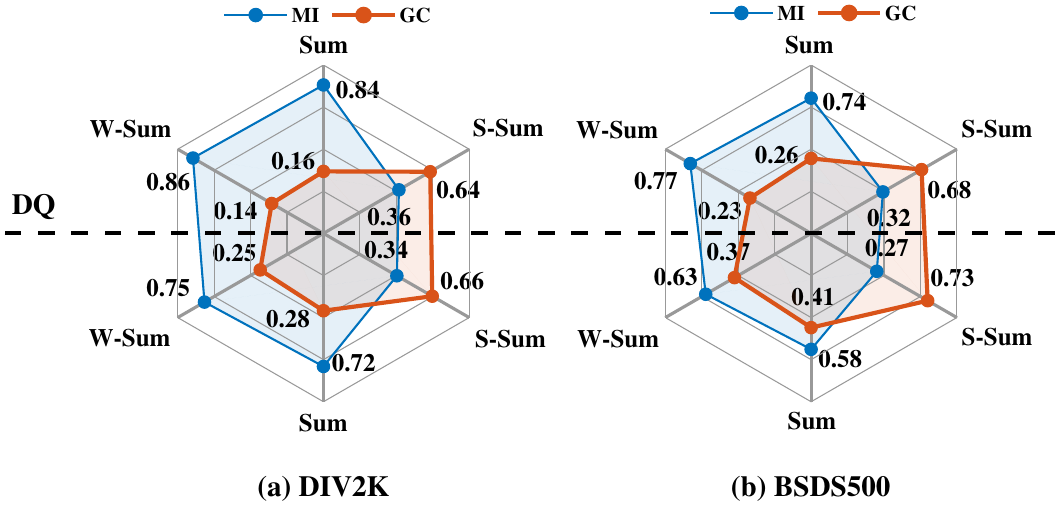}
        \caption{Y}
    \end{subfigure}
    \begin{subfigure}{.33\linewidth}
        \centering
        \includegraphics[trim={9mm 102mm 22mm 109mm}, clip, width=\linewidth]{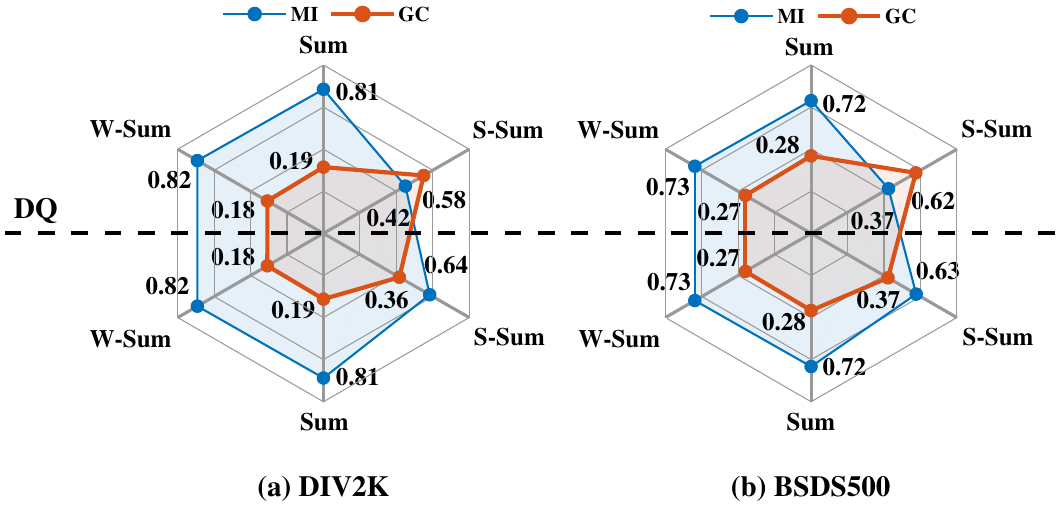}
        \caption{Cb}
    \end{subfigure}
    \begin{subfigure}{.33\linewidth}
        \centering
        \includegraphics[trim={9mm 102mm 22mm 109mm}, clip, width=\linewidth]{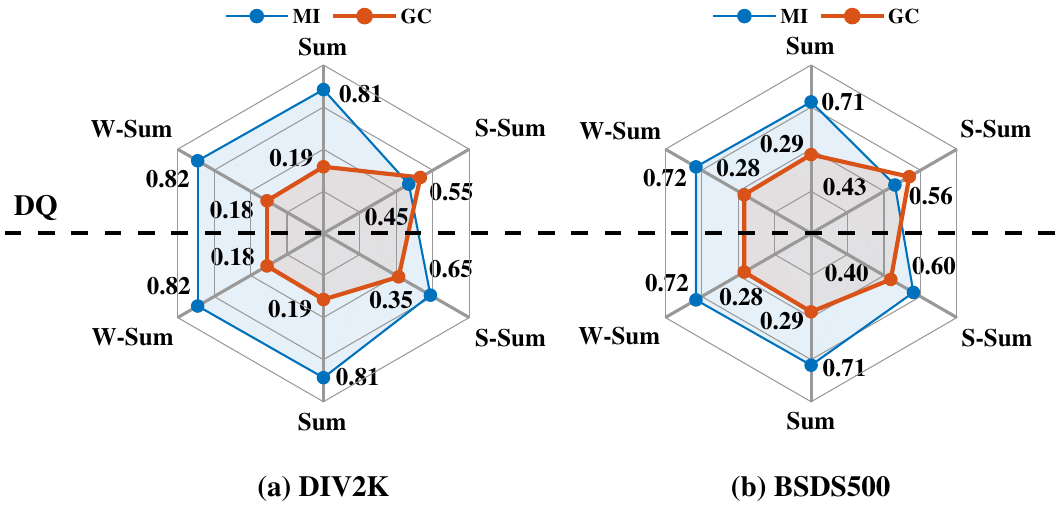}
        \caption{Cr}
    \end{subfigure}
    \vspace{\vspacelength}
    \caption{\textbf{Block-based correlations using different block-based features on the DIV2K and BSDS500 datasets with QF set to 70.}
    Upper: DCT blocks are dequantized before calculating feature values.
    Lower: DCT blocks remain quantized.}
    \label{fig:block_based_corr_supp_70}
\end{figure*}

\begin{figure*}[htbp]
    \centering
    \begin{subfigure}{0.33\linewidth}
        \centering
        \includegraphics[trim={0mm 0mm 0mm 0mm}, clip, width=\linewidth]{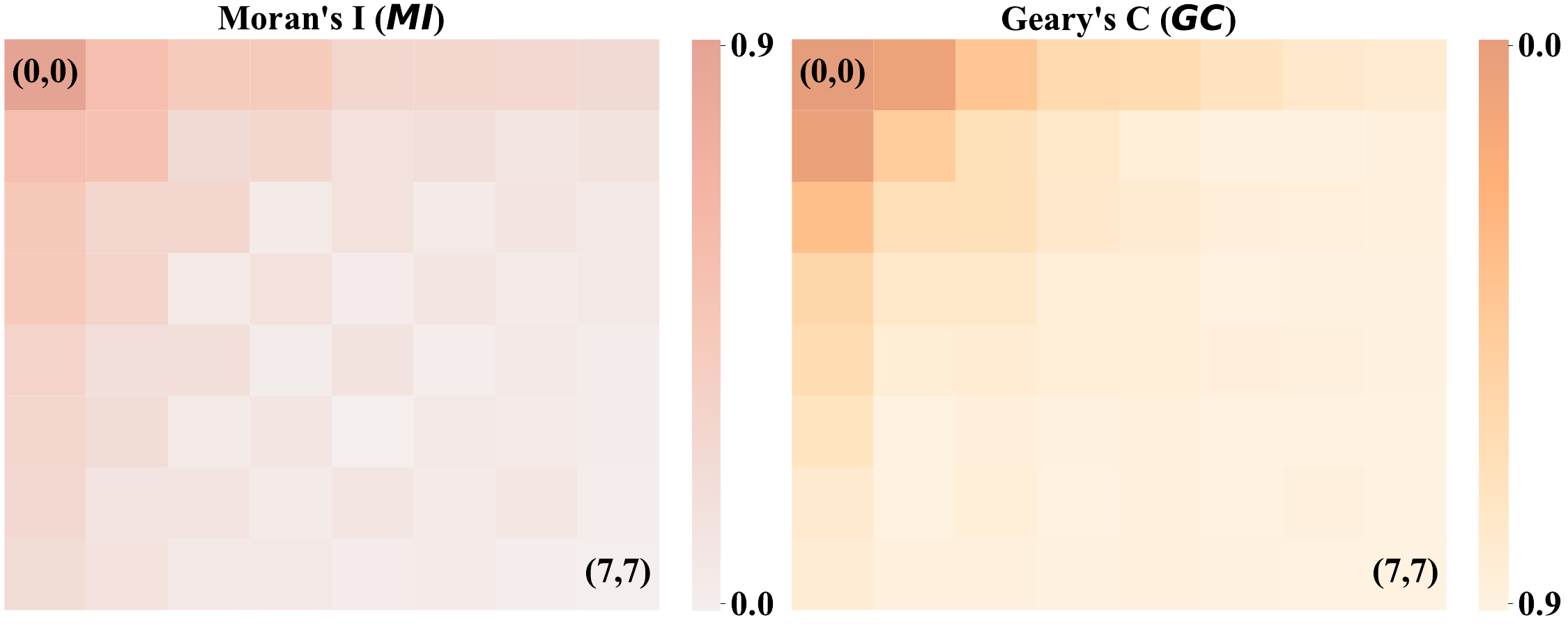}
        \caption{Y of DIV2K}
    \end{subfigure}
    \begin{subfigure}{0.33\linewidth}
        \centering
        \includegraphics[trim={0mm 0mm 0mm 0mm}, clip, width=\linewidth]{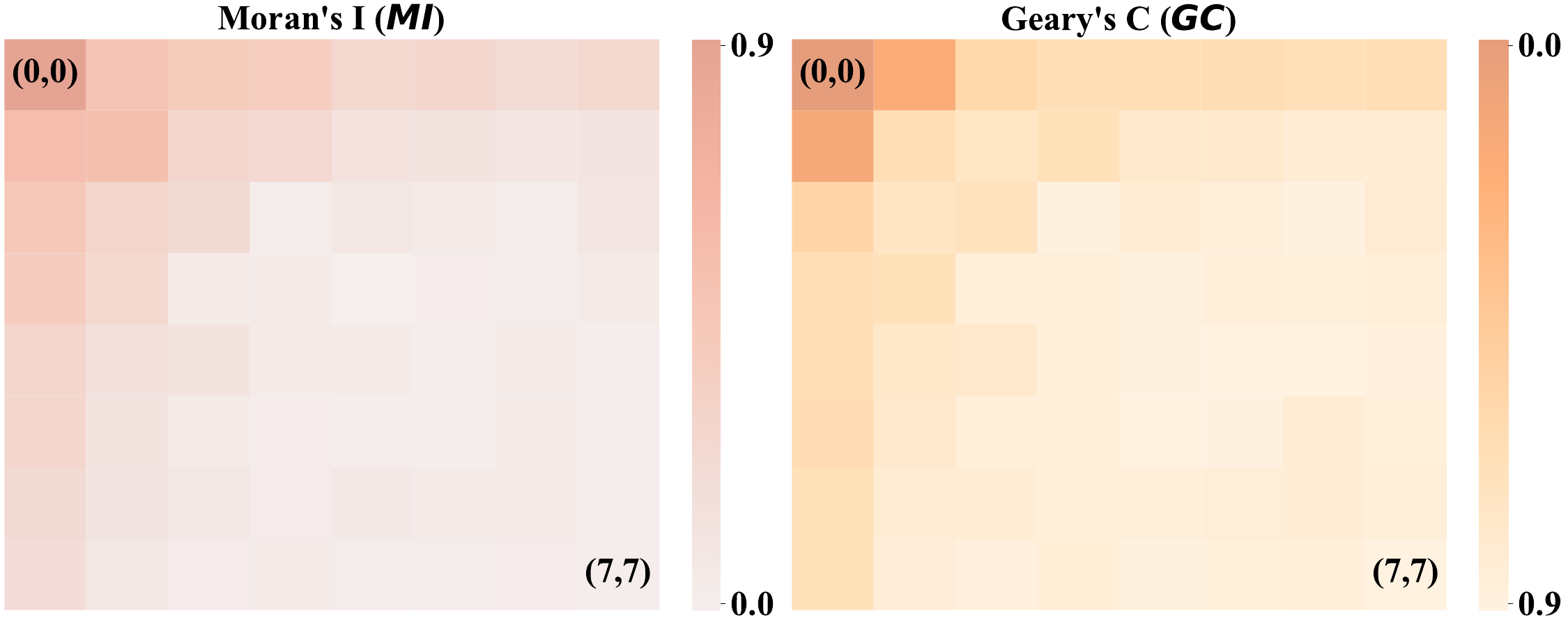}
        \caption{Cb of DIV2K}
    \end{subfigure}
    \begin{subfigure}{0.33\linewidth}
        \centering
        \includegraphics[trim={0mm 0mm 0mm 0mm}, clip, width=\linewidth]{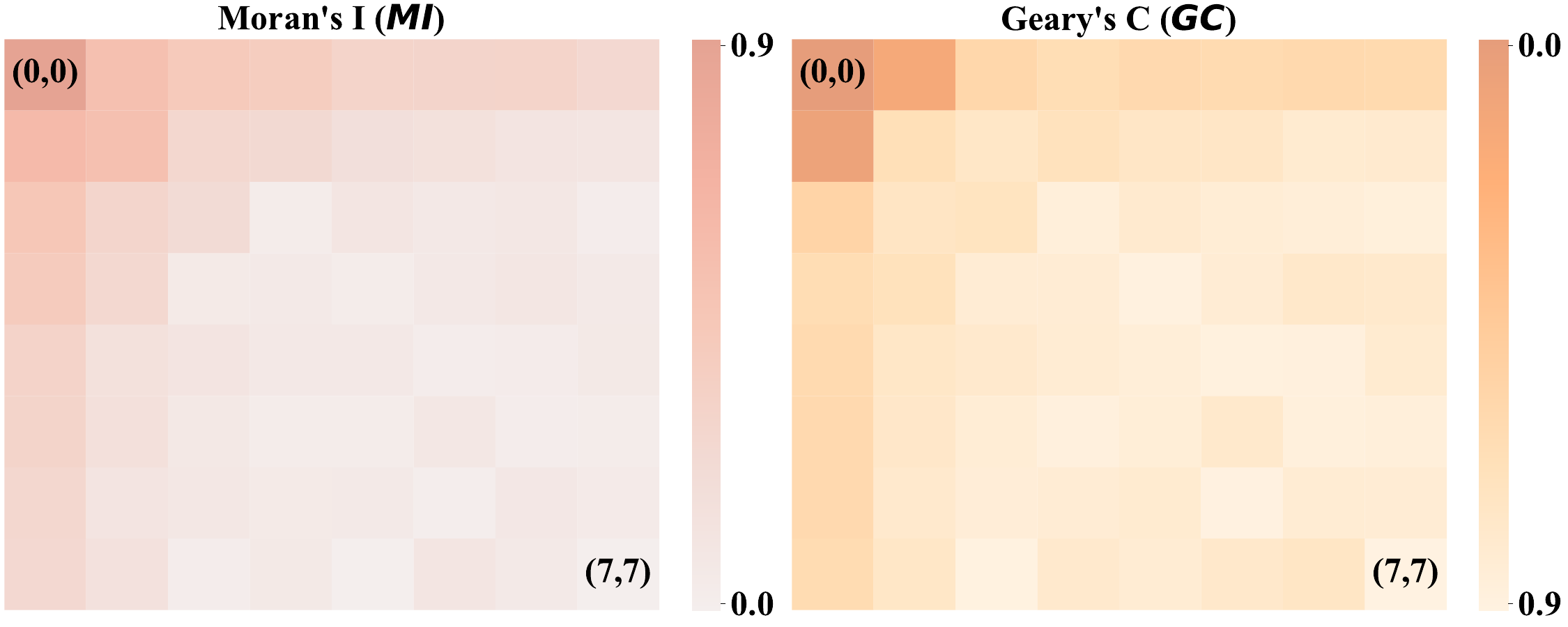}
        \caption{Cr of DIV2K}
    \end{subfigure}
    \begin{subfigure}{0.33\linewidth}
        \centering
        \includegraphics[trim={0mm 0mm 0mm 0mm}, clip, width=\linewidth]{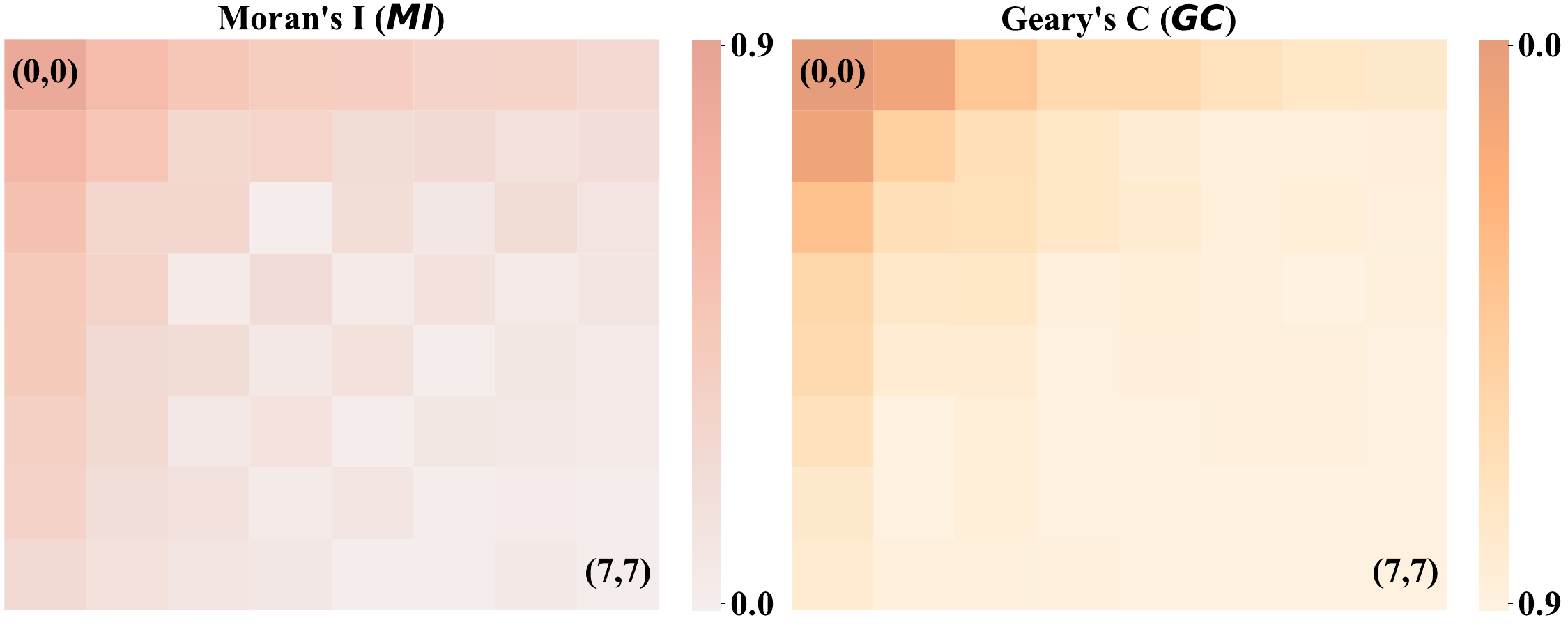}
        \caption{Y of BSDS500}
    \end{subfigure}
    \begin{subfigure}{0.33\linewidth}
        \centering
        \includegraphics[trim={0mm 0mm 0mm 0mm}, clip, width=\linewidth]{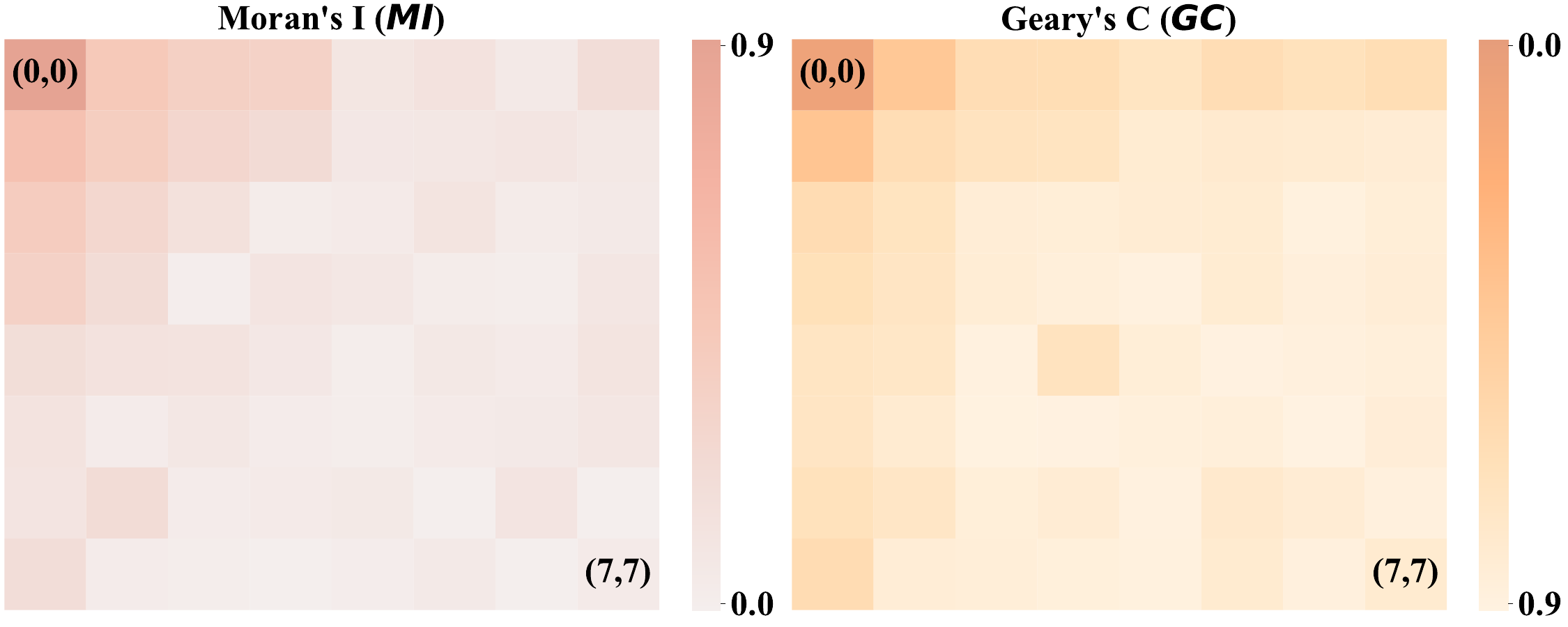}
        \caption{Cb of BSDS500}
    \end{subfigure}
    \begin{subfigure}{0.33\linewidth}
        \centering
        \includegraphics[trim={0mm 0mm 0mm 0mm}, clip, width=\linewidth]{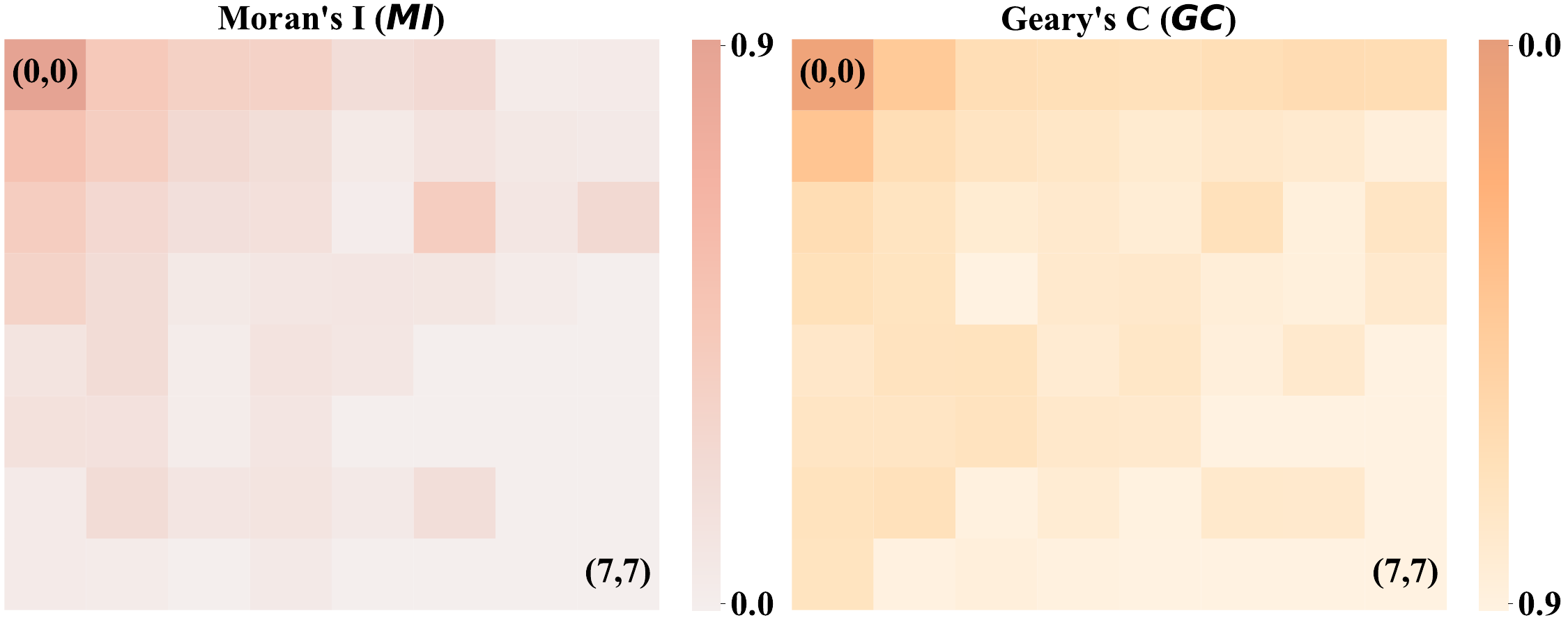}
        \caption{Cr of BSDS500}
    \end{subfigure}
    \vspace{\vspacelength}
    \caption{\textbf{Point-based correlations using coefficient maps on the DIV2K and BSDS500 datasets with QF set to 70.}
    Note that the intensity of heat maps indicate the strength of the correlations.}
    \label{fig:point_based_corr_supp_70}
\end{figure*}

\begin{figure*}[htbp]
    \centering
    \begin{subfigure}{0.33\linewidth}
        \centering
        \includegraphics[trim={9mm 102mm 22mm 109mm}, clip, width=\linewidth]{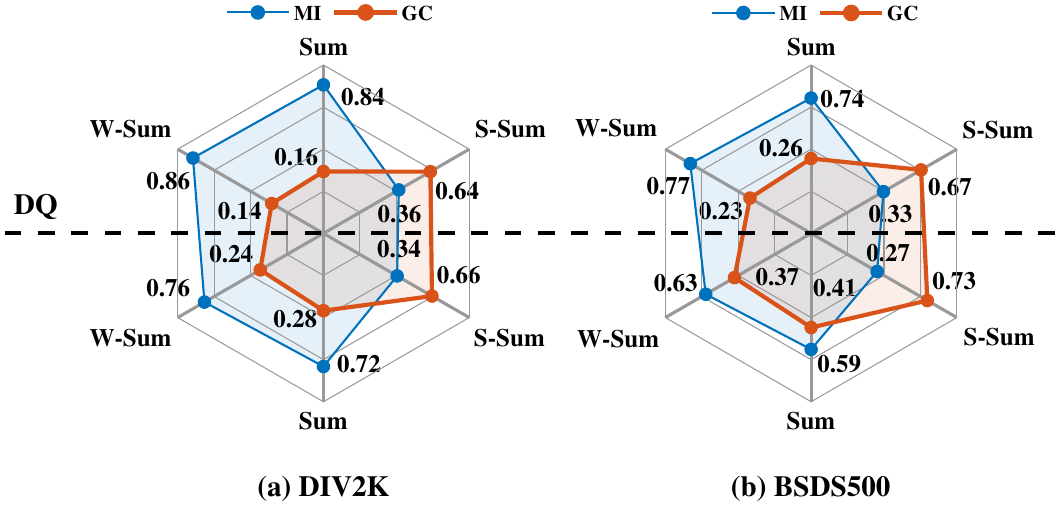}
        \caption{Y}
    \end{subfigure}
    \begin{subfigure}{.33\linewidth}
        \centering
        \includegraphics[trim={9mm 102mm 22mm 109mm}, clip, width=\linewidth]{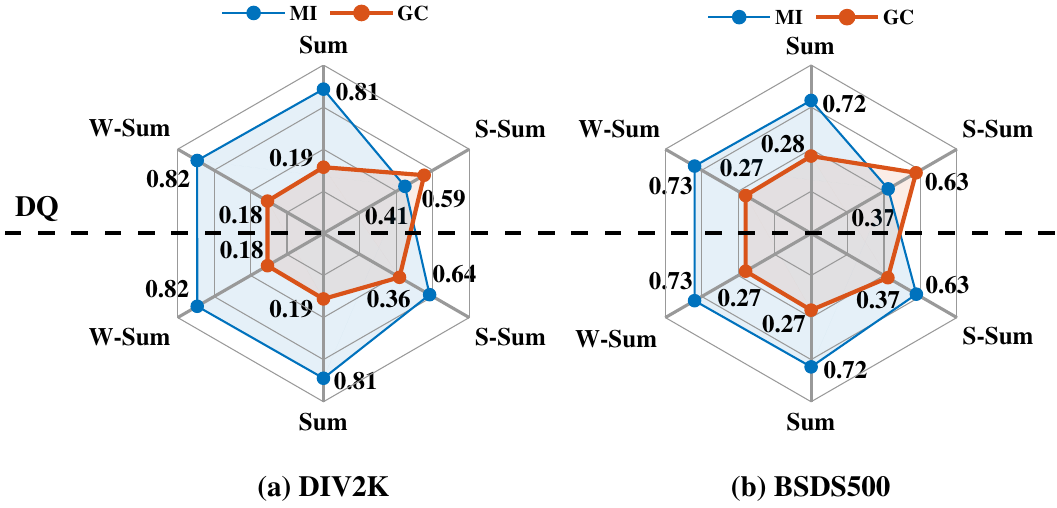}
        \caption{Cb}
    \end{subfigure}
    \begin{subfigure}{.33\linewidth}
        \centering
        \includegraphics[trim={9mm 102mm 22mm 109mm}, clip, width=\linewidth]{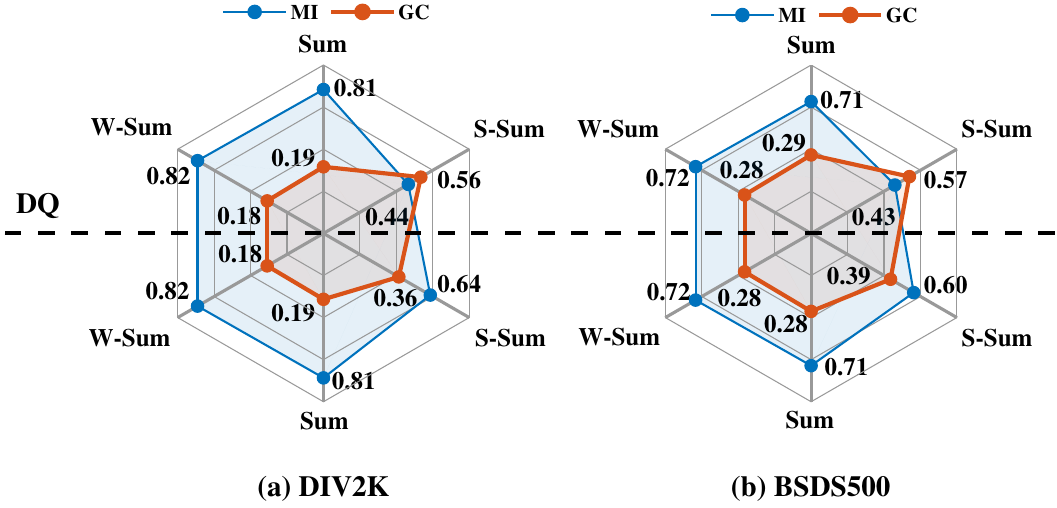}
        \caption{Cr}
    \end{subfigure}
    \vspace{\vspacelength}
    \caption{\textbf{Block-based correlations using different block-based features on the DIV2K and BSDS500 datasets with QF set to 80.}
    Upper: DCT blocks are dequantized before calculating feature values.
    Lower: DCT blocks remain quantized.}
    \label{fig:block_based_corr_supp_80}
\end{figure*}

\begin{figure*}[htbp]
    \centering
    \begin{subfigure}{0.33\linewidth}
        \centering
        \includegraphics[trim={0mm 0mm 0mm 0mm}, clip, width=\linewidth]{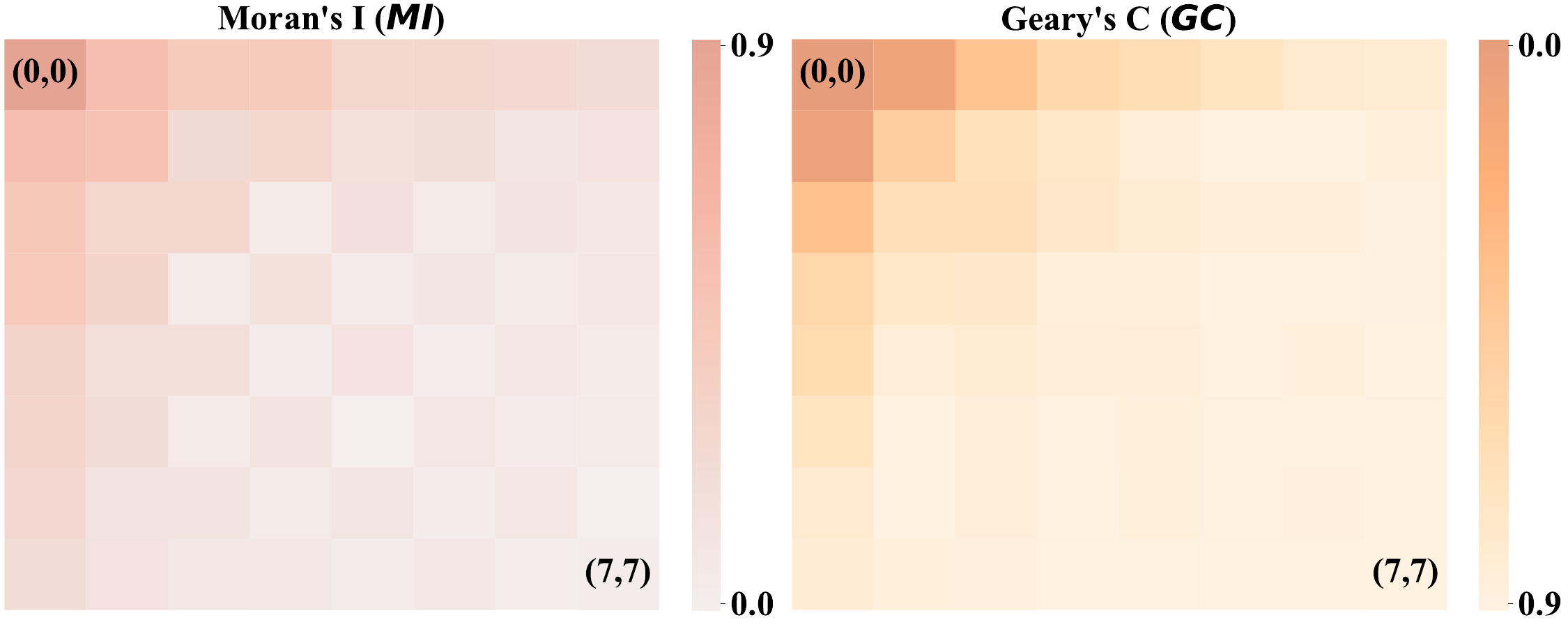}
        \caption{Y of DIV2K}
    \end{subfigure}
    \begin{subfigure}{0.33\linewidth}
        \centering
        \includegraphics[trim={0mm 0mm 0mm 0mm}, clip, width=\linewidth]{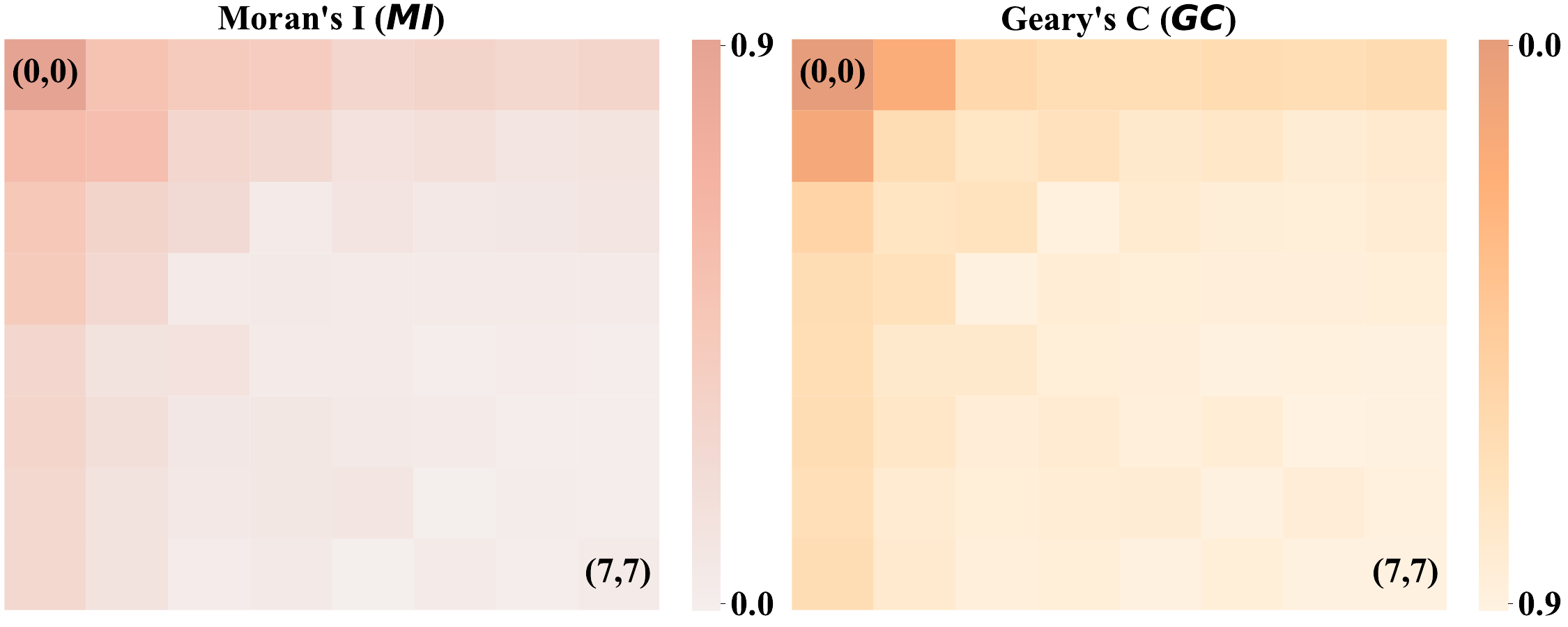}
        \caption{Cb of DIV2K}
    \end{subfigure}
    \begin{subfigure}{0.33\linewidth}
        \centering
        \includegraphics[trim={0mm 0mm 0mm 0mm}, clip, width=\linewidth]{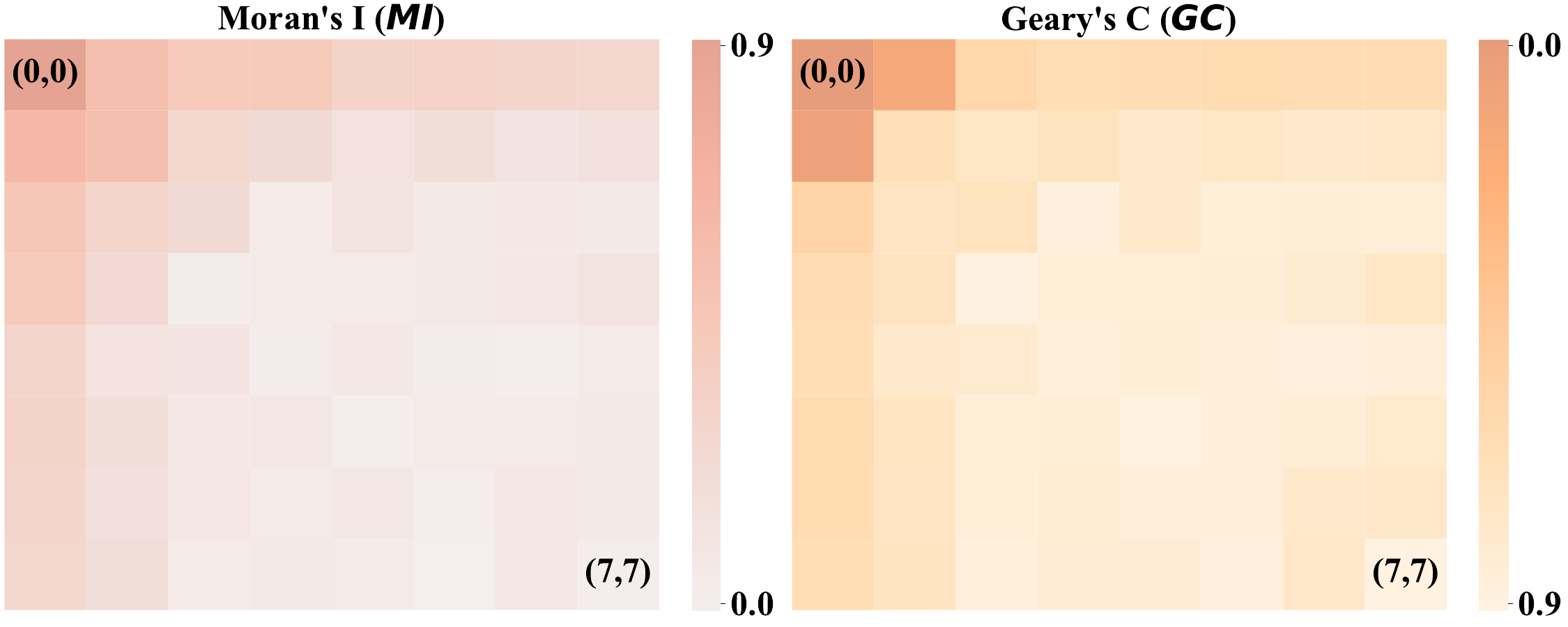}
        \caption{Cr of DIV2K}
    \end{subfigure}
    \begin{subfigure}{0.33\linewidth}
        \centering
        \includegraphics[trim={0mm 0mm 0mm 0mm}, clip, width=\linewidth]{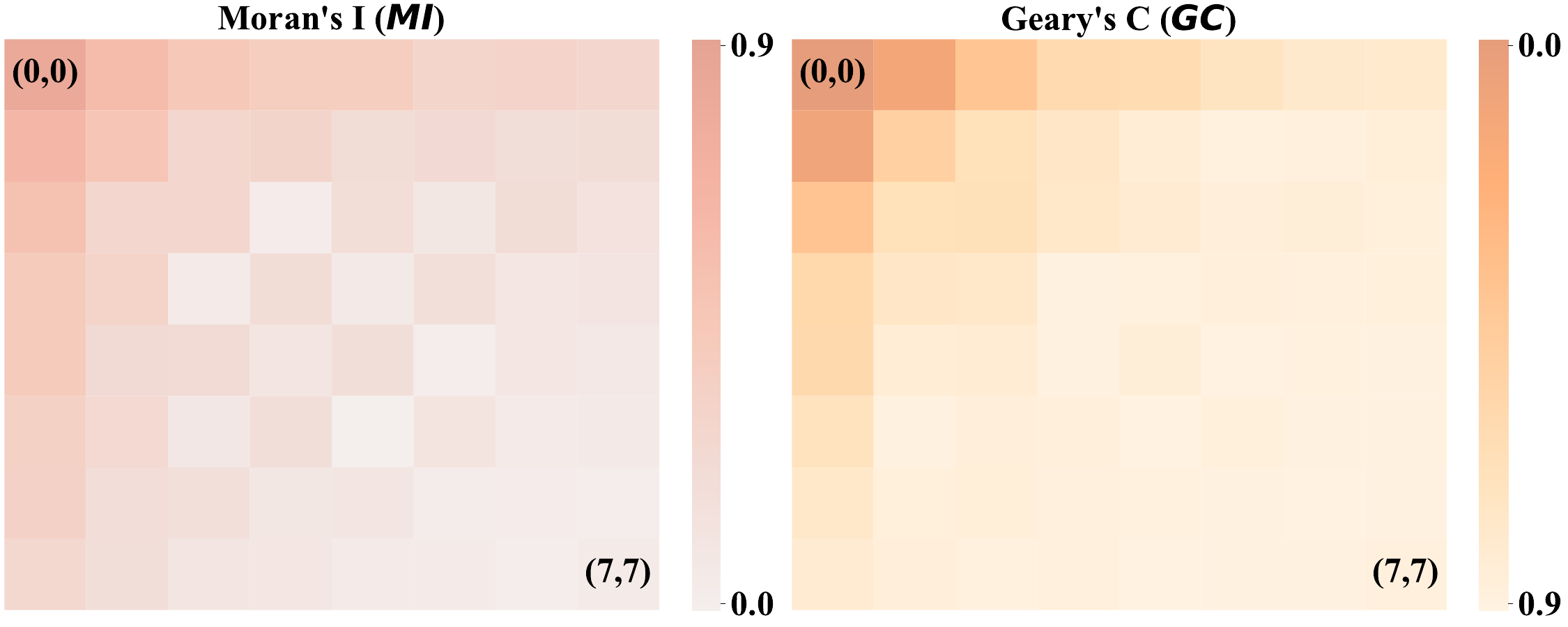}
        \caption{Y of BSDS500}
    \end{subfigure}
    \begin{subfigure}{0.33\linewidth}
        \centering
        \includegraphics[trim={0mm 0mm 0mm 0mm}, clip, width=\linewidth]{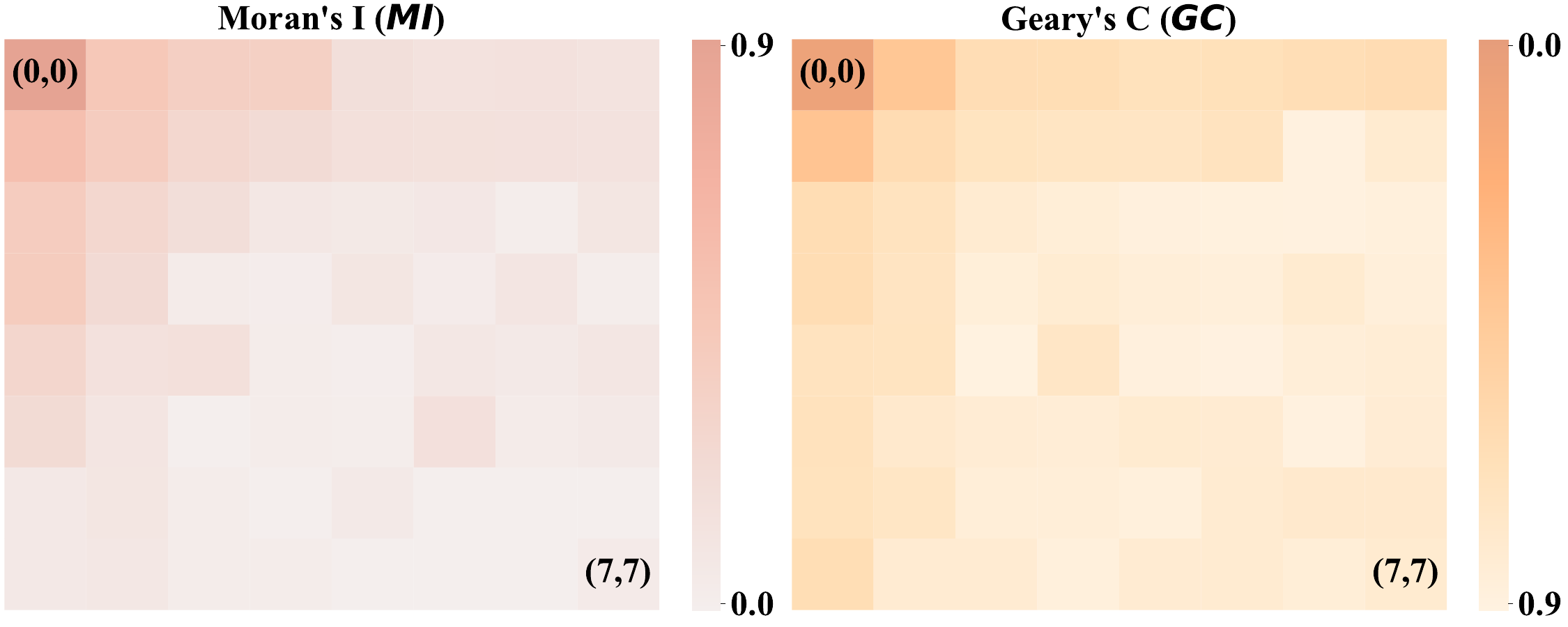}
        \caption{Cb of BSDS500}
    \end{subfigure}
    \begin{subfigure}{0.33\linewidth}
        \centering
        \includegraphics[trim={0mm 0mm 0mm 0mm}, clip, width=\linewidth]{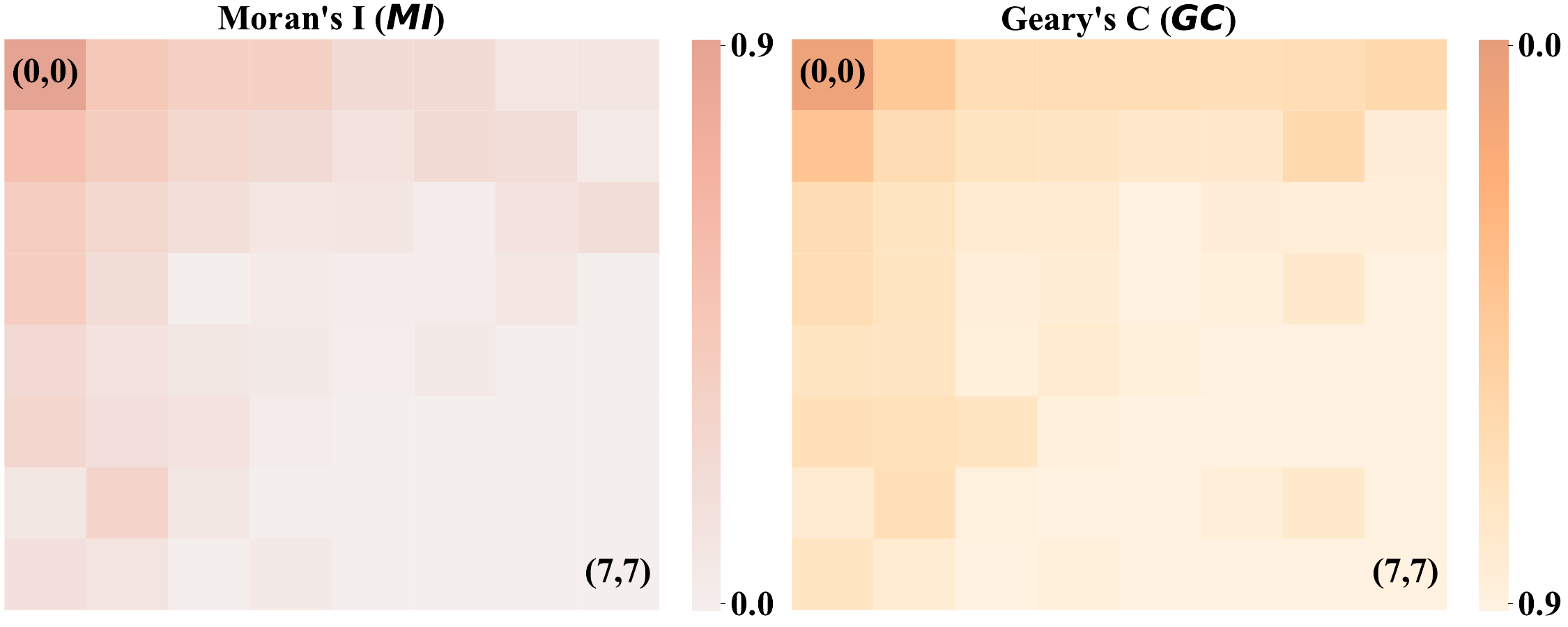}
        \caption{Cr of BSDS500}
    \end{subfigure}
    \vspace{\vspacelength}
    \caption{\textbf{Point-based correlations using coefficient maps on the DIV2K and BSDS500 datasets with QF set to 80.}
    Note that the intensity of heat maps indicate the strength of the correlations.}
    \label{fig:point_based_corr_supp_80}
\end{figure*}

\begin{figure*}[htbp]
    \centering
    \begin{subfigure}{0.33\linewidth}
        \centering
        \includegraphics[trim={9mm 102mm 22mm 109mm}, clip, width=\linewidth]{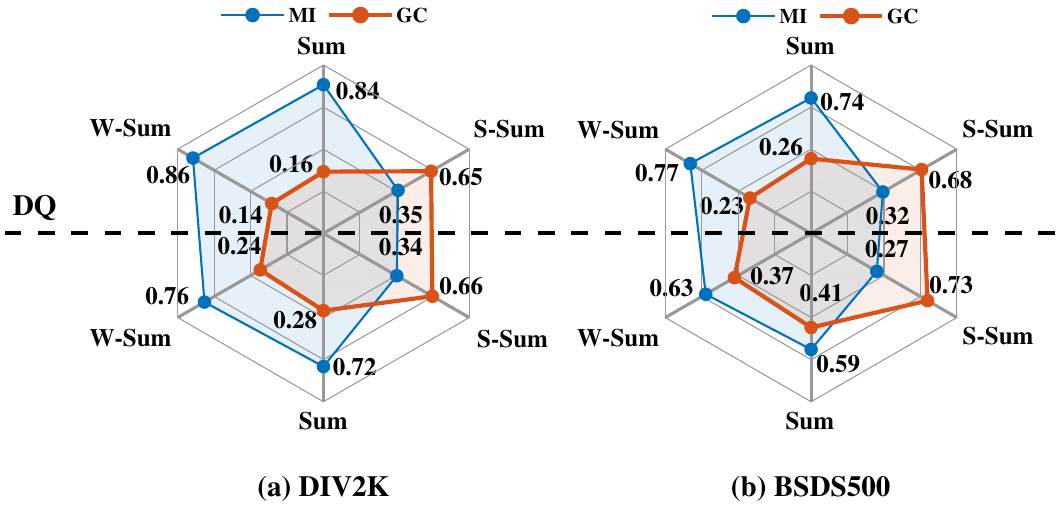}
        \caption{Y}
    \end{subfigure}
    \begin{subfigure}{.33\linewidth}
        \centering
        \includegraphics[trim={9mm 102mm 22mm 109mm}, clip, width=\linewidth]{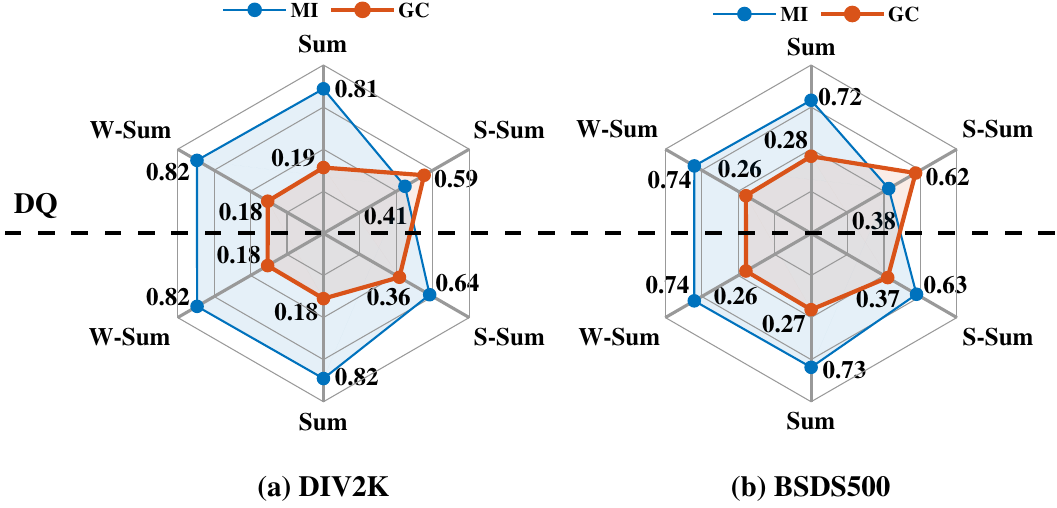}
        \caption{Cb}
    \end{subfigure}
    \begin{subfigure}{.33\linewidth}
        \centering
        \includegraphics[trim={9mm 102mm 22mm 109mm}, clip, width=\linewidth]{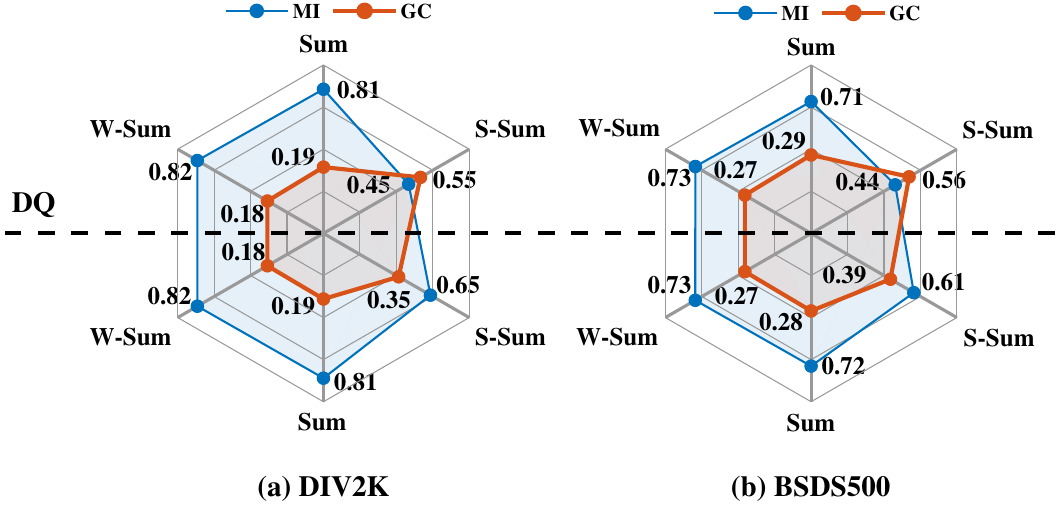}
        \caption{Cr}
    \end{subfigure}
    \vspace{\vspacelength}
    \caption{\textbf{Block-based correlations using different block-based features on the DIV2K and BSDS500 datasets with QF set to 90.}
    Upper: DCT blocks are dequantized before calculating feature values.
    Lower: DCT blocks remain quantized.}
    \label{fig:block_based_corr_supp_90}
\end{figure*}

\begin{figure*}[htbp]
    \centering
    \begin{subfigure}{0.33\linewidth}
        \centering
        \includegraphics[trim={0mm 0mm 0mm 0mm}, clip, width=\linewidth]{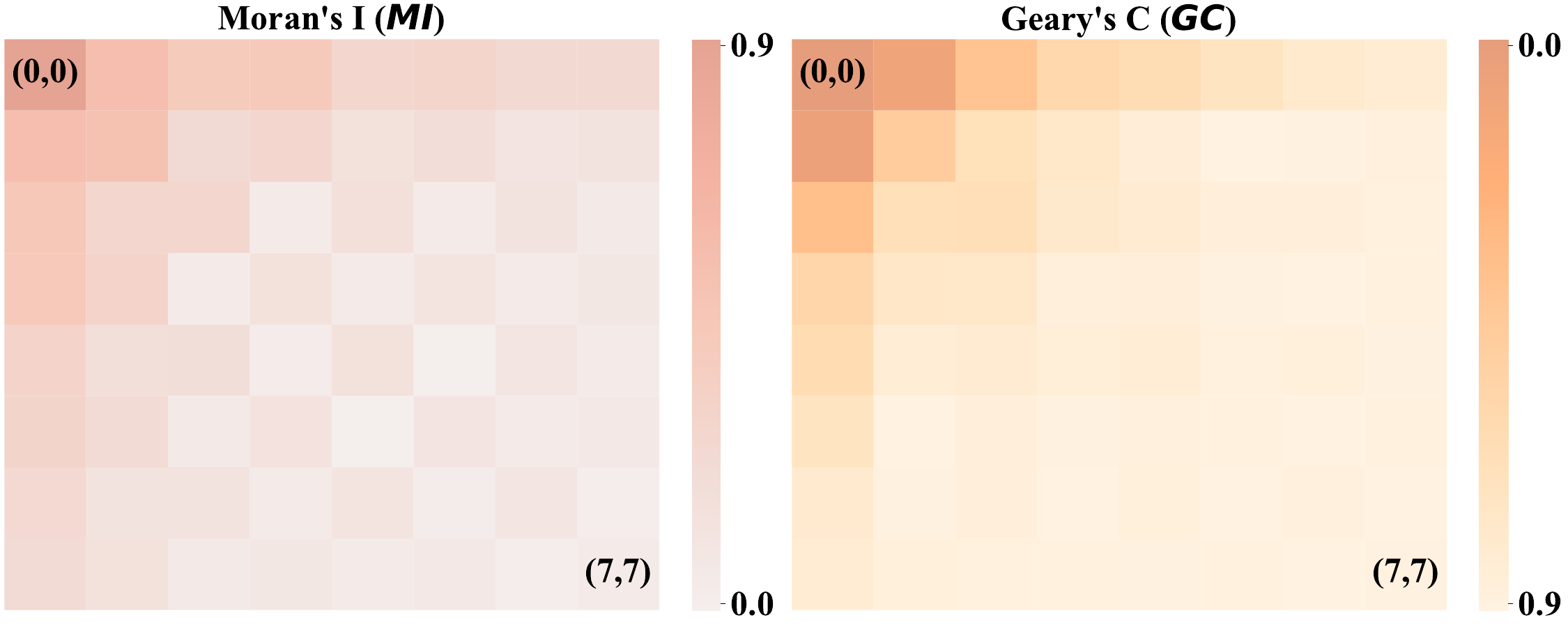}
        \caption{Y of DIV2K}
    \end{subfigure}
    \begin{subfigure}{0.33\linewidth}
        \centering
        \includegraphics[trim={0mm 0mm 0mm 0mm}, clip, width=\linewidth]{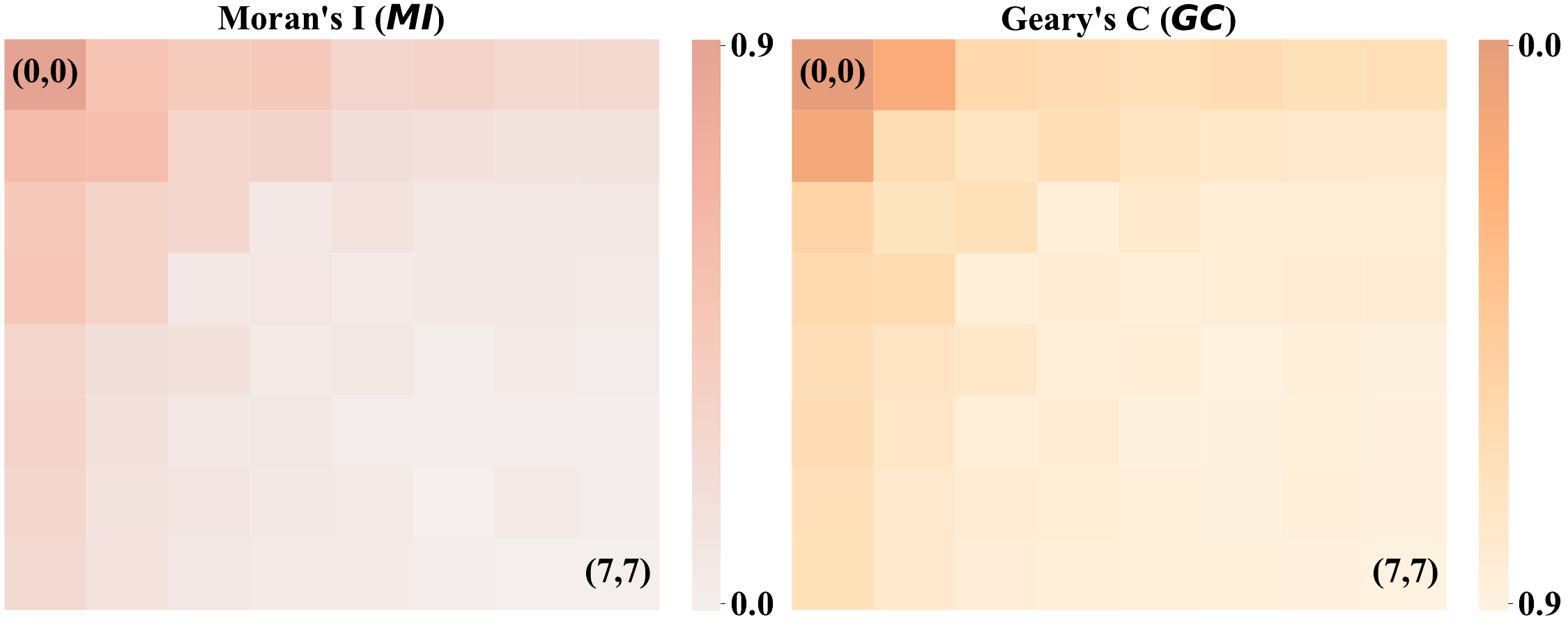}
        \caption{Cb of DIV2K}
    \end{subfigure}
    \begin{subfigure}{0.33\linewidth}
        \centering
        \includegraphics[trim={0mm 0mm 0mm 0mm}, clip, width=\linewidth]{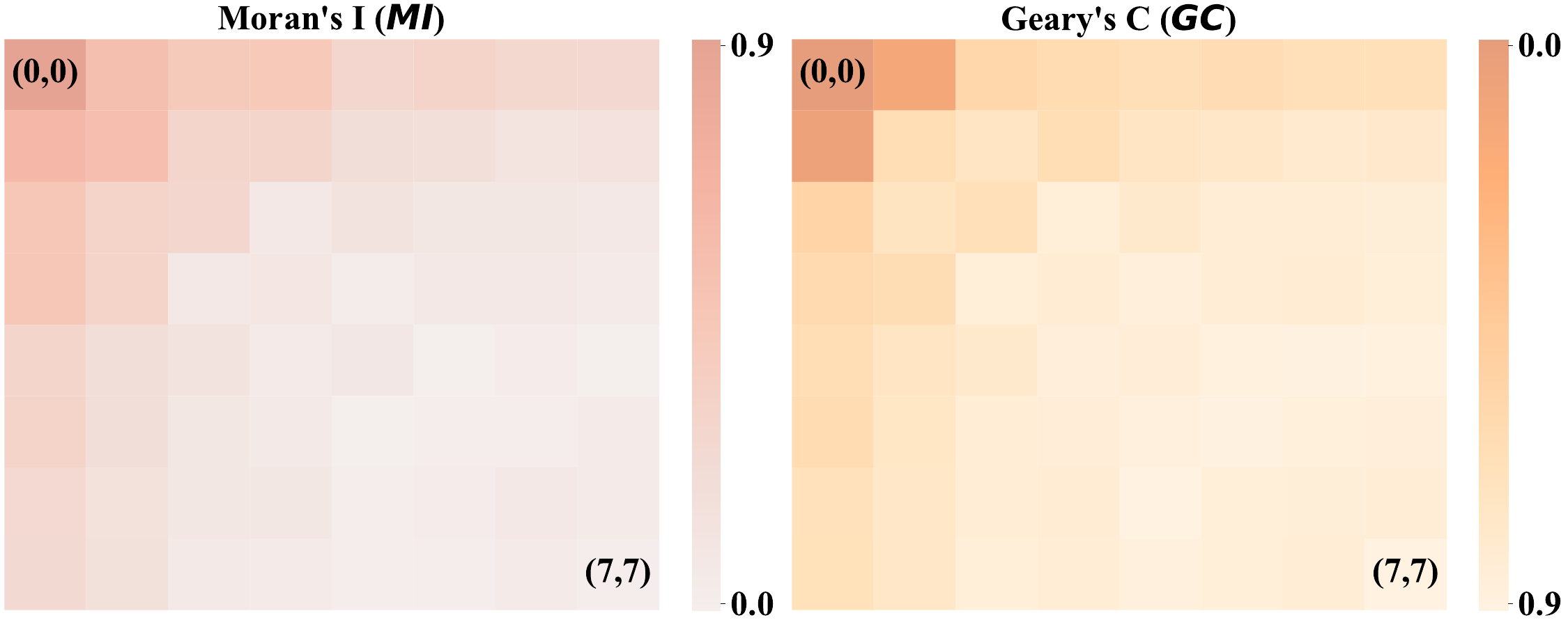}
        \caption{Cr of DIV2K}
    \end{subfigure}
    \begin{subfigure}{0.33\linewidth}
        \centering
        \includegraphics[trim={0mm 0mm 0mm 0mm}, clip, width=\linewidth]{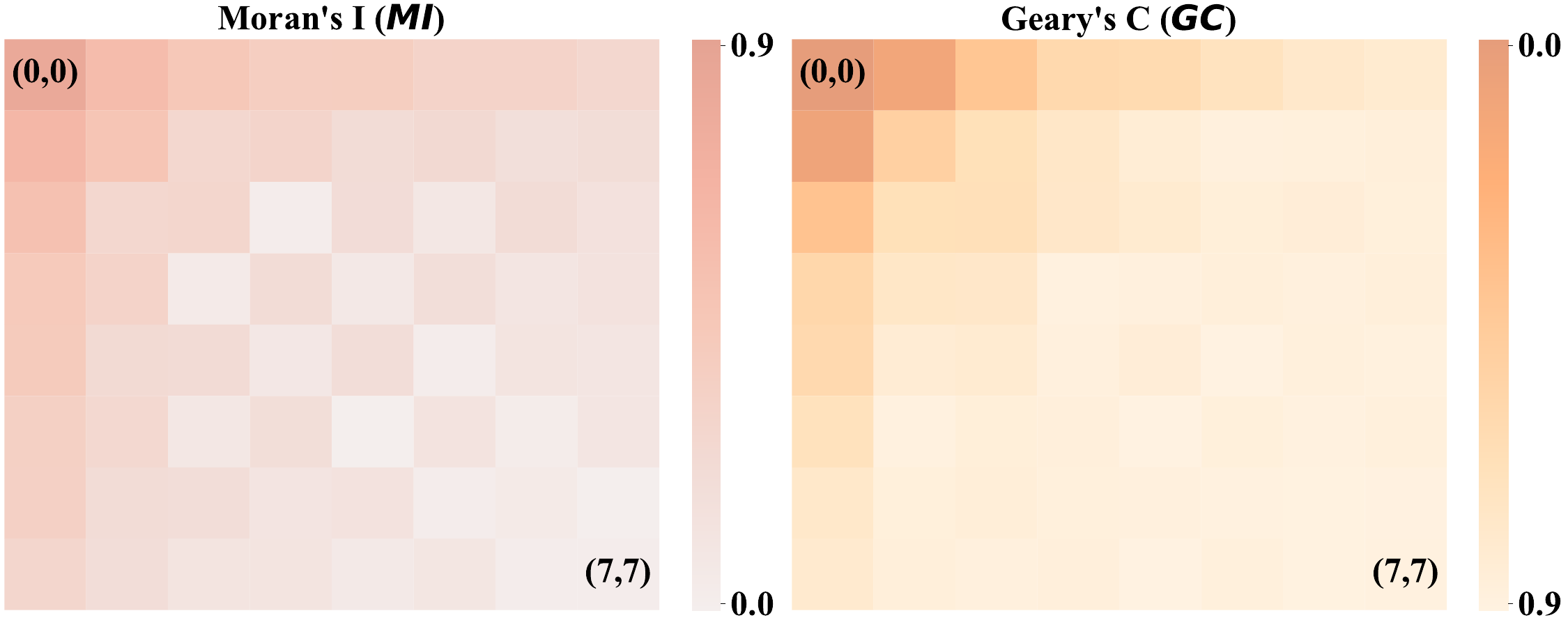}
        \caption{Y of BSDS500}
    \end{subfigure}
    \begin{subfigure}{0.33\linewidth}
        \centering
        \includegraphics[trim={0mm 0mm 0mm 0mm}, clip, width=\linewidth]{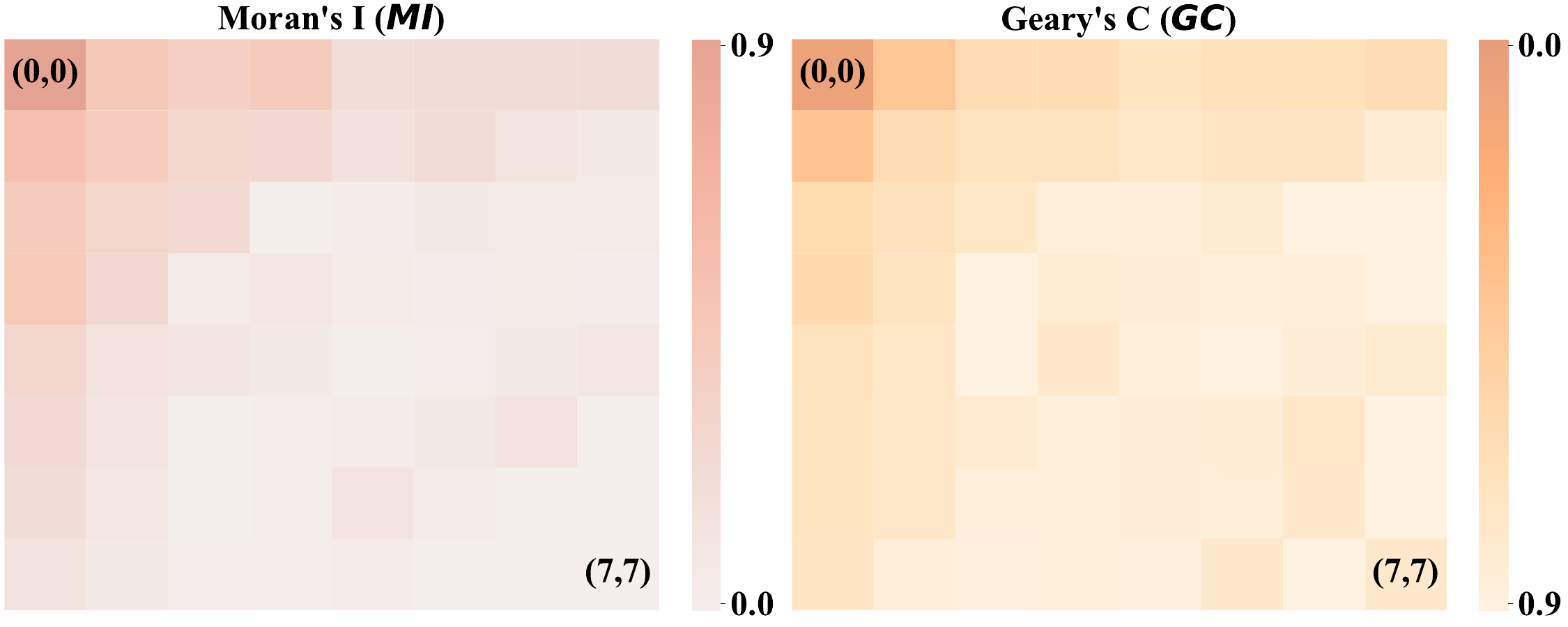}
        \caption{Cb of BSDS500}
    \end{subfigure}
    \begin{subfigure}{0.33\linewidth}
        \centering
        \includegraphics[trim={0mm 0mm 0mm 0mm}, clip, width=\linewidth]{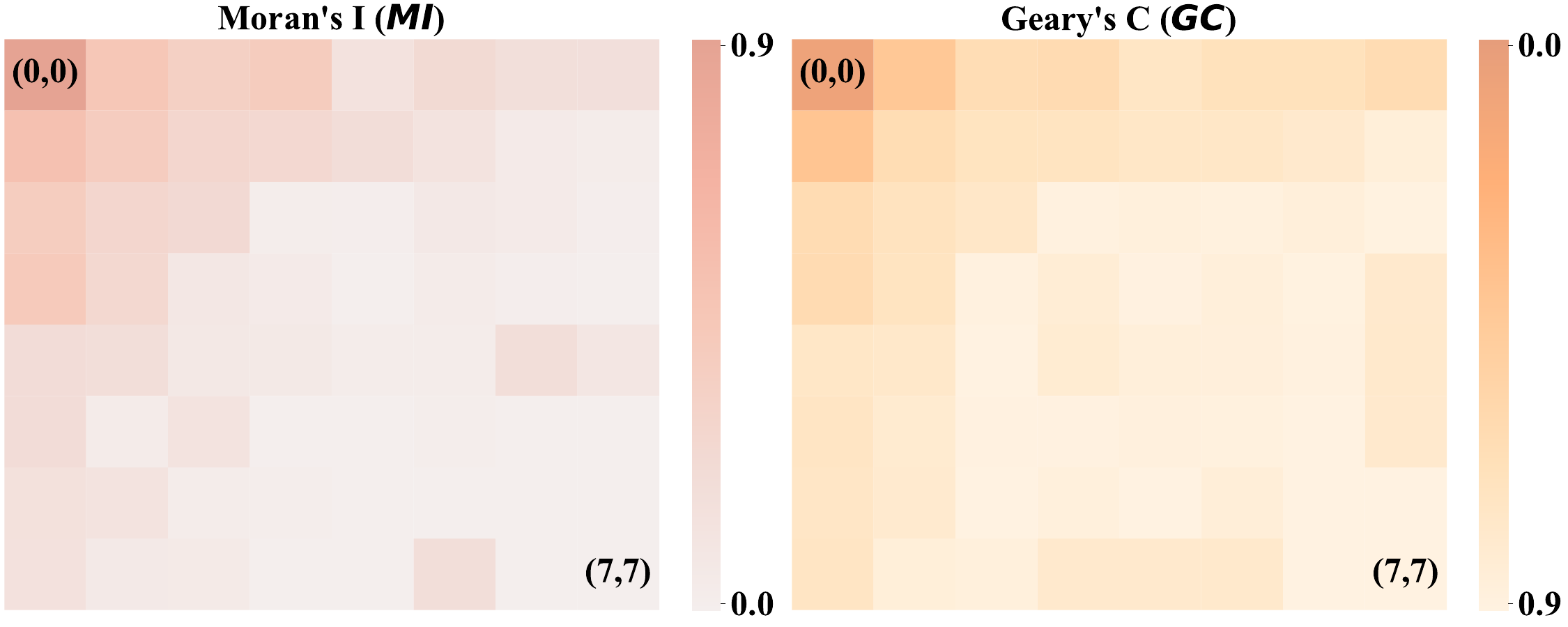}
        \caption{Cr of BSDS500}
    \end{subfigure}
    \vspace{\vspacelength}
    \caption{\textbf{Point-based correlations using coefficient maps on the DIV2K and BSDS500 datasets with QF set to 90.}
    Note that the intensity of heat maps indicate the strength of the correlations.}
    \label{fig:point_based_corr_supp_90}
\end{figure*}

\begin{figure*}[htbp]
    \centering
    \begin{subfigure}{0.33\linewidth}
        \centering
        \includegraphics[trim={9mm 102mm 22mm 109mm}, clip, width=\linewidth]{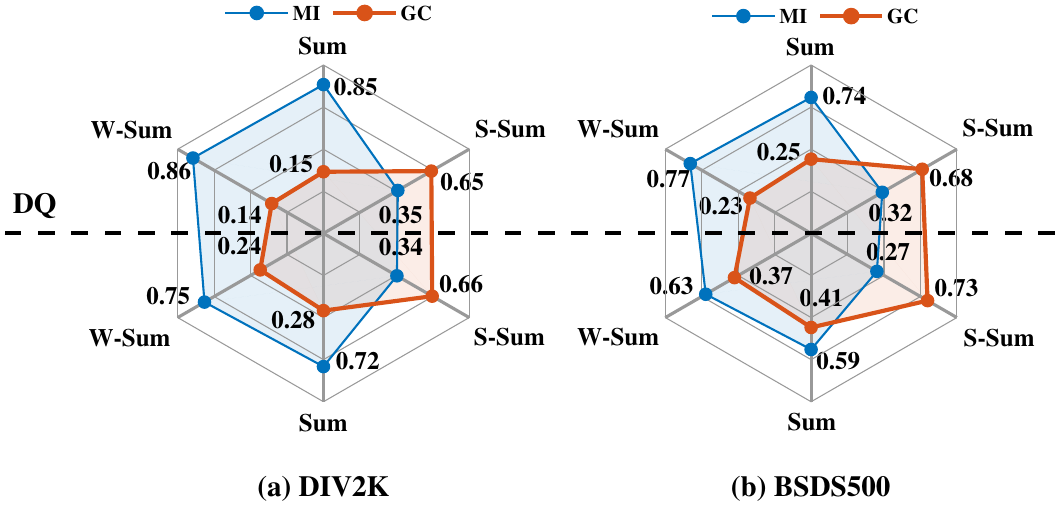}
        \caption{Y}
    \end{subfigure}
    \begin{subfigure}{.33\linewidth}
        \centering
        \includegraphics[trim={9mm 102mm 22mm 109mm}, clip, width=\linewidth]{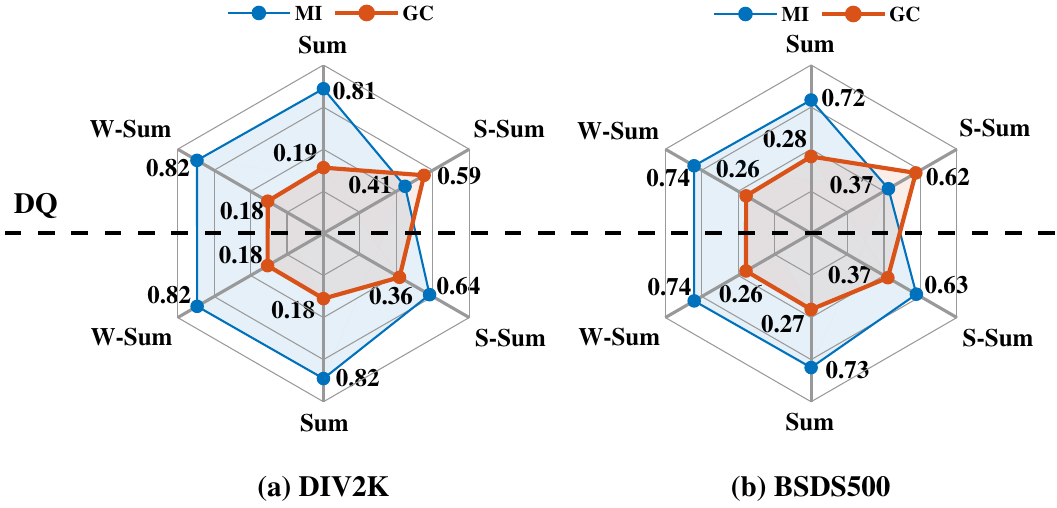}
        \caption{Cb}
    \end{subfigure}
    \begin{subfigure}{.33\linewidth}
        \centering
        \includegraphics[trim={9mm 102mm 22mm 109mm}, clip, width=\linewidth]{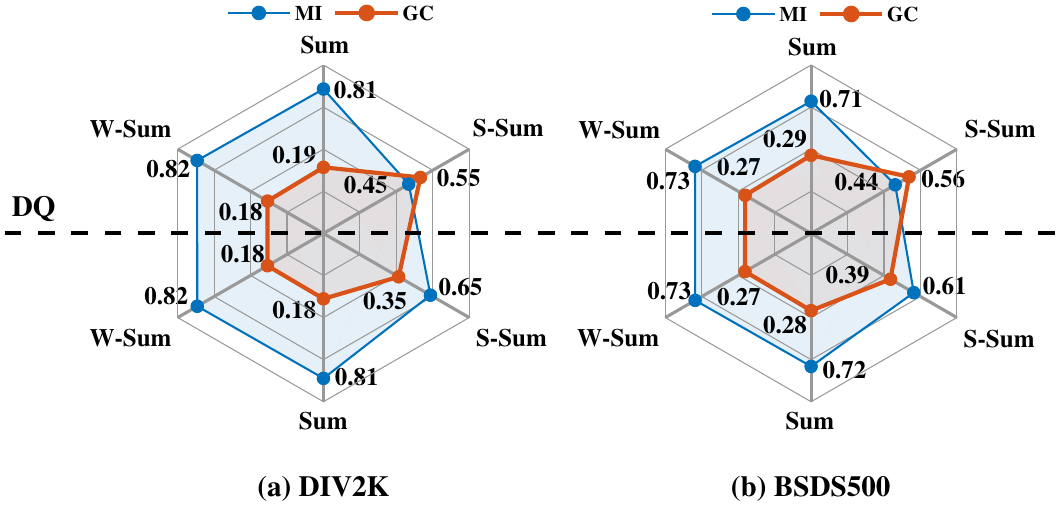}
        \caption{Cr}
    \end{subfigure}
    \vspace{\vspacelength}
    \caption{\textbf{Block-based correlations using different block-based features on the DIV2K and BSDS500 datasets with QF set to 100.}
    Upper: DCT blocks are dequantized before calculating feature values.
    Lower: DCT blocks remain quantized.}
    \label{fig:block_based_corr_supp_100}
\end{figure*}

\begin{figure*}[htbp]
    \centering
    \begin{subfigure}{0.33\linewidth}
        \centering
        \includegraphics[trim={0mm 0mm 0mm 0mm}, clip, width=\linewidth]{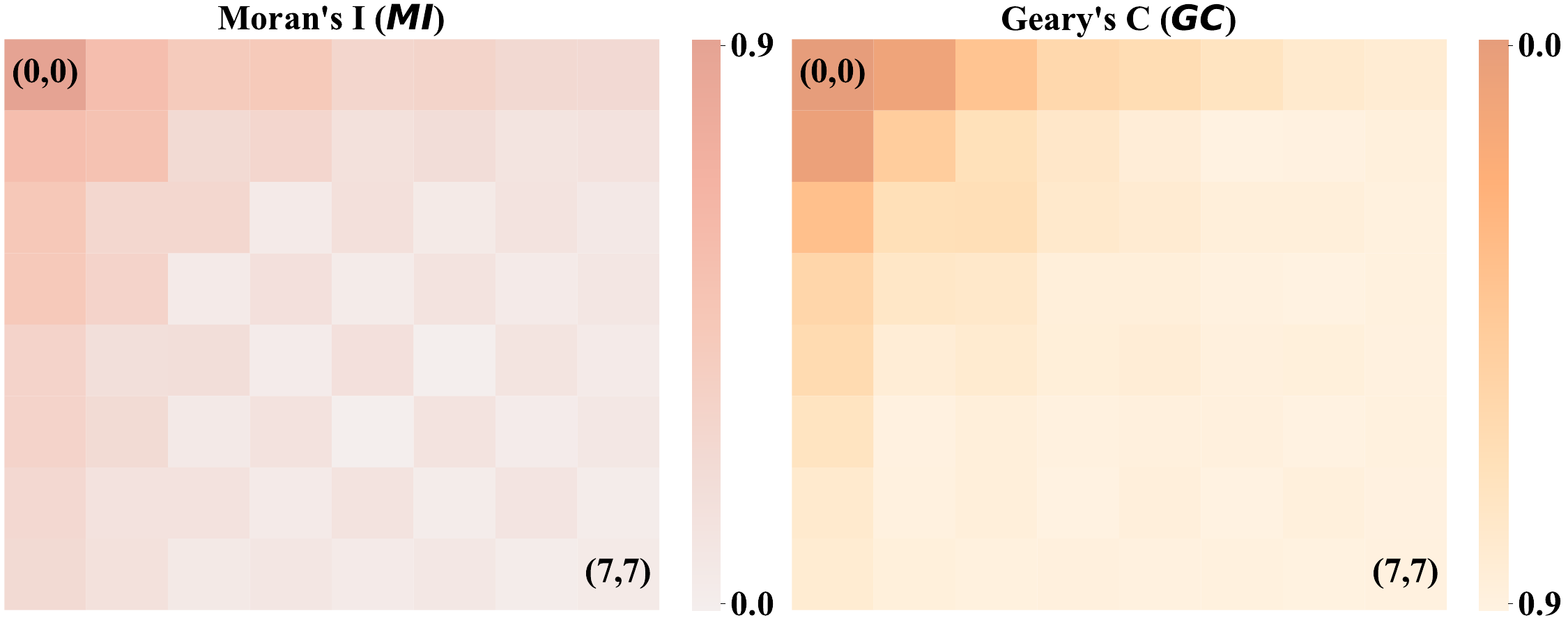}
        \caption{Y of DIV2K}
    \end{subfigure}
    \begin{subfigure}{0.33\linewidth}
        \centering
        \includegraphics[trim={0mm 0mm 0mm 0mm}, clip, width=\linewidth]{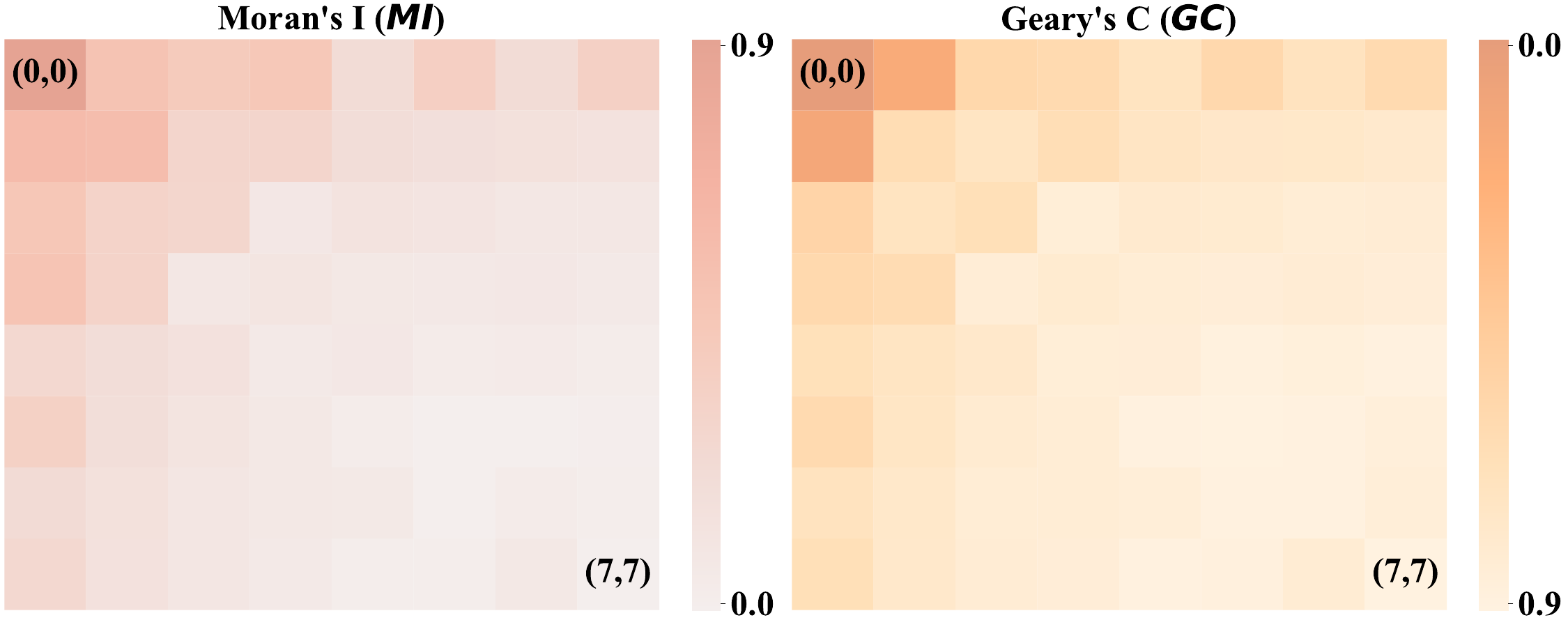}
        \caption{Cb of DIV2K}
    \end{subfigure}
    \begin{subfigure}{0.33\linewidth}
        \centering
        \includegraphics[trim={0mm 0mm 0mm 0mm}, clip, width=\linewidth]{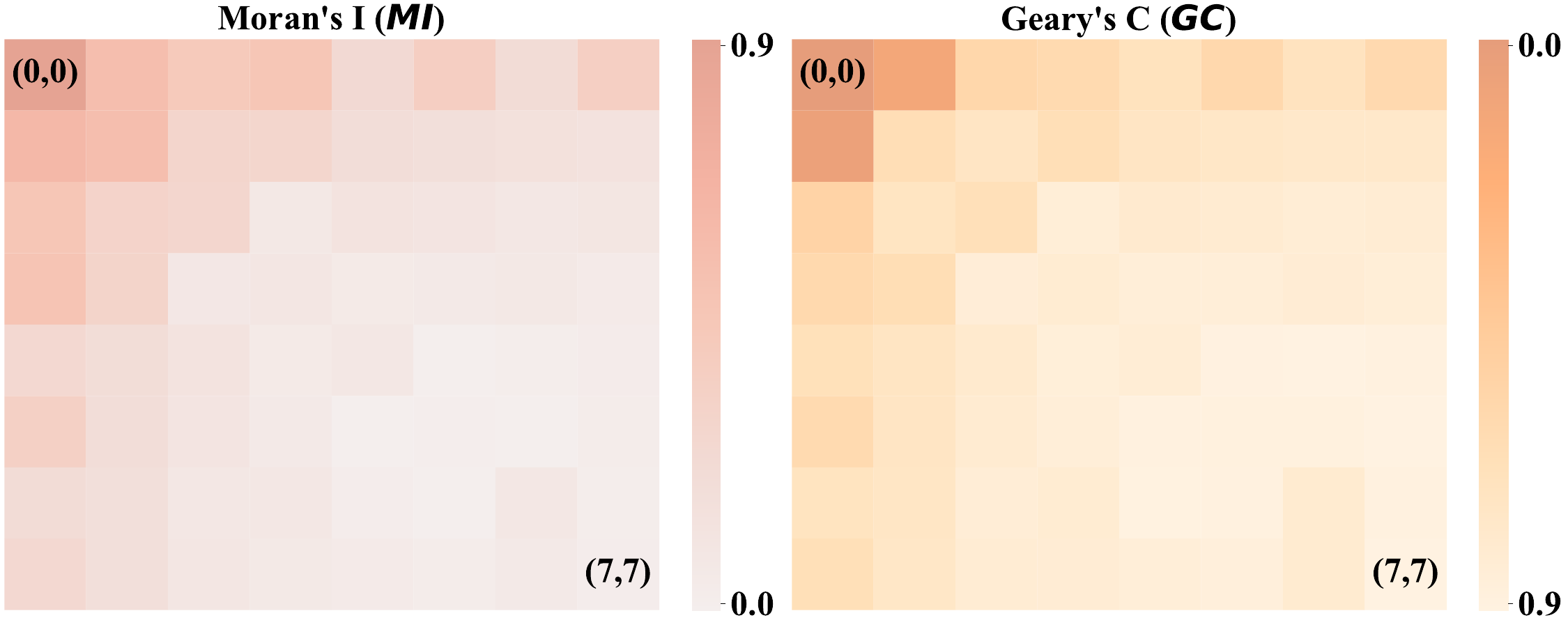}
        \caption{Cr of DIV2K}
    \end{subfigure}
    \begin{subfigure}{0.33\linewidth}
        \centering
        \includegraphics[trim={0mm 0mm 0mm 0mm}, clip, width=\linewidth]{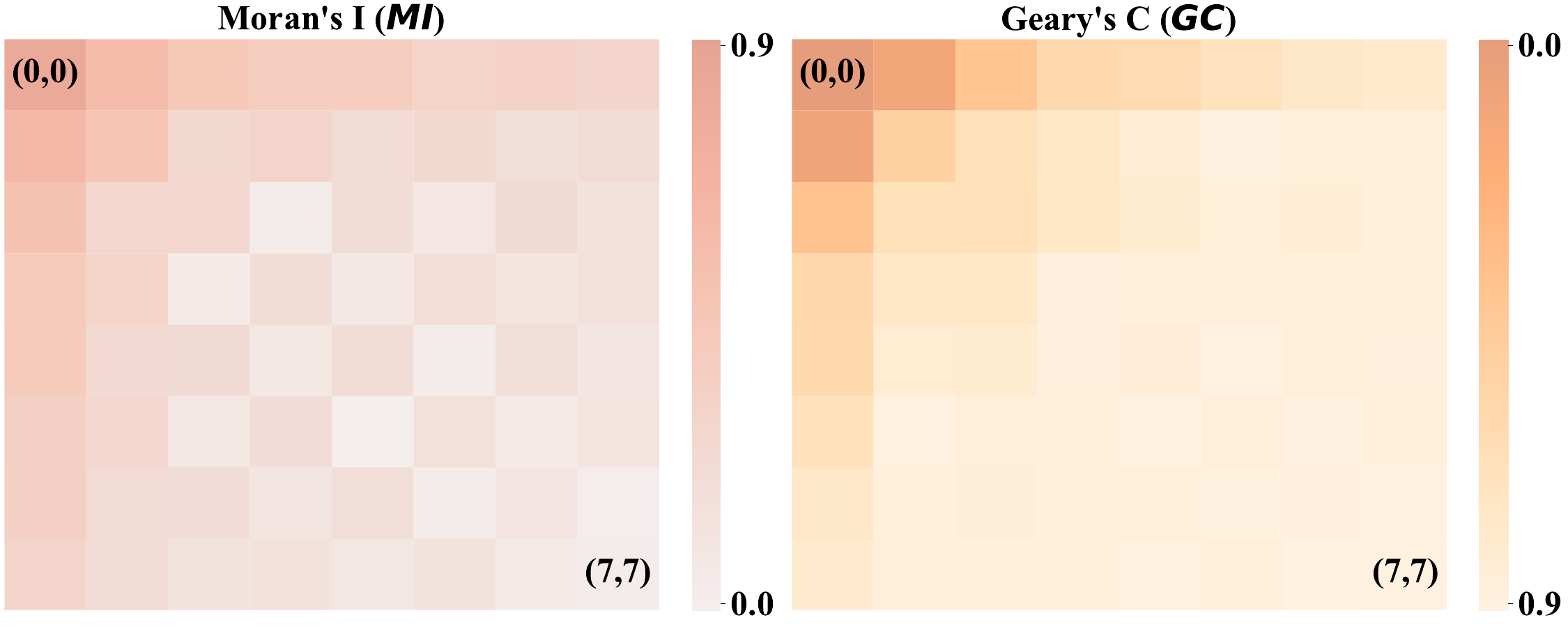}
        \caption{Y of BSDS500}
    \end{subfigure}
    \begin{subfigure}{0.33\linewidth}
        \centering
        \includegraphics[trim={0mm 0mm 0mm 0mm}, clip, width=\linewidth]{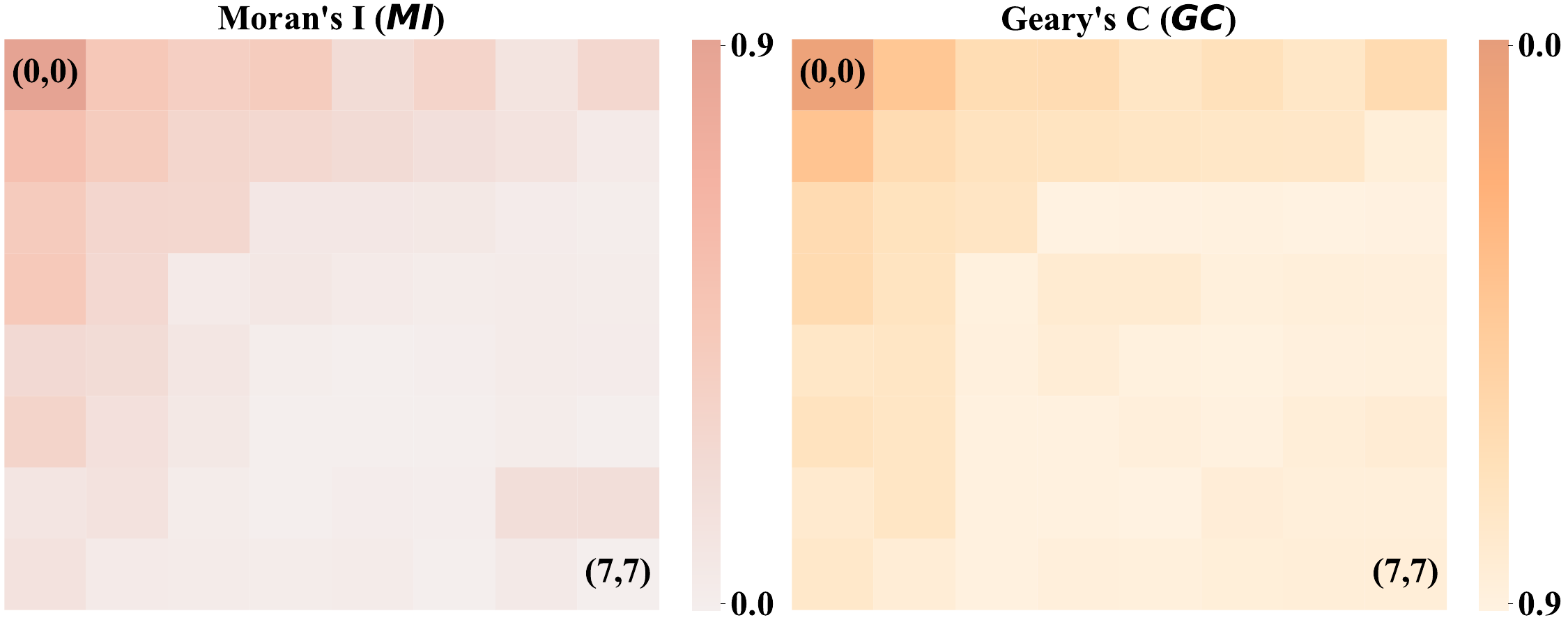}
        \caption{Cb of BSDS500}
    \end{subfigure}
    \begin{subfigure}{0.33\linewidth}
        \centering
        \includegraphics[trim={0mm 0mm 0mm 0mm}, clip, width=\linewidth]{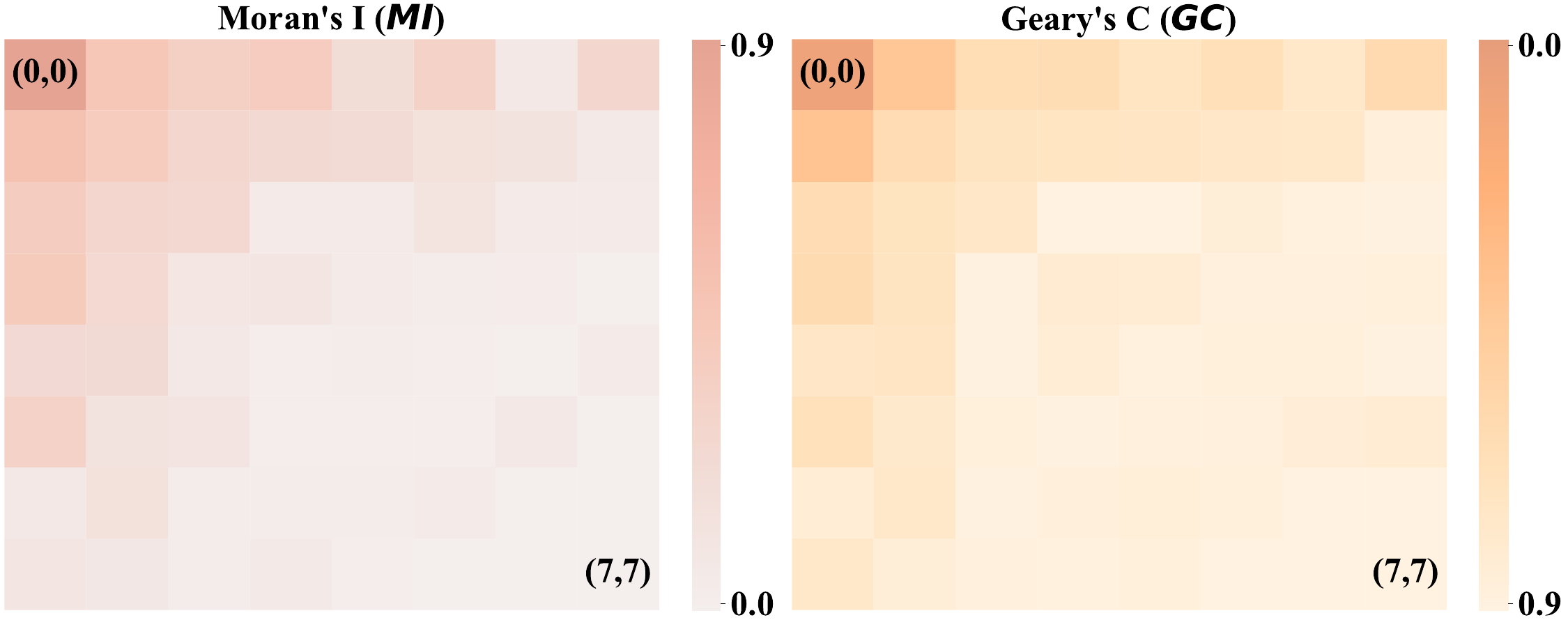}
        \caption{Cr of BSDS500}
    \end{subfigure}
    \vspace{\vspacelength}
    \caption{\textbf{Point-based correlations using coefficient maps on the DIV2K and BSDS500 datasets with QF set to 100.}
    Note that the intensity of heat maps indicate the strength of the correlations.}
    \label{fig:point_based_corr_supp_100}
\end{figure*}

In the main paper, we highlighted two critical correlations within the DCT coefficients of the luminance components of JPEG images at a QF of 50.
This section extends our observations to include both the luminance and chroma components across QF values ranging from 10 to 100.
\cref{tab:corre_supp} demonstrates the consistently strong correlations present within pixel values of both luminance and chroma components across all QF values, whereas the DCT coefficients exhibit relatively weaker spatial correlations.
\crefrange{fig:block_based_corr_supp_10}{fig:point_based_corr_supp_100} illustrate the substantial block-based and point-based correlations within DCT coefficients, further reinforcing the consistency of our findings presented in the main paper.

% QF=10

% QF=20

% QF=30

% QF=40

% QF=50

% QF=60

% QF=70

% QF=80

% QF=90

% QF=100

\end{document}